\documentclass[twocolumn,hyperpdf,amsmath,amssymb,aps,prd,10pt,superscriptaddress,nofootinbib,noeprint,preprintnumbers,floatfix]{revtex4-1}
\usepackage{graphicx, color}
\usepackage[letterspace=-10]{microtype} 
\usepackage{bm, amsmath, amsfonts}
\usepackage{makecell,multirow, tabularx, dcolumn}
\usepackage{afterpage} 
\usepackage{float}
\usepackage[utf8]{inputenc} 
\usepackage{hyperref}

\definecolor{jlab_red}{RGB}{192,39,45}
\definecolor{jlab_orange}{RGB}{249,102,0}
\definecolor{jlab_blue}{RGB}{47,122,121}
\definecolor{jlab_green}{RGB}{65,125,10}

\newcommand{\Kiso}[0]{\mathcal{K}_{3,{\sf iso}}}
\newcommand{\cK}[0]{\mathcal K}
\newcommand{\cM}[0]{\mathcal M}
\newcommand{\cD}[0]{\mathcal D}

\newcommand{\df}[0]{\mathrm{df}}

\newcommand{\ERrowNPCOV}[6]{{#1}${}_2$ & ${#2}$ & \ \ $ -1/\A + r_0 p^2/2$ \ \  &  \parbox[c][1.0cm][c]{5.0cm}{\begin{equation*} \begin{aligned}  m_\pi \A  & = {#3} \\[-3pt] m_\pi r_0 & = {#4}   \end{aligned}   \  \bigg [ \begin{aligned} \!  1.0 \ &   {#6}  \\[-3pt]   & \phantom{+}1.0  \end{aligned} \bigg ] \end{equation*} } & ${#5}$}

\newcommand{\AFProwNPCOV}[6]{{#1}${}_2$ & ${#2}$ & \ \ $-A(c_0,p)/\A$ \ \  &     \parbox[c][1.0cm][c]{5.0cm}{\begin{equation*} \begin{aligned}  m_\pi \A  & = {#3} \\[-3pt] m_\pi c_0 & = {#4}   \end{aligned}   \  \bigg [ \begin{aligned}  \!  1.0 \ &   {#6}  \\[-3pt]   & \phantom{+}1.0  \end{aligned} \bigg ] \end{equation*} }     & ${#5}$}
\newcommand{\ArowNPCOV}[6]{{#1}${}_2$ & ${#2}$ & \ \ $A(1,p)(-1/\A + c_0 p^2)$ \ \  &     \parbox[c][1.0cm][c]{5.0cm}{\begin{equation*} \begin{aligned}  m_\pi \A  & = {#3} \\[-3pt] m_\pi c_0 & = {#4}   \end{aligned}   \  \bigg [ \begin{aligned} \!   1.0 \ &   {#6}  \\[-3pt]   & \phantom{+}1.0  \end{aligned} \bigg ] \end{equation*} }  & ${#5}$}

\newcommand{\ERrowRvalNPCOV}[6]{ {#1} & ${#2}$ & \ \ $ -1/\A + r_0 p^2/2$ \ \  &    \parbox[c][1.0cm][c]{5.0cm}{\begin{equation*} \begin{aligned}  m_\pi \A  & = {#3} \\[-3pt] m_\pi r_0 & = {#4}   \end{aligned}   \  \bigg [ \begin{aligned}  \! 1.0 \ &   {#6}  \\[-3pt]   & \phantom{+}1.0  \end{aligned} \bigg ] \end{equation*} }  & ${#5}$}

\newcommand{\A}[0]{a_0}
\newcommand{\SecIntro}[0]{\emph{Introduction}}
\newcommand{\SecKMat}[0]{\emph{Analyzing the finite-volume spectra}}
\newcommand{\SecSpec}[0]{\emph{Spectral Determination}}

\renewcommand{\vec}[0]{\boldsymbol}
\hypersetup{%
pdftitle = {The energy-dependent $\pi^+ \pi^+ \pi^+$ scattering amplitude from QCD},
pdfsubject = {QCD},
pdfauthor = {Hadron Spectrum Collaboration},
colorlinks = {true},
filecolor = {black},
linkcolor = {jlab_blue},
menucolor = {black},
citecolor = {jlab_green},
urlcolor = {jlab_green},
}{}

\begin{document}

\preprint{CERN-TH-2020-147, JLAB-THY-20-3242}

\title{
The energy-dependent $\pi^+ \pi^+ \pi^+$ scattering amplitude from QCD
}
\author{Maxwell T. Hansen}
\email[]{maxwell.hansen@cern.ch}
\affiliation{Theoretical Physics Department, CERN, 1211 Geneva 23, Switzerland}
\author{Raul~A.~Brice\~{n}o}
\email{rbriceno@jlab.org}
\affiliation{\lsstyle Thomas Jefferson National Accelerator Facility, 12000 Jefferson Avenue, Newport News, VA 23606, USA}
\affiliation{\lsstyle Department of Physics, Old Dominion University, Norfolk, VA 23529, USA}
\author{Robert~G.~Edwards}
\email{edwards@jlab.org}
\affiliation{\lsstyle Thomas Jefferson National Accelerator Facility, 12000 Jefferson Avenue, Newport News, VA 23606, USA}
\author{Christopher~E.~Thomas}
\email{c.e.thomas@damtp.cam.ac.uk}
\affiliation{DAMTP, University of Cambridge, Wilberforce Road, Cambridge, CB3 0WA, UK}
\author{David~J.~Wilson}
\email{d.j.wilson@damtp.cam.ac.uk}
\affiliation{DAMTP, University of Cambridge, Wilberforce Road, Cambridge, CB3 0WA, UK}

\collaboration{for the Hadron Spectrum Collaboration}
\date{\today}

\begin{abstract}
Focusing on three-pion states with maximal isospin ($\pi^+\pi^+\pi^+$), we present the first non-perturbative determination of an energy-dependent three-hadron scattering amplitude from first-principles QCD. The calculation combines finite-volume three-hadron energies, extracted using numerical lattice QCD, with a relativistic finite-volume formalism, required to interpret the results. To fully implement the latter, we also solve integral equations that relate an intermediate three-body K matrix to the physical three-hadron scattering amplitude. The resulting amplitude shows rich analytic structure and a complicated dependence on the two-pion invariant masses, represented here via Dalitz-like plots of the scattering rate.
\end{abstract}

\maketitle

\noindent
\SecIntro~---~The three-body problem lies at the core of a broad range of outstanding questions in quantum chromodynamics (QCD). The largest uncertainty in QCD-based structure calculations of light nuclei, for example, is the estimate of the three-nucleon force (see Ref.~\cite{Piarulli:2017dwd}). In addition, many QCD resonances have significant branching fraction to channels with three or more hadrons. The Roper resonance, for example, has defied simple quark-model descriptions, due in part to its nature as a broad resonance with a $\sim 30\%$ branching fraction to $N\pi\pi$. A rigorous QCD calculation would elucidate the role of non-perturbative dynamics in the Roper's peculiar properties, e.g.~the fact that it has a lower mass than the negative-parity ground state, which seems unnatural from the perspective of the quark model~\cite{Isgur:1977ef,Isgur:1978wd}. 

As a necessary step towards studying a broad class of three-hadron systems, in this work we present the first study of an energy-dependent three-body scattering amplitude from QCD. This non-perturbative result is achieved by the coalescence of three novel techniques: a calculation of finite-volume three-hadron energies based in numerical lattice QCD, a relativistic finite-volume formalism to relate the energies to K matrices, and a numerical evaluation of corresponding integral equations to convert the latter into the three-hadron scattering amplitude. The theoretical basis required to achieve these final two steps was derived in Refs.~\cite{Hansen:2014eka,Hansen:2015zga}.%
\footnote{%
A large body of work has investigated general methods for relating finite-volume energies to scattering amplitudes for both two- and three-body states. See Refs.~%
\cite{Luscher:1990ux, Rummukainen:1995vs, Bedaque:2004kc, Kim:2005gf, Fu:2011xz, Leskovec:2012gb, Gockeler:2012yj, He:2005ey, Lage:2009zv, Bernard:2010fp, Doring:2011vk, Doring:2011nd, Agadjanov:2013kja, Doring:2012eu, Hansen:2012tf, Briceno:2012yi, Guo:2012hv,Briceno:2017max,Briceno:2014oea}
and Refs.~%
\cite{Romero-Lopez:2019qrt, Briceno:2018aml, Briceno:2017tce, Hansen:2019nir, Briceno:2018mlh, Blanton:2019igq, Hammer:2017uqm, Hammer:2017kms, Briceno:2012rv, Polejaeva:2012ut,Mai:2017bge, Guo:2017ism, Guo:2017crd, Guo:2018xbv, Romero-Lopez:2018rcb,Jackura:2019bmu,Briceno:2019muc,Hansen:2020zhy,Blanton:2020jnm,Blanton:2020gha,Beane:2020ycc}%
, respectively.}

This work considers the scattering of three-pion states with maximal isospin $(I = 3)$ in QCD with three dynamical quarks ($N_f = 2 + 1$): two degenerate light quarks, with heavier-than-physical mass corresponding to a pion mass $m_\pi \approx 391 \, \text{MeV}$, and a strange quark. 
This channel offers an optimal benchmark case, since both the maximal-isospin three-pion system and its two-pion subsystem are expected to be weakly interacting and non-resonant.

Many numerical studies of three-hadron states have been published over the last decade, ranging from early work deriving and fitting large-volume expansions of the three-pion ground state \cite{Beane:2007es,Detmold:2008fn,Detmold:2012wc} to more recent results using quantization conditions to study ground \cite{Mai:2018djl} and excited states \cite{Blanton:2019vdk,Mai:2019fba,Guo:2020kph}, with the latter set each analyzing the lattice QCD spectrum published in Ref.~\cite{Horz:2019rrn}. Independent sets of finite-volume energies have also been calculated and analyzed in Refs.~\cite{Culver:2019vvu} and \cite{Fischer:2020jzp}. The present investigation goes beyond this previous work, by providing the first complete numerical determination of physical scattering amplitudes for three-body systems. 

In the following, we first discuss our determination of two- and three-pion finite-volume energies, before describing the fits used to relate these to infinite-volume K matrices. The latter then serve as inputs to known integral equations, which we solve numerically to extract the $3\pi^+ \to 3\pi^+$ scattering amplitude. Additional details of the analysis are discussed in the supplemental material.

\begin{figure}
\hspace{0pt} \includegraphics[width=1.0\columnwidth]{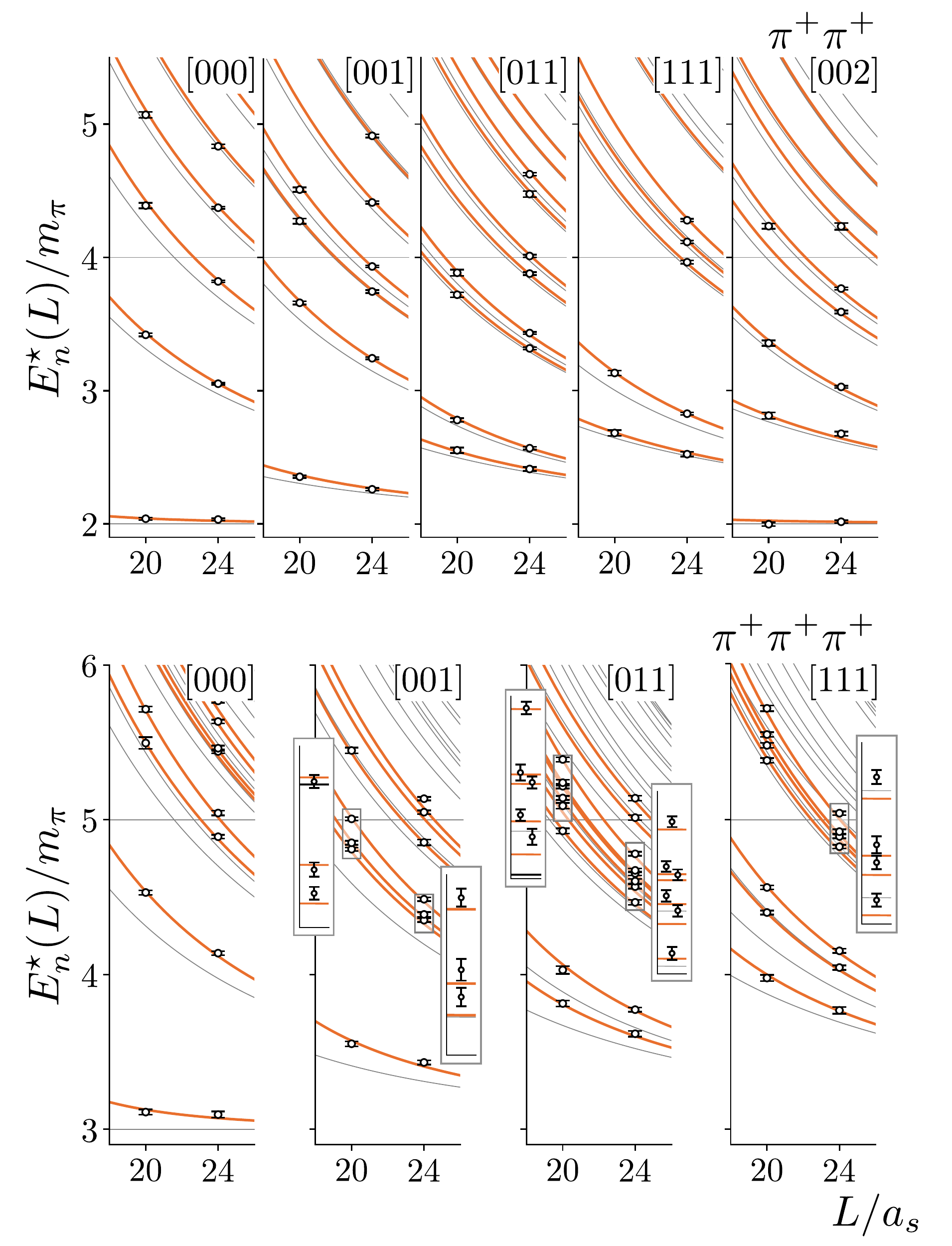}
\caption{The $\pi^+ \pi^+$ and $\pi^+ \pi^+ \pi^+$ finite-volume spectra in the center-of-momentum frame for the relevant finite-volume irreps with various overall momenta, as explained in the text. Points are computed energy levels on the two volumes with error bars showing statistical uncertainties. Each rectangular insert shows a vertical zoom of the region indicated by the small neighboring rectangle. Grey curves are the ``non-interacting'' finite-volume energies, i.e.~the energies in the absence of any interactions between pions. Orange curves are predictions from the finite-volume formalism based only on the two-particle scattering length, given in Eq.~\eqref{eq:bestfit} (here with the local three-body interaction set to zero).}
\label{fig:FVspec}
\end{figure}

\bigskip
\noindent
\SecSpec~---~Figure~\ref{fig:FVspec} summarizes the two- and three-pion finite-volume spectra calculated in this work.\footnote{Two-pion energies on the larger volume have already appeared in Ref.~\cite{Dudek:2012gj}.}

Computations were performed on anisotropic lattices which have a temporal lattice spacing, $a_t$, finer than the spatial lattice spacing, $a_s$ ($a_t = a_s/\xi$ with $\xi =3.444(6)$~\cite{Dudek:2012gj}). Two lattice ensembles were used, differing only in the volume: $(L/a_s)^3 \times (T/a_t) = 20^3 \times 256$ (with 256 gauge-field configurations) and $24^3 \times 128$ (with 512 configurations). We use $2+1$ flavors of dynamical clover fermions, with three-dimensional stout-link smearing in the fermion action, and a tree-level Symanzik-improved gauge action. The bare parameters and basic lattice properties are detailed in Refs.~\cite{Edwards:2008ja,Lin:2008pr}. 
Setting the scale via ${a_t^{-1}= m_\Omega^{\mathrm{exp}} \, (a_t m_\Omega^{\mathrm{latt}}} )^{-1}$,%
\footnote{where $a_t m_\Omega^{\mathrm{latt}} = 0.2951(22)$ was measured in Ref.~\cite{Edwards:2011jj} and $m_\Omega^{\mathrm{exp}}$ is the experimentally determined $\Omega$ baryon mass from Ref.~\cite{Zyla:2020zbs}}
and combining with $a_t m_\pi =0.06906(13)$~\cite{Dudek:2012gj} and $a_t m_K = 0.09698(9)$~\cite{Wilson:2014cna}, yields $m_\pi \approx 391 \, \text{MeV}$ and $m_K \approx 550 \, \text{MeV}$. The values of $a_t m_\pi$ and $\xi$ translate into spatial extents of $m_\pi L = 4.76$ and $m_\pi L = 5.71$ for the two ensembles.

The spectrum of energies in a finite volume is discrete and each energy level provides a constraint on the scattering amplitudes at the corresponding center-of-momentum energy. To obtain more constraints, we compute spectra for systems with overall zero and non-zero momentum, $\vec{P}$. Momenta are quantized by the cubic spatial boundary conditions, $\vec{P} = \tfrac{2\pi}{L}(n_1, n_2, n_3)$, where $\{n_i\}$ are integers, and we write this using a shorthand notation as $[n_1 n_2 n_3]$.

In this work we restrict attention to $S$-wave scattering. The reduced symmetry of a cubic lattice means that total angular momentum, $J$, is not a good quantum number and instead channels are labelled by the irreducible representation (irrep, $\Lambda$) of the octahedral group with parity for $\vec{P} = \vec{0}$ or the relevant subgroup that leaves $\vec P$ invariant for $\vec{P} \neq \vec{0}$~\cite{Johnson:1982yq,Moore:2005dw}. We consider the relevant irreps which contain $J=0$: $A_1^- (A_1^+)$ for $\pi \pi \pi$ ($\pi \pi$) at rest and $A_2 (A_1)$ for $\pi \pi \pi$ ($\pi \pi$) with non-zero $\vec P$. Isospin, $I$, and $G$-parity, $G$, are good quantum numbers in our lattice formulation; these distinguish the two-pion ($I^G =2^+$) and three-pion ($I^G=3^-$) channels.
We neglect higher partial waves here, in particular the two-particle $D$-wave which mixes with the $S$-wave in the finite-volume energies. As described in Ref.~\cite{Dudek:2012gj}, a nonzero $D$-wave interaction can be extracted, in particular if aided by the consideration of other, non-trivial finite-volume irreps, but has a small influence on the two-pion energies considered here.\footnote{There is, in principle, a systematic uncertainty associated with neglecting the $D$-wave contribution. Given the consistency of our results with Ref.~\cite{Dudek:2012gj}, this appears to be below the statistical uncertainty in the present fits. See also Secs.~VIII A and B of that work for more discussion.}  %

To reliably extract the finite-volume energies we have computed two-point correlation functions using a large basis of appropriate interpolating operators. From these, the spectra are determined using the variational method~\cite{Michael:1985ne,Luscher:1990ck,Blossier:2009kd}, with our implementation described in Refs.~\cite{Dudek:2007wv,Dudek:2010wm}. This amounts to calculating a matrix of correlation functions,
\begin{equation}
G_{ij}(t) = \langle \mathcal O^{\vphantom{\dagger}}_i(t) \mathcal O^\dagger_j(0) \rangle \,,
\end{equation}
and diagonalizing $M(t,t_0) = G(t_0)^{-1/2} \cdot G(t) \cdot G(t_0)^{-1/2}$ for a fixed $t_0$. One can show that the corresponding eigenvalues satisfy $\lambda_n(t,t_0) \to e^{- E_n(L) (t-t_0)}$, where $E_n(L)$ is the $n$'th energy level with overlap to some of the operators in the basis. This basic methodology has been applied to a wide range of two-hadron scattering observables for several phenomenologically interesting channels~\cite{Dudek:2012gj, Dudek:2012xn, Briceno:2015dca, Briceno:2016kkp, Briceno:2016mjc, Dudek:2016cru, Moir:2016srx, Woss:2016tys, Briceno:2017qmb, Woss:2018irj,Wilson:2015dqa,Cheung:2016bym, Wilson:2019wfr}.  See Sec.~\ref{app:suppspec} of the supplemental material for some example plots of $\lambda_n(t,t_0)$.

In order to robustly interpolate the two- and three-pion energy eigenstates we use operators with two- and three-meson-like structures in the appropriate irrep, constructed from products of single-meson-like operators projected to definite spatial momentum. The latter are built from linear combinations, chosen to optimize overlap to the single-pion states, of fermion bilinears of the form, $\bar{\psi}\Gamma {D} \dots {D} \psi$, where $\psi$ is a quark field and $D$ is a discretized covariant derivative. Details of these operator constructions are given in Sec.~\ref{app:operators} of the supplemental material with further details relevant to the three-meson-like operators presented in Ref.~\cite{Woss:2019hse}. Using such a wide variety of optimized operators, and especially multi-hadron operators with momentum-projected single-hadron components, allows one to minimize excited state contamination and extract the energies reliably and precisely from small values of $t$. This approach is made feasible due to the distillation method~\cite{Peardon:2009gh} which we employ to efficiently compute the numerous quark-field Wick contractions that are required. We use 128 distillation vectors for the $20^3$ ensemble and 162 for the $24^3$.

Returning to the two- and three-pion spectra summarized in Fig.~\ref{fig:FVspec}, we observe a one-to-one correspondence between the computed energy levels and the non-interacting energies in all panels, with the computed values slightly higher in energy than the non-interacting levels.  This suggests that the system is weakly interacting and repulsive in both the two- and three-hadron sectors.

\begin{figure}
\hspace{0pt} \includegraphics[width=1.0\columnwidth]{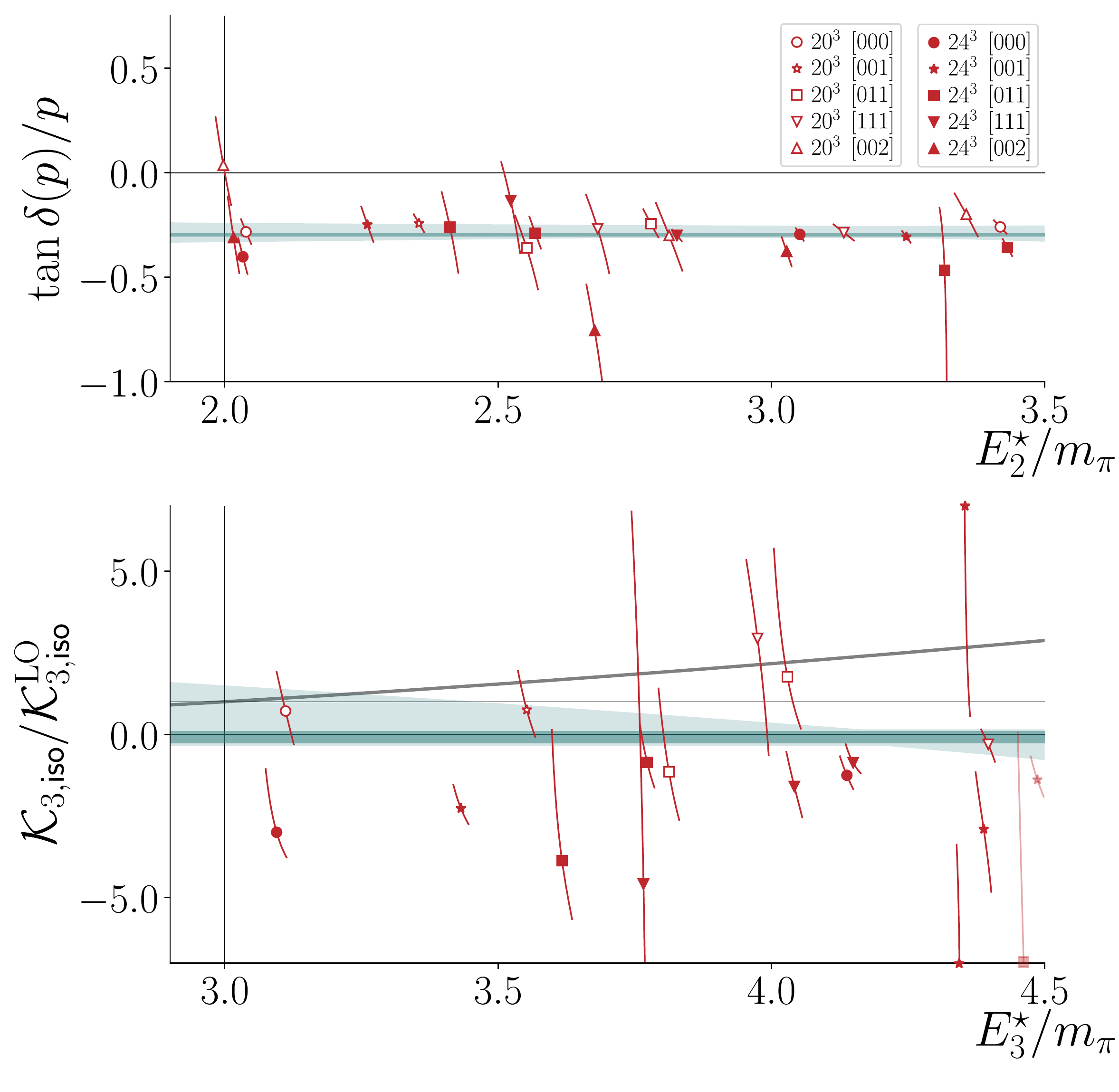}
\caption{Example of data and fits for $\mathcal K_2$ and $\Kiso$, as described in the text. The red points are given by substituting finite-volume energies into $-1/F(E_2, \vec P, L)$ and $-1/F_{3,{\sf iso}}(E_3, \vec P, L)$ for the two- and three-particle energies, respectively, with the volume and $\vec P$ indicated in the legend. A symbol appearing at the very top or bottom represents a case where the central value falls outside the plotted region. The dark cyan bands represent the fit shown in Eq.~\eqref{eq:bestfit} while the lighter bands show the spread covered by the various fits described in the supplemental material.
For the bottom panel we normalize to $m_\pi^2 \Kiso^{\text{LO}} =  4608 \, \pi^2 (m_\pi \A)^2$, with $m_\pi \A$ taken from Eq.~\eqref{eq:bestfit}. This simple relation between $\Kiso$ and the two-particle scattering length holds at leading order in chiral perturbation theory at threshold, as was first derived in Ref.~\cite{Blanton:2019vdk}. The grey curve gives the full leading-order prediction, which is linear in $E_3^{\star 2}$.}
\label{fig:KvE}
\end{figure}

\bigskip
\noindent
\SecKMat~---~We now describe our method for determining two- and three-body K matrices from the extracted finite-volume energies, beginning with an overview of scattering observables:

The two-pion scattering amplitude is defined as the connected part of the overlap between an incoming $\pi^+ \pi^+$ asymptotic state (with momenta $\vec p, - \vec p$) to an outgoing $\pi^+ \pi^+$ state (with $\vec p', - \vec p'$). Without loss of generality, here we have assumed the center-of-momentum frame. We also define $p = \vert \vec p \vert = \vert \vec p' \vert$, where we have used that the magnitudes must be equal to satisfy energy conservation. In addition, $s_2 \equiv E_2^{\star 2} \equiv 4( p^2 + m_\pi^2)$ defines the squared center-of-momentum frame energy. The only remaining degree of freedom is the scattering angle between $\vec p$ and $\vec p'$. In this work we focus on the $S$-wave scattering amplitude, denoted $\mathcal M_2$, in which this angle is integrated to project onto zero-angular-momentum states.
Finally we recall the simple relation between $\mathcal M_2$ and the K matrix in the elastic region, $\mathcal K_2^{-1} = \text{Re} \, \mathcal M_2^{-1}$.%
\footnote{The imaginary part of $\mathcal M_2^{-1}$ is completely fixed by unitarity so that $\mathcal K_2$ is the only part free to depend on the dynamics of the system. We work with the simple phase space factor, proportional to the momentum magnitude. See, e.g., Ref.~\cite{Briceno:2017max} for more details.} In contrast to $\mathcal M_2$, $\mathcal K_2$ is real for real $s_2$ and is meremorphic in a region of the complex $s_2$ plane around $s_2 = 4m_\pi^2$. In this work we also consider an analogous, three-body K matrix, introduced in Ref.~\cite{Hansen:2014eka} and denoted by $\cK_{\text{df},3}$.

In the two-pion sector, in the case that the $S$-wave interactions are dominant, the scalar-irrep finite-volume energies satisfy the quantization condition~\cite{Luscher:1990ux, Rummukainen:1995vs, Kim:2005gf}, 
\begin{align}
\mathcal{K}_{2}(E^\star_2)+F^{-1}(E_2, \vec P,L)=0 \,,
\label{eq:QC2}
\end{align}
where $E_2^\star \equiv \sqrt{E_2^2 - \vec P^2}$ is the center-of-momentum energy and $F(E_2, \vec P,L)$ is a known geometric function.
For the three-body sector, we use the isotropic approximation of the general formalism derived in Ref~\cite{Hansen:2014eka}, which takes an analogous form, now for pseudoscalar-irrep energies
\begin{align}
\Kiso(E^\star_3)+F^{-1}_{\rm 3, {\sf {iso}}}[\mathcal K_2](E_3, \vec P,L)=0 \,,
\label{eq:QC3}
\end{align}
where the notation is meant to stress that $F_{\rm 3, {\sf {iso}}}[\mathcal K_2](E_3, \vec P,L)$ is a functional of $\mathcal K_2(E_2^\star)$. $F_{\rm 3, {\sf {iso}}}$ is defined in Eq.~(39) of Ref.~\cite{Hansen:2014eka}. Here $\Kiso$ is the component of $\cK_{\df,3}$ that only depends on the total three-hadron energy, i.e.~is ``isotropic''. Equation~\eqref{eq:QC3} holds only when $\cK_{\df,3}$ is well approximated to be isotropic and our fits give evidence that this is a good approximation for this system.

Combining these two conditions with the energies plotted in Fig.~\ref{fig:FVspec} allows one to constrain both the two- and three-hadron K matrices. One strategy is to fit a parametrization of $\mathcal K_2$ and use this to determine the energy dependence of $\Kiso$ as summarized in Fig.~\ref{fig:KvE}. An alternative approach is to parametrize both K matrices and fit these simultaneously to the entire set of finite-volume energies. A detailed discussion with a wide range of fits is given in Sec.~\ref{app:Kfits} of the supplemental material. Both strategies give consistent results and the key message is that the full data set is well described by a constant $\Kiso$ that is consistent with zero, together with the leading-order effective range expansion: $\tan \delta(p) = - \A p$ with $\mathcal K_2(E^\star_2) = - 16 \pi E_2^\star \tan \delta(p)/p$. Here the second equation defines the $S$-wave scattering phase shift, $\delta(p)$, and the first defines the scattering length, $a_0$. Our best fit, performed simultaneously to all spectra shown in Fig.~\ref{fig:FVspec} but with a cutoff in the center-of-momentum frame energies included,\footnote{This fit is denoted by B${}_{2+3}$ in Sec.~\ref{app:Kfits} of the supplemental material. As explained there, the fitted data includes all two-pion energies below $E^\star_{2,{\sf cut}} = 3.4 m_\pi$ and all three-pion energies below $E^\star_{3,{\sf cut}} = 4.4 m_\pi$, with both cutoffs applied to energies in the center-of-momentum frame.} yields
\begin{equation}
\label{eq:bestfit} 
\begin{aligned}  
m_\pi \A & = 0.296 \pm 0.008 \\[-3pt] m_\pi^2 \Kiso & = -339 \pm 770   
\end{aligned}   \  
\bigg [ \begin{aligned}  \! 1.0 \ &   0.6   \\[-3pt]   & 1.0  \end{aligned} \bigg ]  \,,
\end{equation} 
with a $\chi^2$ per degree-of-freedom of $64.5/(37-2) = 1.84$. The square-bracketed matrix gives the correlation between the two fit parameters.
This is consistent with the previous determination of the scattering length at this pion mass, presented in Ref.~\cite{Dudek:2012gj}, and is also the value used to generate the orange curves in Fig.~\ref{fig:FVspec} (together with $\Kiso=0$). In Fig.~\ref{fig:KvE} we illustrate the same fit using the darker cyan curves. In addition, we include the lighter bands as a systematic uncertainty, estimated from the spread of various constant and linear fits, as detailed in Sec.~\ref{app:Kfits} of the supplemental material.

\vspace{2mm}
\emph{$3\pi^+$ scattering amplitude}~---~Following the relativistic integral equations presented in Ref.~\cite{Hansen:2015zga}, we can write the $J=0$ and $\Kiso = 0$ amplitude as follows:
\begin{multline}
\cM^{(u,u)}_3( p, k) 
=
- \cM_2( E_{2,p}^{\star}) G_{\sf s}( p, k) \cM_2( E_{2,k}^{\star})
\\[3pt]
- \cM_2( E_{2,p}^{\star}) 
\int_{k'} 
\! 
G_{\sf s}( p, k')
\cM^{(u,u)}_3( k', k) \,,
\label{eq:M3_Kdf0}
\end{multline}
where $\int_k \equiv \int d k \,k^2/[(2 \pi)^2  \omega_k]$ and we have introduced
\begin{align}
G_{\sf s}( p, k )
& \equiv - \frac{H(p,k)}{4 pk}
\log\left[
\frac{\alpha(p,k)-2pk +i\epsilon}
{\alpha(p,k)+2pk+i\epsilon}
\right] \,, \\[5pt]
\alpha(p,k) & \equiv (E_3-\omega_k-\omega_p)^2-p^2-k^2-m^2 
\,.
\end{align}

$\cM_2$ is the $S$-wave two-particle scattering amplitude, introduced above, which depends on the invariant $ E_{2,k}^{\star 2} \equiv (E_3 - \omega_k)^2 - k^2 $, with $\omega_k = \sqrt{k^{2} + m^2}$. The $(u,u)$ superscript emphasizes that specific spectator momenta, $k$ and $p$, are singled out in the initial and final states respectively. The function $G_{\sf s}$ encodes the spectator exchange, projected to the $S$-wave. It inherits a scheme dependence through the smooth cutoff function $H$, defined in Eqs.~(28) and (29) of Ref.~\cite{Hansen:2014eka}. This scheme dependence is matched by that inside of $\Kiso$ such that the resulting scattering amplitude is universal.

\begin{figure}
\begin{center}
\includegraphics[width=0.46\textwidth]{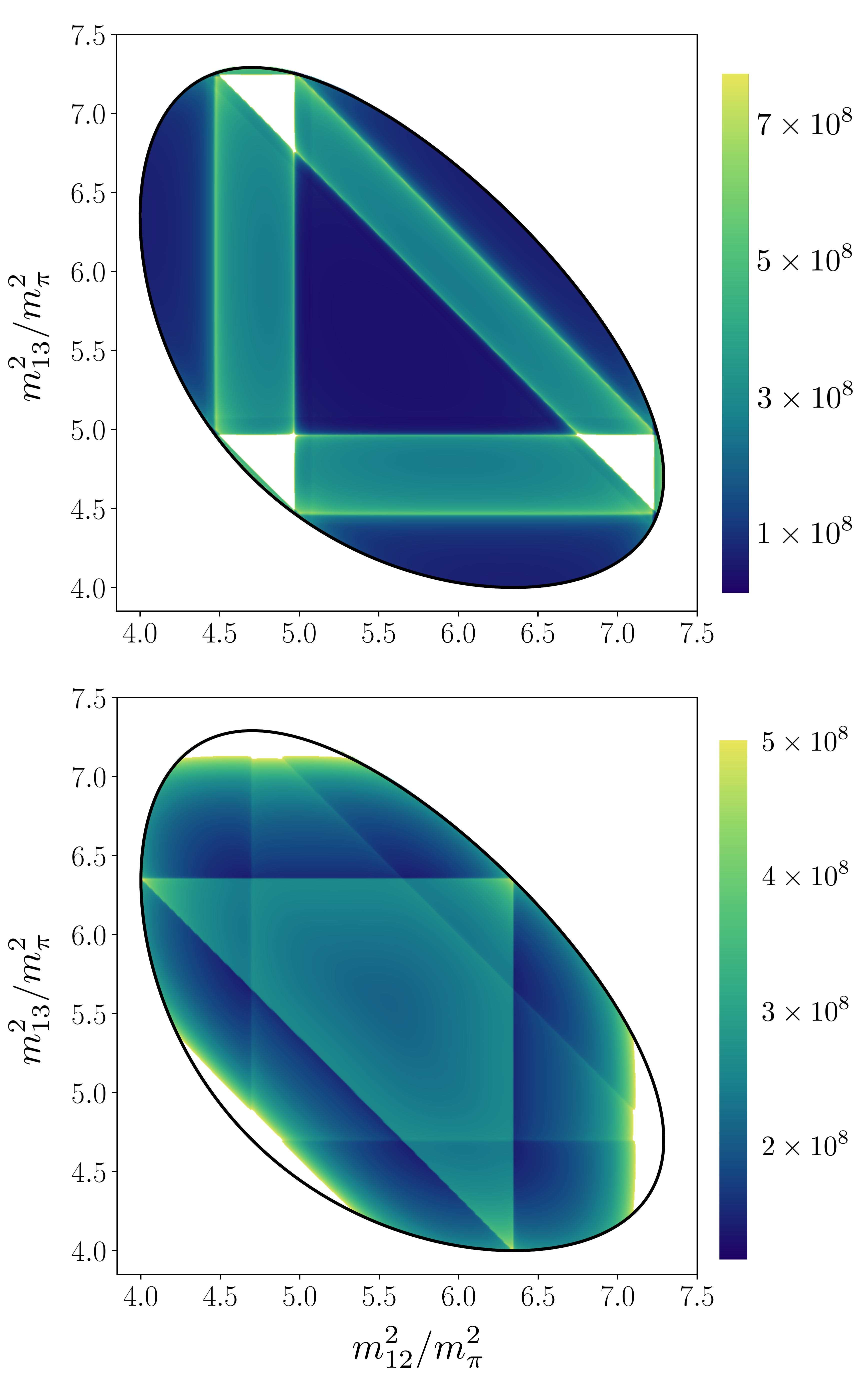}
\caption{\emph{Top:} Dalitz-like plot of $m_\pi^4 \vert \mathcal M_3 \vert^2$ for $\sqrt{s_3} = 3.7m$ with final kinematics fixed to $\{\vec p'^2_1, \, \vec p'^2_2\} =\{0.01 m_\pi^2 , \, 0.7 m_\pi^2 \} \, \Longrightarrow \, \{m'_{12}, \, m'_{13}\} = \{2.1m_\pi, \, 2.25m_\pi\}$. \emph{Bottom:} Same total energy, now with incoming and outgoing kinematics set equal, as discussed in the text.}
\label{fig:3piAmplitudesVSm}
\end{center}
\end{figure}

To use Eq.~\eqref{eq:M3_Kdf0} in practice, one requires a parameterization for $\cM_2$. As described in the previous section, the $\pi^+ \pi^+$ system is well described using the leading order effective range expansion for $\cM_2$, 
\begin{align}
\cM_2(E_2^{\star}) &= 
\frac{16 \pi E_2^{\star}}{-1/\A - i \sqrt{E_2^{\star2} /4-m_\pi^2}} 
\,.
\label{eq:ERE}
\end{align}

Following the derivation of Ref.~\cite{Hansen:2015zga}, the final step is to symmetrize with respect to the spectators, to reach
\begin{equation}
\cM_3(s_3, m'^2_{12}, m'^2_{13}, m^2_{12}, m^2_{13} ) = \sum_{\vec p_i \in \mathcal P_p} \sum_{\vec k \in \mathcal P_k} \cM^{(u,u)}_3( p, k) \,,
\end{equation}
where $\mathcal P_p = \{\vec p, \vec a' , - \vec p - \vec a' \}$ and $\mathcal P_k = \{\vec k, \vec a , - \vec k - \vec a \}$. We have presented the left-hand side as a function of the five Lorentz invariants that survive after truncating to $J=0$ in both the two and three particle sector: the squared three-hadron center-of-momentum frame energy, $s_3$, as well as two pion-pair invariant masses for each of the initial and final states. These are defined by introducing the notation $ \{\vec k, \vec a , - \vec k - \vec a \} =  \{\vec p_1, \vec p_2 , \vec p_3 \}$, then for example
\begin{equation}
m_{12}^2 = (p_1 +p_2)^2= (E_3^\star - [{m_\pi^2 + \vec p_3^2}]^{1/2}) - \vec p_3^2 \,,
\end{equation}
where the middle expression depends on on-shell four-vectors with $p_1^2 = m_\pi^2$.

In the top panel of Fig.~\ref{fig:3piAmplitudesVSm} we show a Dalitz-like plot of $\vert \cM_3 \vert^2$ as a function of $(m_{12}, m_{13})$, with all other kinematics fixed as indicated in the caption. In a usual Dalitz description, the incoming energy is fixed by the decaying particle so that only the outgoing kinematics can vary, whereas here we simply fix the other kinematics.  The inputs to this plot are the best-fit scattering length, given in Eq.~\eqref{eq:bestfit}, together with $\Kiso=0$. The bottom panel of Fig.~\ref{fig:3piAmplitudesVSm} shows the same $\sqrt{s_3}$ but varies incoming and outgoing kinematics according to $m_{12}=m_{12}'$ and $m_{13} = m_{13}'$.

Additional details concerning the $S$-wave integral equations are presented in Secs.~\ref{app:int_eq} and \ref{sec:NumSolveM3} of the supplemental material, where we also describe the propagation of the uncertainties of $m_\pi \A$ and $\Kiso$ into the predicted amplitude.%
\footnote{See also Ref.~\cite{Jackura:2018xnx} for more details on expressing the three-particle amplitude via a truncated partial wave series and Ref.~\cite{Sadasivan:2020syi} for a discussion of integral equations and their solutions in a resonant three-hadron channel.}

\bigskip
\noindent
\emph{Summary}~---~In this work we have presented the first lattice QCD determination of the energy-dependent three-to-three scattering amplitude for three pions with maximal isospin. The calculation proceeded in three steps: \emph{(i)} determining finite-volume energies with $\pi^+\pi^+\pi^+$ quantum numbers, \emph{(ii)} using the framework of Ref.~\cite{Hansen:2014eka} to extract two- and three-body K matrices from these, and \emph{(iii)} applying the results of Ref.~\cite{Hansen:2015zga} to convert these to the three-hadron scattering amplitude, by solving known integral equations. The three steps are summarized, respectively, by Figs.~\ref{fig:FVspec}, \ref{fig:KvE} and \ref{fig:3piAmplitudesVSm} of the text.

Having established this general workflow, it is now well within reach to rigorously extract three-hadron resonance properties from lattice QCD calculations. In particular the formalism has recently been extended to three-pion states with any value of isospin in Ref.~\cite{Hansen:2020zhy}. This should enable studies, for example, of the $\omega$, $h_1$ and $a_1$ resonances. The main outstanding challenges here include rigorous resonant parametrizations of the intermediate three-body K matrix, as well as a better understanding of the analytic continuation required to identify the resonance pole position.

\acknowledgments{The authors would like to thank M.~Bruno, J.~Dudek, A.~Jackura, L.~Leskovec and A.~Rodas for useful conversations, as well as our other colleagues within the Hadron Spectrum Collaboration. The authors would also like to thank W.~Detmold, F.~Romero-L{\'o}pez and S.~R.~Sharpe for useful discussions and for valuable feedback on a previous version of this manuscript.
RAB, RGE and CET acknowledge support from the U.S.~Department of Energy contract DE-AC05-06OR23177, under which Jefferson Science Associates, LLC, manages and operates Jefferson Lab. RAB also acknowledges support from the USDOE Early Career award, contract de-sc0019229. CET and DJW acknowledge support from the U.K.~Science and Technology Facilities Council (STFC) [grant number ST/P000681/1].  DJW acknowledges support from a Royal Society University Research Fellowship.
MTH, CET and DJW acknowledge the MITP topical workshop ``Scattering Amplitudes and Resonance Properties for Lattice QCD'' for stimulating this project and for hospitality during the initial discussions.
MTH and CET also acknowledge the CERN-TH Institute ``Advances in Lattice Gauge Theory'', which provided the opportunity to make significant progress on this work. CET further acknowledges CERN TH for hospitality and support during a visit in January and February of this year. 
The software codes
{\tt Chroma}~\cite{Edwards:2004sx} and {\tt QUDA}~\cite{Clark:2009wm,Babich:2010mu,QUDA3} were used. 
The authors acknowledge support from the U.S. Department of Energy, Office of Science, Office of Advanced Scientific Computing Research and Office of Nuclear Physics, Scientific Discovery through Advanced Computing (SciDAC) program.
Also acknowledged is support from the U.S. Department of Energy Exascale Computing Project.
This work was also performed on clusters at Jefferson Lab under the USQCD Collaboration and the LQCD ARRA Project.
This research was supported in part under an ALCC award, and used resources of the Oak Ridge Leadership Computing Facility at the Oak Ridge National Laboratory, which is supported by the Office of Science of the U.S. Department of Energy under Contract No. DE-AC05-00OR22725.
This research used resources of the National Energy Research Scientific Computing Center (NERSC), a DOE Office of Science User Facility supported by the Office of Science of the U.S. Department of Energy under Contract No. DE-AC02-05CH11231.
The authors acknowledge the Texas Advanced Computing Center (TACC) at The University of Texas at Austin for providing HPC resources.
Gauge configurations were generated using resources awarded from the U.S. Department of Energy INCITE program at the Oak Ridge Leadership Computing Facility, the NERSC, the NSF Teragrid at the TACC and the Pittsburgh Supercomputer Center, as well as at Jefferson Lab.\vspace{0pt}}

\bibliographystyle{apsrev4-1}
\bibliography{refs}

\appendix
\onecolumngrid

\clearpage

\section*{Supplemental Material}

\subsection{Spectra}
\label{app:suppspec}

In this section we give details concerning the finite-volume spectra described in the main text. %
We focus here on two representative examples for the correlators used to extract the three-pion energies. The quality of two-pion correlators can be inferred from the earlier work presented in Ref.~\cite{Dudek:2012gj}, which includes a partially overlapping data set.

\begin{figure}[H]%
\begin{center}
\includegraphics[width=0.75\textwidth]{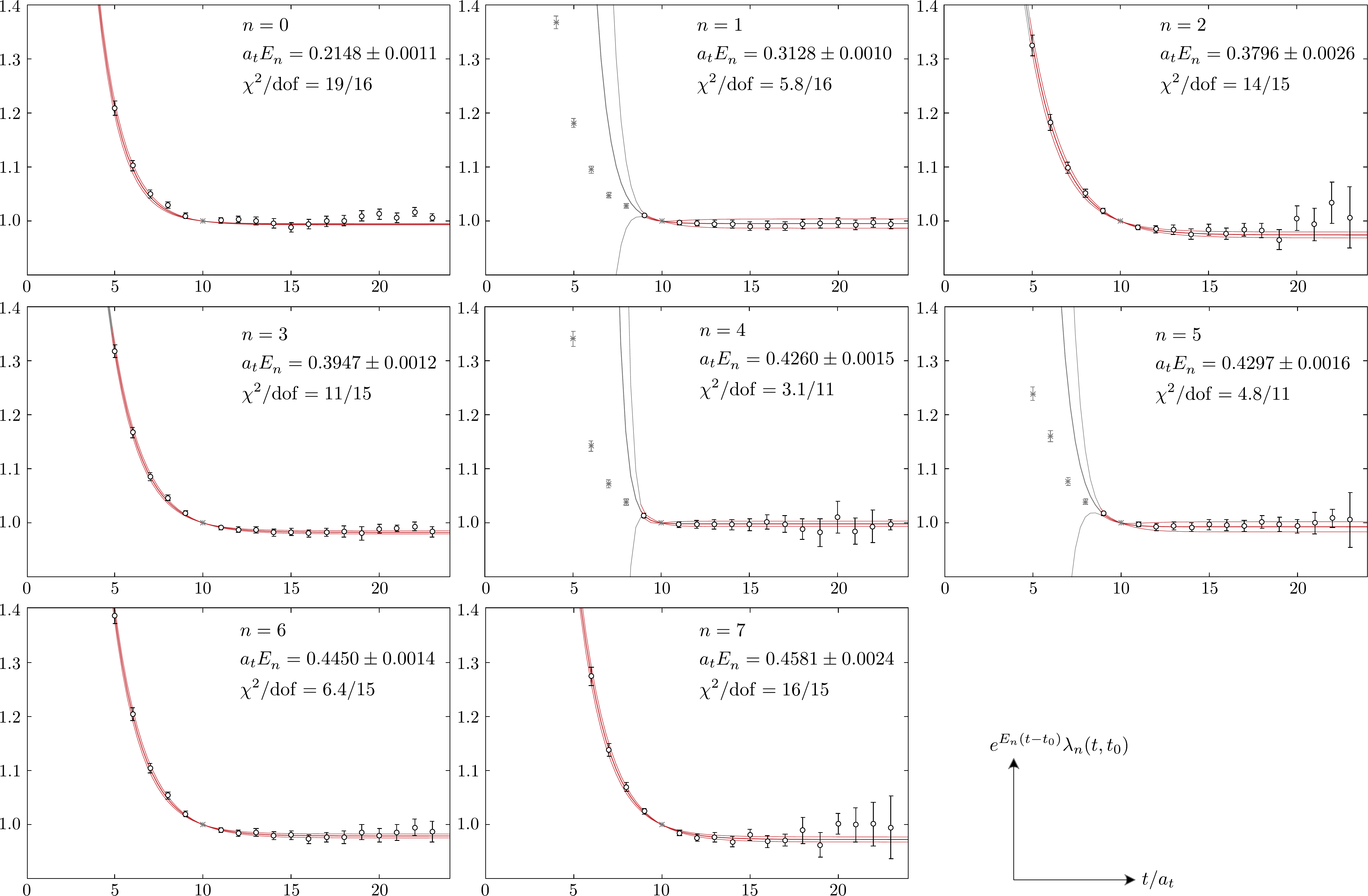}%
\caption{GEVP eigenvalues used to determine the $\vec P = [000]$, $\Lambda = A_1^-$, three-pion spectrum on the  $20^3 \times 256$ ensemble. As explained in the text, the diagonalization is performed with $t_0/a_t=10$ and the eigenvalues are rescaled by their expected large-$t$ fall-off.\label{fig:specL20}}
\end{center}
\end{figure}

As a first example, consider the three-pion spectrum for the $\vec P = [000]$, $\Lambda = A_1^-$ irrep on the $20^3 \times 256$ ensemble. In this case $G_{ij}(t) = \langle \mathcal O_i(t) \mathcal O^\dagger_j(0) \rangle$ is an $8 \times 8$ matrix of correlators, built from the first 8 operators listed in Table~\ref{table:ops:3pi:0} of Sec.~\ref{app:operators}. In Fig.~\ref{fig:specL20} we plot the corresponding eigenvalues, $\lambda_n(t,t_0)$, of the matrix
\begin{equation}
M(t,t_0)  \equiv G(t_0)^{-1/2} \cdot G(t) \cdot G(t_0)^{-1/2} \,,
\end{equation}
entering the generalized eigenvalue problem (GEVP). These are determined for $t_0/a_t = 10$ and are plotted vs.~$t/a_t$ for a range of values both before and after the reference time. To display the eigenvalues in a useful manner, we plot the combination $e^{E_n (t-t_0)} \lambda_n(t,t_0)$, where $E_n$ has been determined from a two-state fit to  $\lambda_n(t,t_0)$
\begin{equation}
\lambda_n(t,t_0) =  (1-A_n) e^{- E_n(t-t_0)} + A_n e^{- E'_n(t-t_0)}\,.
\end{equation}
The quality of the fit is indicated by the $\chi^2/\text{dof}$ in each panel. The plotted combination behaves as expected for a successful GEVP, showing a reasonable plateaux over a range of $t/a_t$.

\bigskip

 This result also exhibits no evidence for thermal states on this lattice, as expected given the length of the temporal extent, $m_\pi T \approx  17.7$. We detour slightly, to explain this point in more detail:
 
 In general, for multi-pion systems, the leading finite-$T$ effects are given by
 \begin{equation}
 \label{eq:thermal}
 G_{ij}(t) = \langle 0 \vert \mathcal O_i(0) \, e^{- \hat H t} \, \mathcal O^\dagger_j(0) \vert 0 \rangle +  e^{- m_\pi (T - t)}   \langle \pi^- \vert \mathcal O_i(0) \, e^{- \hat H t} \, \mathcal O^\dagger_j(0) \vert \pi^- \rangle  + \cdots \,,
 \end{equation}
where $\hat H$ is the Hamiltonian and the ellipsis represents thermal effects falling faster than $e^{- m_\pi T}$. For concreteness, we have assumed that $O^\dagger(0)$ creates three-$\pi^+$ quantum numbers, so that $\mathcal O^\dagger(0) \vert \pi^- \rangle$ has the quantum numbers of two pions with isospin two. A spectral decomposition of Eq.~\eqref{eq:thermal} then yields
 \begin{equation}
 \label{eq:thermal2}
 G_{ij}(t) = \sum_n c^{(n)}_i c^{(n)*}_j e^{- E_n^{\pi \pi \pi} t}   +  e^{- m_\pi (T - t)}  \sum_{n} b^{(n)}_i b^{(n)*}_j e^{- E_n^{\pi \pi} t} + \cdots \,,
 \end{equation}
 where the sum in the first term (second term) runs over all maximal-isospin three-pion (two-pion) finite-volume states with specified $\vec P $.
In the case of $\vec P=[000]$, the two- and three-pion ground states take the form $N m_\pi + \Delta E_{N}$ where $N = 2,3$ and $\Delta E_{N} \sim 1/L^3$ for weakly-interacting systems. Taking the leading ($n=0$) terms of Eq.~\eqref{eq:thermal2} and substituting this scaling then yields
 \begin{equation}
 \label{eq:thermal3}
 G_{ij}(t) =   c^{(0)}_i c^{(0)*}_j e^{-  (3 m_\pi + \Delta E_3)t}   +  b^{(0)}_i b^{(0)*}_j e^{- m_\pi (T + t)}   e^{-    \Delta E_2  t} + \cdots \,.
 \end{equation}

When this same exercise is performed for a \emph{two-pion} correlator, again focusing on the case of $\vec P=[000]$, one finds that the leading thermal contamination is a constant in $t$ in the non-interacting limit. As discussed in Ref.~\cite{Dudek:2012gj} this can thus be removed by applying a shift to the correlator $G_{ij}(t)  \to G_{ij}(t)  - G_{ij}(t + \delta t) $. In the present case, however, the leading contaminations are $t$-dependent. One option is to reweight and shift, i.e.
\begin{equation}
\label{eq:shift}
G_{ij}(t) \ \to \ e^{- m_\pi t} \big ( G_{ij}(t) e^{m_\pi t}  - G_{ij}(t + \delta t) e^{m_\pi (t + \delta t)} \big )\,.
\end{equation}
This approach, already used in Ref.~\cite{Dudek:2012gj} for $\pi \pi$ systems with non-zero total momentum, reduces thermal effects at the cost of generally degrading the signal quality.
Fortunately, for the $20^3 \times 256$ lattice, this is not required. Comparing the leading and subleading terms of Eq.~\eqref{eq:thermal3}, and neglecting the interactions, one finds that the relative
size of the three-pion thermal contamination is
$e^{- m_\pi (T - 2 t)}  $.
Thus, assuming that the relevant matrix elements have the same order of magnitude, for the range of $t$ considered these effects are $ \sim 10^{-7}$ and are safely negligible, despite the high statistical precision of the extracted energies. This concludes our general comments on thermal effects.

\begin{figure}%
\begin{center}
\includegraphics[width=1.0\textwidth]{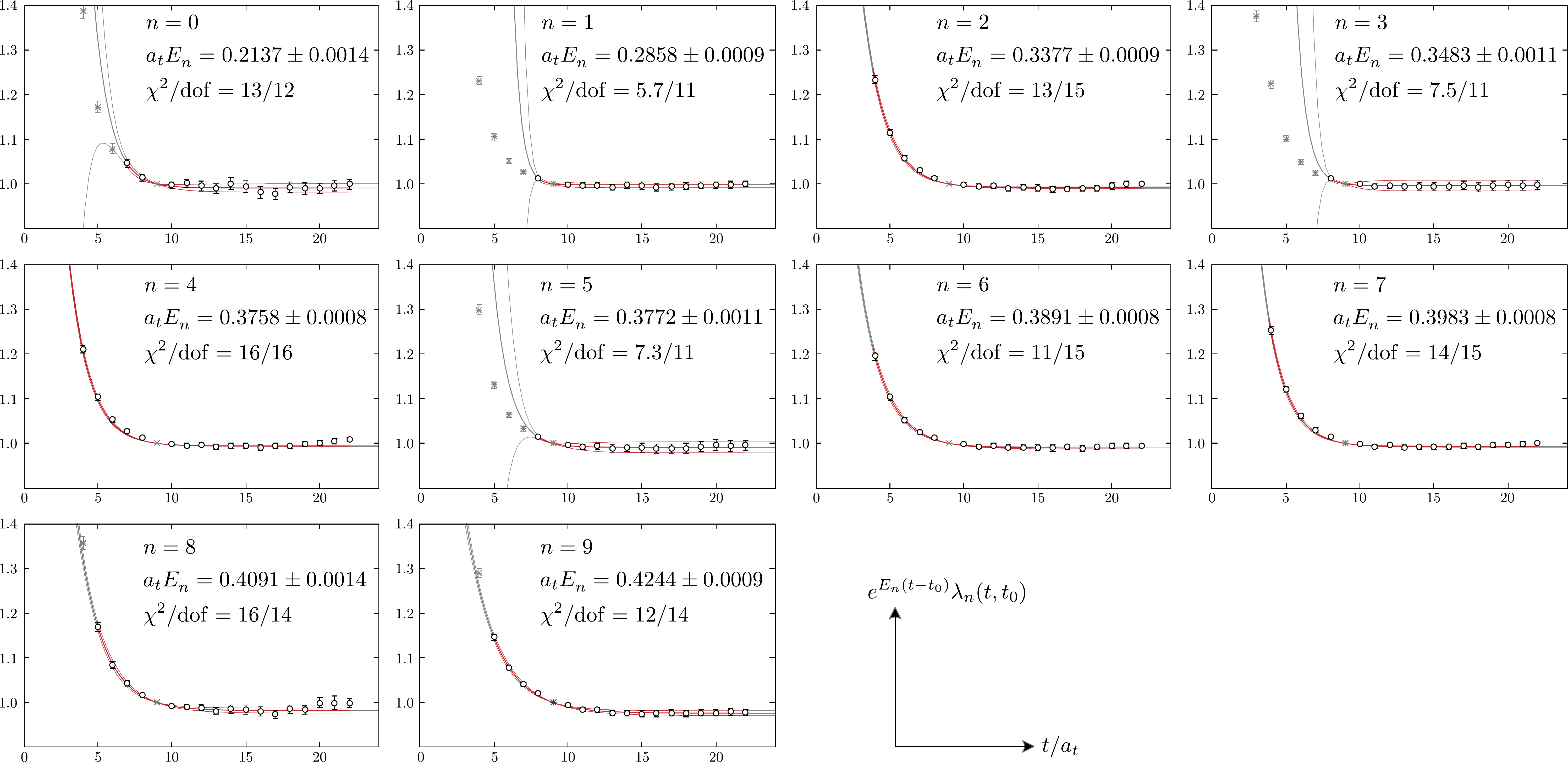}%
\caption{GEVP eigenvalues used to determine the $\vec P = [000]$, $\Lambda = A_1^-$, three-pion spectrum on the  $24^3 \times 128$ ensemble. The diagonalization is performed with $t_0/a_t=9$ and the eigenvalues are rescaled as in Fig.~\ref{fig:specL20}.\label{fig:specL24}}
\end{center}
\end{figure}

\bigskip

As a second example, in Fig.~\ref{fig:specL24} we consider the same three-pion quantum numbers ($\vec P = [000]$, $\Lambda = A_1^-$) on the $24^3 \times 128$ ensemble. Because the larger spatial volume lowers the value of the $n$th level, here we include 2 additional operators to better absorb the excited states. As with the previous example, the $\chi^2/\text{dof}$ and the plotted curves provide strong evidence of a successful GEVP extraction. For this case, $e^{- m_\pi (T - 2 t)} \sim 10^{-3}$ so that finite-$T$ effects potentially present a more significant issue. As a result we have also considered shifting and reweighting, as summarized by Eq.~\eqref{eq:shift}, in our various fits. However, across all values of $\vec P$, we find that more stable fits are %
achieved via the unmodified correlators, relying on the basis of operators to push the extraction to earlier times and examining the resulting GEVP eigenvalues. This is in contrast to the $24^3 \times 128$ two-pion fits, where the extractions are improved by shifting in certain cases, as described in Ref.~\cite{Dudek:2012gj}.

\subsection{K-matrix fits}
\label{app:Kfits}

In this subsection we give additional details concerning the K-matrix fits, summarized in the main text. We present four basic types of fits:
\begin{enumerate}
\item Fitting only the $\pi^+ \pi^+$ spectra to various choices of $p \cot \delta(p)$ [Table~\ref{tab:pipi}].%
\item Fitting only the $\pi^+ \pi^+ \pi^+$ spectra to various choices of $p \cot \delta(p)$, with $\Kiso = 0$ [Table~\ref{tab:pipipiK2}].
\item Fitting only the $\pi^+ \pi^+ \pi^+$ spectra to various choices of $\Kiso$, with $p \cot \delta(p)$ fixed by independent $\pi^+  \pi^+$ fits [Table~\ref{tab:pipipiWithK2Fixed}].
\item Fitting all spectra simultaneously to various choices of $p \cot \delta(p)$ and $\Kiso$ [Table~\ref{tab:simult}].
\end{enumerate}

Here $\delta(p)$ is the $S$-wave, $\pi^+ \pi^+$ scattering phase shift, related to the scattering amplitude via
\begin{equation}
\mathcal M_2(E_2^\star) = \frac{16 \pi E_2^\star}{p \cot \delta(p) - i p}   \,,
\end{equation}
where $p^2 = E_2^{\star2}/4 - m_\pi^2$. One standard parametrization of the scattering amplitude follows from the effective range expansion
\begin{equation}
p \cot \delta(p) = - \frac{1}{a_0} + \frac12 r_0 p^2 + \mathcal O(p^4) \,,
\end{equation}
and below we present fits to the leading term as well as to the leading two terms.

{
\renewcommand{\arraystretch}{1.7}

\begin{table}
\begin{center}
\begin{tabular}{c | c | c | c | c }
\ \ \ Fit \ \ \ & \ \ $E^\star_{2,\text{cut}}$ \ \ & \ \ \ $p\, \text{cot}\delta(p)$ \ \ \ & \ \ \ \text{fit\ result} \ \ \ & \ \ $\chi^2/\text{dof}$\ \ \\[0pt] \hline \hline
A${}_2$  &  $4.0m_\pi$ & \ $ -1/\A$ \ & \ \ $m_\pi \A = 0.278 \pm 0.007$ \ \ & \ $89.8/(32-1) = 2.90 $ \ \\[0pt]
B${}_2$  &  $3.4m_\pi$ & \ $ -1/\A$ \ & \ \ $m_\pi \A = 0.292 \pm 0.010$ \ \ & \ $26.9/(21-1) = 1.35 $ \ \\[0pt] \hline
\ERrowNPCOV{C}{4.0m_\pi}{0.317 \pm 0.015}{-0.39 \pm 0.12}{79.6/(32-2) = 2.65}{-0.9}\\[0pt]
\ERrowNPCOV{D}{3.4m_\pi}{0.258 \pm 0.018}{0.68 \pm 0.33}{22.7/(21-2) = 1.20}{-0.9} \\[0pt]  \hline
\AFProwNPCOV{E}{4.0m_\pi}{0.355 \pm 0.021 }{11.2 \pm 2.1}{96.7/(32-2) = 3.22}{-0.8} \\[0pt]
\AFProwNPCOV{F}{3.4m_\pi}{0.260 \pm 0.035}{3.7 \pm 1.1}{22.7/(21-2) = 1.20}{-0.9} \\[0pt] \hline
\ArowNPCOV{G}{4.0m_\pi}{0.223 \pm 0.019}{-2.88 \pm 0.19}{70.1/(32-2) = 2.34}{\phantom{+}0.9} \\[0pt]
\ArowNPCOV{H}{3.4m_\pi}{0.184 \pm 0.022}{-2.2\pm0.4}{26.5/(21-2) = 1.40}{\phantom{+}0.96} \\[0pt] \hline \hline
\hspace{1.2pt} \ I${}_2$ $[r=0.01\ (9/32)]$\ \,  &  $4.0m_\pi$ & \ $  -1/\A$ \ & \ \ $m_\pi \A = 0.292 \pm 0.008$ \ \ & \ $40.5/(32-1) = 1.31$ \ \\[0pt]
  \ERrowRvalNPCOV{ J${}_2$  $[r=0.01\ (9/32)]$}{4.0m_\pi}{0.300 \pm 0.016}{-0.08 \pm 0.14}{40.3/(32-2) = 1.34}{-0.8}
\end{tabular}
\caption{Summary of fits to $\pi^+ \pi^+$ finite-volume energies, for various choices of $p \cot \delta(p)$. All values of total momentum $\vec P$ (from $[000]$ to $[002]$) and both volumes ($20^3$ and $24^3$) are used in each fit. The entries below the lower double horizontal line are determined using a regularized covariance matrix as explained in the text. The function $A(c_0,p) = c_0 m_\pi \sqrt{p^2 + m_\pi^2} /(2 p^2 +  c_0 m_\pi^2)$ encodes the effect of the Adler zero, with $c_0=1$ corresponding to the pole position from leading-order chiral perturbation theory. The columns are understood as follows: ``Fit'' gives a label to the fit (and defines the regularized covariance matrix for the final two fits); ``$E^\star_{2,\text{cut}}$'' gives the two-particle center-of-momentum frame energy cutoff (i.e.~only points with central values below this threshold enter the fit);  ``$p\, \text{cot}\delta(p)$'' indicates the fit function; ``fit result'' displays the extracted parameters and their correlation; ``$\chi^2/\text{dof}$'' gives the value of $\chi^2(\{\eta_i\})$ (evaluated at the best fit parameters) divided by the number of degrees of freedom.\label{tab:pipi}}
\end{center}
\end{table}

}

In the case of $\pi^+ \pi^+$ scattering, chiral perturbation theory predicts the Adler zero, which leads to a pole in $p \cot \delta(p)$, limiting the range of convergence for the effective range expansion. This motivates the alternative form
\begin{equation}
p \cot \delta(p) = A(c,p) \Big [ \! - \frac{1}{a_0} + c' p^2 + \mathcal O(p^4) \Big ] \,,
\end{equation}
where we have introduced
\begin{equation}
A(c,p) \equiv \frac{c \, m_\pi \sqrt{p^2 + m_\pi^2}}{ 2 p^2 +  c \, m_\pi^2 } \,.
\end{equation}
The function $A(c,p)$ is chosen to match the analytic structure predicted by leading-order chiral perturbation theory (both the energy numerator and the pole in the denominator) and is normalized so that $A(c,0) = 1$. The leading-order prediction for the pole position corresponds to $c=1$ and in the following we present fits both with $c$ fixed and allowed to vary.

As we explain in detail in Sec.~\ref{app:int_eq} below, the three-particle scattering amplitude, $\mathcal M_3$, is determined from the two-particle scattering amplitude together with a local three-particle K matrix, first introduced in Ref.~\cite{Hansen:2014eka} and denoted by $\mathcal K_{\text{df},3}$. As already described in the main text, we work here in the isotropic approximation, for which this three-particle K matrix reduces to a simple function of the total center-of-momentum frame energy, denoted $\Kiso(E_3^\star)$. This quantity admits an expansion similar to the effective range expansion
\begin{equation}
\Kiso(E_3^\star) = c_1/m_\pi^2 + c_2 \Delta/m_\pi^4 + \mathcal O(\Delta^2) \,,
\end{equation}
where $\Delta \equiv E_3^{\star 2} - 9 m_\pi^2$. The fits presented below take either the first or else the first two terms in this expansion.

\bigskip

In each case, the fit is performed by minimizing the $\chi^2(\{\eta_i\})$, where
\begin{equation}
\chi^2(\{\eta_i\}) \equiv [\boldsymbol {\mathcal E}_{\sf d} - \boldsymbol {\mathcal E}(\{\eta_i\}) ] \cdot C^{-1} \cdot [\boldsymbol {\mathcal E}_{\sf d} -\boldsymbol {\mathcal E}(\{\eta_i\}) ]^{\text{T}} \,.
\end{equation}
Here $\boldsymbol {\mathcal E}_{\sf d}$ is a vector built from two and three-particle energies extracted from the lattice calculation, $C$ is the covariance matrix, and $\boldsymbol {\mathcal E}(\{\eta_i\})$ is a vector of solutions to the two- and three-particle quantization conditions. The solved energies depend on $\{\eta_i\}$, which stands for all two- and three-particle K-matrix parameters, over which the minimization is performed.

In certain cases the low-lying eigenvalues of $C$ cannot be estimated reliably and, if underestimated, can lead to artificially enhanced eigenvalues in $C^{-1}$, and therefore inflated values for $\chi^2(\{ \eta_i \})$. To study this problem we have also considered an alternative method in which the low lying eigenvalues of $C$ are adjusted. To do so, one first diagonalizes $C$
\begin{equation}
C = R^{\text{T}} \cdot D \cdot R \,,
\end{equation}
where $R$ is an orthogonal matrix of eigenvectors and $D$ a diagonal matrix of eigenvalues. We label the eigenvalues (ordered from smallest to largest) by $\lambda_1, \cdots, \lambda_N$ and note that these are positive. We assume also that the rows and columns of $R$ are organized such that
\begin{equation}
D = \text{diag}[\lambda_1, \lambda_2, \cdots, \lambda_N] \,.
\end{equation}

$\chi^2(\{ \eta_i \})$ may be poorly estimated if there is a large hierarchy between the smallest and largest eigenvalues, $\lambda_1$ and $\lambda_N$, respectively. 
This motivates the definition
\begin{equation}
C_r = R^{\text{T}} \cdot D_r \cdot R \,,
\end{equation}
where $D_r$ is a diagonal matrix defined as
\begin{equation}
D_r = \text{diag}\Big [\text{max}[ \lambda_1, r \lambda_N ], \ \text{max}[ \lambda_2, r \lambda_N ], \ \cdots, \ \text{max}[ \lambda_N, r \lambda_N ] \Big] \,.  %
\end{equation}
Note, if $r=0$, then $D = D_r \ \Longrightarrow \ C = C_r$. As this parameter is increased, the lowest eigenvalues are adjusted to some fixed fraction of the largest value. This approach defines a new test statistic and, in principle, one can sample its corresponding distribution to define $p$-values and assess the quality of fits. This goes beyond the scope of this work and we only perform the modified fits as a cross check to show that the extracted fit parameters are robust under these regularizations of the covariance matrix. Such fits are reported in the tables of this section with labels of the form $[r = 0.01\ (n/m)]$ where $r$ indicates the adjustment parameter and $(n/m)$ indicates the number of eigenvalues that have been changed versus the total number. We have additionally re-done many of the fits summarized Tables \ref{tab:pipi}-\ref{tab:simult} with the pion mass, $m_\pi$, and the anisotropy, $\xi$, varied by one standard deviation. We find in all cases that the effect of this shift is well below the statistical uncertainties on the extracted fit parameters.

{\renewcommand{\arraystretch}{1.7}

\begin{table}[H]
\begin{center}
\begin{tabular}{c | c | c | c | c }
\ \ \ Fit \ \ \  & \ \ $E^\star_{3,\text{cut}}$ \ \ & \ \ \ $p \cot \delta(p)$ \ \ \ & \ \ \ \text{fit\ result} \ \ \ & \ \ $\chi^2/\text{dof}$\ \ \\  \hline \hline
A${}_{3({\sf K}2)}$  & $4.4m_\pi$ & \ $  -1/a_0$ \ & \ \ $m_\pi \A = 0.293 \pm 0.011$ \ \ & \ $ 31.5/(16-1) = 2.10$ \ \\ \hline \hline
B${}_{3({\sf K}2)}$  $[r=0.01\ (5/16)]$\   & $4.4m_\pi$ & \ $  -1/a_0$ \ & \ \ $m_\pi \A = 0.298 \pm 0.014$ \ \ & \ $ 24.5/(16-1) = 1.63$ \ 
\end{tabular}
\caption{Summary of fits to $\pi^+ \pi^+ \pi^+$ finite-volume energies, for $p \cot \delta(p)=-1/a_0$ with $\Kiso = 0$ fixed. All values of total momentum $\vec P$ (from $[000]$ to $[111]$) and both volumes ($20^3$ and $24^3$) are used in each fit.  Columns as in Table \ref{tab:pipi}, with ``$E^\star_{3,\text{cut}}$'' indicating the three-particle center-of-momentum frame energy cutoff. \label{tab:pipipiK2}}
\end{center}
\end{table}
}

{\renewcommand{\arraystretch}{1.7}
\begin{table}[H]
\begin{center}
\begin{tabular}{c | c| c | c | c | c  }
\ \ \ Fit \ \ \ & \ \ $p \cot \delta(p)$ \ \  & \ \ $E^\star_{3,\text{cut}}$ \ \ & \ \ \ $\Kiso$ \ \ \ & \ \ \ \text{fit\ result} \ \ \ & $\chi^2/\text{dof}$ \\  \hline \hline
A${}_{3({\sf K}3)}$ & \ $-1/a_0$ $($B${}_2)$ \  & $4.4m_\pi$ & \ $ c_1/m_\pi^2$ \ & \ $c_1 = -253 \pm 874$ \ & \ $31.4/(16-1) = 2.10$ \ \\[2pt]
B${}_{3({\sf K}3)}$ & \ $-1/a_0$ $($B${}_2)$ \  & $4.4m_\pi$ & \  $  c_1/m_\pi^2 + c_2 \Delta/m_\pi^4 $ \ &   \parbox[c][1.0cm][c]{4.5cm}{\begin{equation*} \begin{aligned}  c_1 & = 5039 \pm 1731 \\[-3pt] c_2 & = -637 \pm 92   \end{aligned}   \  \bigg [ \begin{aligned}   \!1.0  & \,   -0.8   \\[-3pt] \! \phantom{1.0}   & \phantom{++}\!1.0  \end{aligned} \bigg ] \end{equation*} }
 & \ $25.9/(16-2) = 1.85$ \    \\  \hline \hline  
C${}_{3({\sf K}3)}$ $[r=0.01 \ (5/16) ]$ & \ $-1/a_0$ $($B${}_2)$ \  & $4.4m_\pi$ & \ \ $  c_1/m_\pi^2$ \ \ & \ $c_1 = -524 \pm 892$ \ & \ $24.4/(16-1) = 1.63$ \ %
\end{tabular}
\caption{Summary of fits to $\pi^+ \pi^+ \pi^+$ finite-volume energies, for various choices of $\Kiso$ (with $p \cot \delta(p)$ given by fit B${}_2$ of Table \ref{tab:pipi}). Columns as in Tables \ref{tab:pipi} and \ref{tab:pipipiK2}, with ``$\Kiso$'' indicating the fit function used and $\Delta \equiv E_3^{\star2} - 9m_\pi^2$ encoding a linear-dependence in the squared center-of-momentum frame energy.
\label{tab:pipipiWithK2Fixed}}
\end{center}
\end{table}
}

{\renewcommand{\arraystretch}{2.0}
\begin{table}[H]
\begin{center}
\begin{tabular}{c   | c | c | c | c | c | c }
\ \ \ Fit \ \ \ & \ \ $E^\star_{2,\text{cut}}$ \ \ & \ \ $E^\star_{3,\text{cut}}$ \ \ & \ \ \  $p \cot \delta(p)$ \ \ \  & \ \ \ $\Kiso$ \ \ \ & \ \ \ \text{fit\ result} \ \ \ & \ \ $\chi^2/\text{dof}$ \ \ \\ \hline \hline
A$_{2+3}$  & $3.4m_\pi$ & $4.4m_\pi$ & \ $ -1/a_0$ \ & \ $0$ \ & \ \ $m_\pi \A = 0.300 \pm 0.007$ \ \ & \ $ 64.7/(37-1) = 1.80$ \ \\
B$_{2+3}$ &  $3.4m_\pi$ & $4.4m_\pi$  &  \ $-1/a_0$ \  & \ $c_1/m_\pi^2$ \ &   \parbox[c][1.0cm][c]{4.5cm}{\begin{equation*} \begin{aligned}  m_\pi a_0 & = 0.296 \pm 0.008 \\[-3pt] c_1 & = -339 \pm 770   \end{aligned}   \  \bigg [ \begin{aligned}  \! 1.0 \ &   0.6   \\[-3pt]   & 1.0  \end{aligned} \bigg ] \end{equation*} } &  $64.5/(37-2) = 1.84$ \\[0pt] \hline   \hline 
C${}_{2+3}$ $[r=0.005\ (11/37)]$   & $3.4m_\pi$ & $4.4m_\pi$ & \ $ -1/a_0$ \ & \ $0$ \ & \ \ $m_\pi \A = 0.297 \pm 0.008$ \ \ & \ $ 50.9/(37-1) = 1.42$ \ \\
   D${}_{2+3}$ $[r=0.005\ (11/37)]$   &  $3.4m_\pi$ & $4.4m_\pi$  &  \ $-1/a_0$ \  & \ $c_1/m_\pi^2$ \ &   \parbox[c][1.0cm][c]{4.5cm}{\begin{equation*} \begin{aligned}  m_\pi a_0 & = 0.293 \pm 0.010 \\[-3pt] c_1 & = -426 \pm 814   \end{aligned}   \  \bigg [ \begin{aligned} \!  1.0 \ &   0.7   \\[-3pt]   & 1.0  \end{aligned} \bigg ] \end{equation*} } &  $50.7/(37-2) = 1.45$ 
\end{tabular}
\caption{Summary of combined fits to both two- and three-pion energies. Columns as in Tables \ref{tab:pipi}-\ref{tab:pipipiWithK2Fixed}.\label{tab:simult}}
\end{center}
\end{table}
}

\subsection{Details of the three-particle integral equations}
\label{app:int_eq}

In this subsection we give additional details concerning Eq.~\eqref{eq:M3_Kdf0} of the main text and give details on its numerical implementation. We begin by reviewing the integral equations presented in Ref.~\cite{Hansen:2015zga}. As the results of the fits summarized in the main text~(and detailed in the previous subsection) are consistent with $\Kiso = 0$, we focus here on the case of a weak three-body interaction, keeping only the linear contribution in this term. We begin with the unsymmetrized three-body scattering amplitude 
\begin{align}
\cM^{(u,u)}_3(\vec p, \vec k) & = \cD^{(u,u)}(\vec p, \vec k) + \mathcal E^{(u)}(\vec p) \Kiso \mathcal E^{(u)}(\vec k) + \mathcal O(\Kiso^2) \,,
\label{eq:M3uu}
\end{align}
where the superscripts indicate the lack of exchange symmetry, and $\vec k$ and $ \vec p$ specify the momenta of the so-called spectator particles in the initial and final state, respectively. In general, the factors appearing in Eq.~\eqref{eq:M3uu} carry angular momentum indices. However, as discussed in the main text, the $\pi^+\pi^+$ system at low energies is dominated by the $S$-wave component. Thus, we restrict attention here to $\pi^+ \pi^+$ with zero angular momentum, such that $\cM^{(u,u)}_3(\vec p, \vec k) $ and $\cD^{(u,u)}(\vec p, \vec k)$ are simple functions, with no implicit indices. 

In this limiting case, $\cD^{(u,u)}(\vec p, \vec k)$ satisfies the implicit equation
\begin{align}
\cD^{(u,u)}(\vec p, \vec k) &= - \cM_2( E_{2,p}^{\star}) G^\infty(\vec p, \vec k) \cM_2( E_{2,k}^{\star}) - \cM_2( E_{2,p}^{\star}) \int \! \frac{d^3 \vec k'}{(2 \pi)^3 2 \omega_{k'}} G^\infty(\vec p, \vec k') \cD^{(u,u)}(\vec k', \vec k) \,,
\label{eq:Dmaster}
\end{align}
where $ E_{2,k}^{\star 2} \equiv (E - \omega_k)^2 - k^2 $ is the center-of-momentum energy for the non-spectator pair (with $k = \vert \boldsymbol k \vert$ defined in the three-particle zero-momentum frame). In words, the unsymmetrized amplitude $\cD^{(u,u)}$ can be evaluated by solving an integral equation depending only on the two-particle scattering amplitude $\cM_2( E_{2,p}^{\star})$ and the exchange propagator 
\begin{equation}
\label{eq:G}
G^\infty(\vec p, \vec k) \equiv \frac{H( p, k)}{b_{pk}^2 - m^2 + i \epsilon} \,.
\end{equation}
Here we have suppressed the $\pi$ subscript on the mass and have introduced $b^2_{pk} \equiv (E^\star - \omega_p - \omega_k)^2 - (\vec p + \vec k)^2 $ as well as $H( p, k)$, a cut-off function built into the relation between finite-volume energies and $\Kiso$, as well as that between $\Kiso$ and the physical scattering amplitude. The definition used here is
\begin{equation}
\label{eq:cutoff} 
H( p, k) \equiv J\big ( E_{2,k}^{\star 2} / [2 m]^2 \big )\, J\big ( E_{2,p}^{\star 2} / [2 m]^2 \big ) \,,
\qquad \qquad J(x) \equiv
\begin{cases}
0 \,, & x \le 0 \,; 
\\ 
\exp \left( - \frac{1}{x} \exp \left [-\frac{1}{1-x} \right] \right ) \,, 
& 0<x \le 1 \,; 
\\ 
1 \,, & 1<x \,.
\end{cases}
\end{equation}
(See also Refs.~\cite{Hansen:2014eka,Hansen:2015zga} for more discussion on this technical aspect.) Finally $ \mathcal E^{(u)}(\vec p) $ is a function closely related to $ \cD^{(u,u)}(\vec p, \vec k) $ and defined, in a specific limiting case, in Eq.~\eqref{eq:EEdef} below. In this work it is only relevant to demonstrate that $\Kiso$ contributes negligibly to both the central value and uncertainty of the three-hadron scattering amplitude, as we describe in the following section.

\begin{figure} 
\centering
\includegraphics[width=0.45\textwidth]{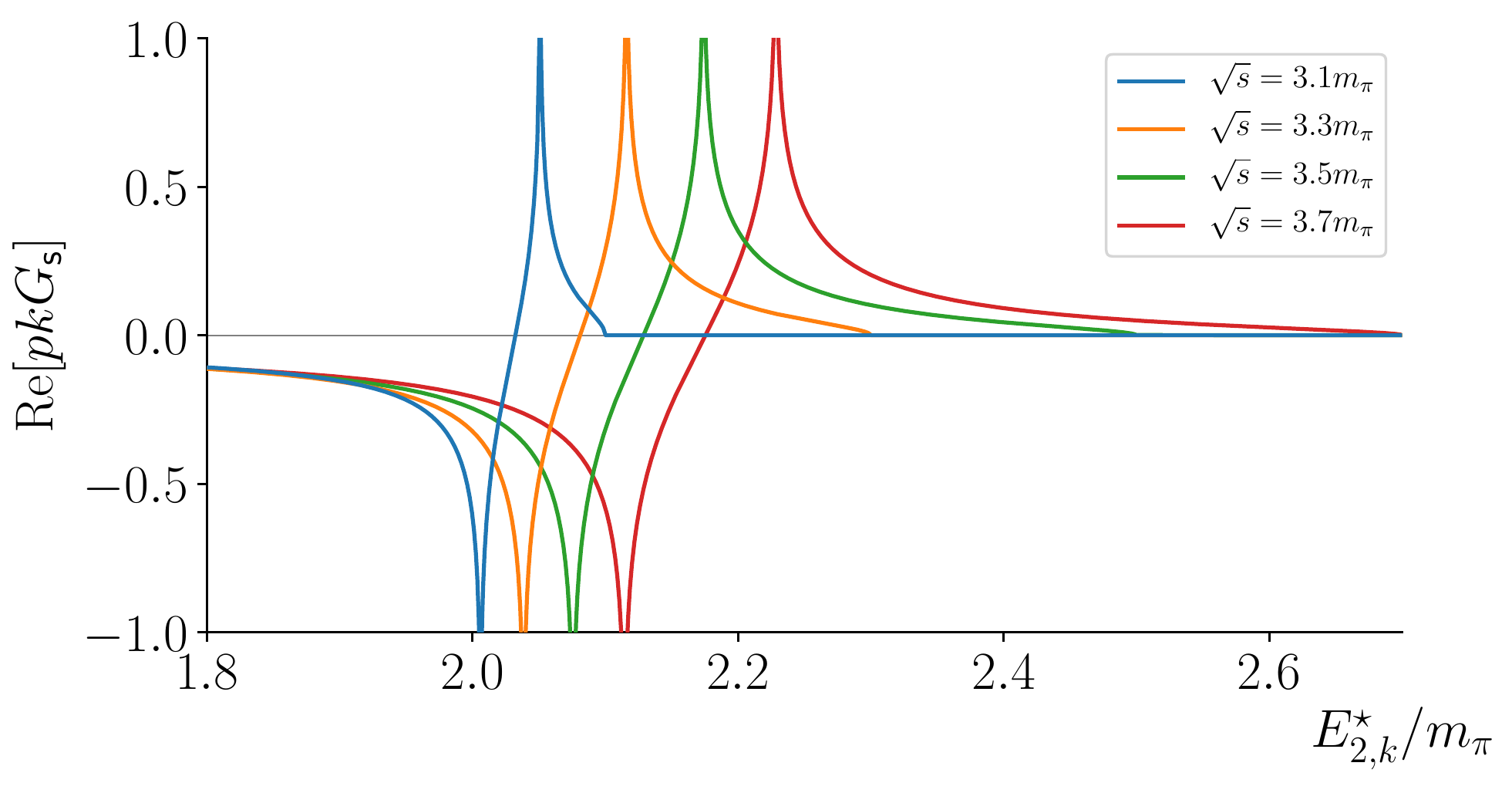} \includegraphics[width=0.45\textwidth]{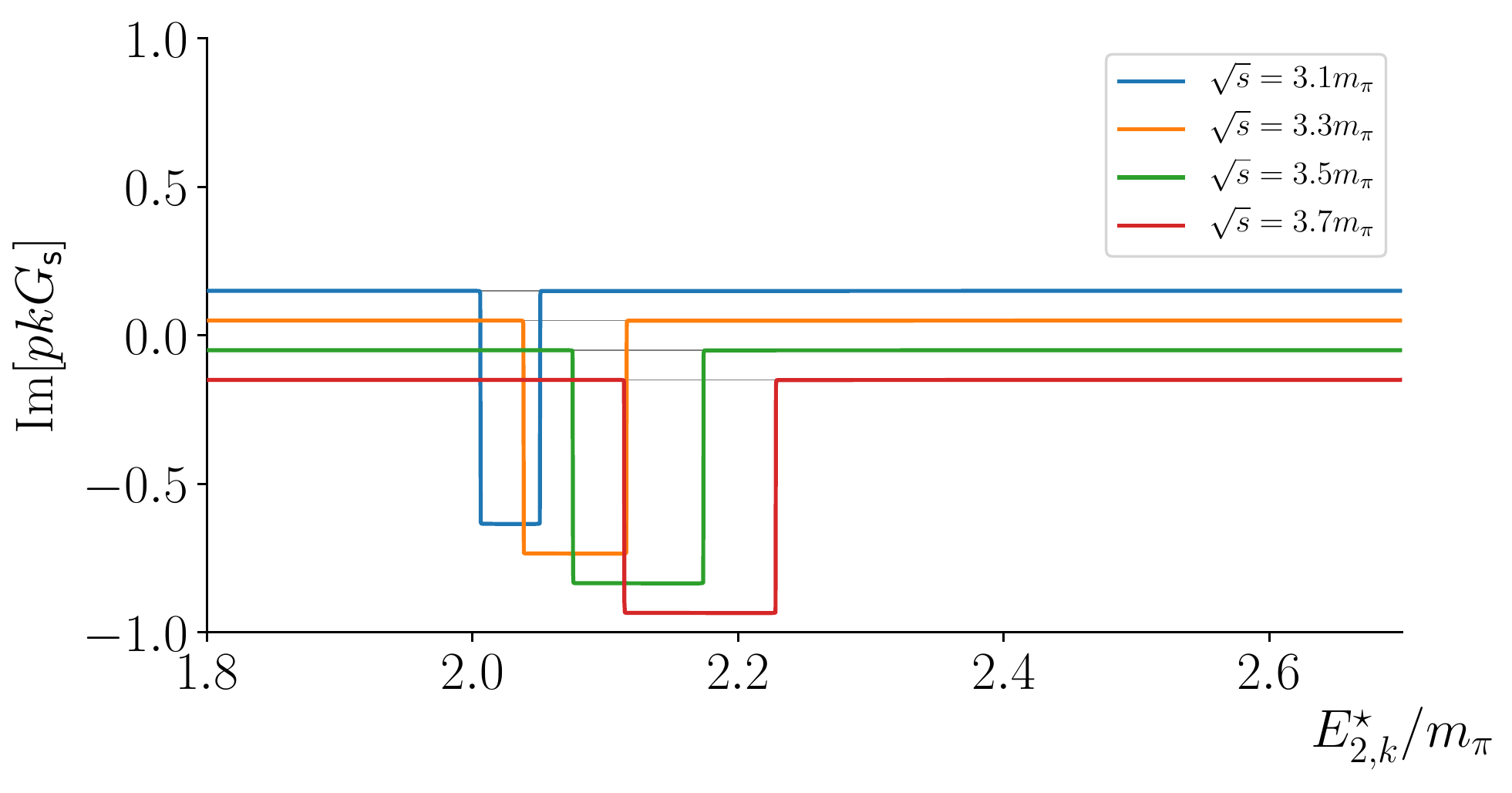}
\caption{Plots of the kernel function $G_{\sf s}(p,k,\epsilon)$ [defined Eq.~\eqref{eq:GS}] vs $E_{2,k}^\star$, here for four different energies as shown in the legend and with $p=0.1 m_\pi$ and $\epsilon \to 0^+$. For $\text{Im}[p \, k \, G_{\sf s}(p,k,\epsilon)]$, in the right panel, we have vertically off-set the curves for improved readability. }
\label{fig:Kernel_plots_E_dep}
\end{figure}

The next step is to project the remaining directional freedom within $\mathcal D^{(u,u)}$ onto vanishing three-particle angular momentum, i.e.~to $J=0$. Defining
\begin{align}
\cD^{(u,u)}_{{\sf s}}( p, k) 
&\equiv
\int \frac{d\Omega_{\hat{k}}}{4\pi} \frac{\,d\Omega_{\hat{p}}}{4\pi} \cD^{(u,u)}(\vec p, \vec k) \,,
\end{align}
one can show that $\cD_{\sf s}$ satisfies a one-dimensional integral equation of the form
\begin{align}
\cD^{(u,u)}_{{\sf s}}( p, k) &= - 
\cM_2( E_{2,p}^{\star}) G_{\sf s}( p, k,\epsilon) \cM_2( E_{2,k}^{\star}) 
- \cM_2( E_{2,p}^{\star}) 
\int_0^{k_{\sf max}} \frac{k'^2\,dk'}{(2 \pi)^2 \omega_{k'}} 
\,
G_{\sf s}( p, k',\epsilon)
\cD^{(u,u)}_{\sf s}( k', k),
\label{eq:Swaveproj_D}
\end{align}
where 
\begin{align}
G_{\sf s}( p, k,\epsilon)
& \equiv \int \frac{d\Omega_{\hat{p}}}{4\pi} \frac{\,d\Omega_{\hat{k}}}{4\pi}
G^\infty(\vec p, \vec k)
=- \frac{H(p,k)}{4 pk}
\log\left[
\frac{2pk - (E-\omega_k-\omega_p)^2 + p^2 + k^2 + m^2 - i\epsilon}
{-2pk - (E-\omega_k-\omega_p)^2 + p^2 + k^2 + m^2 - i\epsilon}
\right]
\,.
\label{eq:GS}
\end{align}
In Fig.~\ref{fig:Kernel_plots_E_dep} we plot $G_{\sf s}$ for a range of kinematic values. Setting $\Kiso = 0$ and combining Eqs.~\eqref{eq:M3uu}, \eqref{eq:Swaveproj_D} and \eqref{eq:GS}, we arrive at Eq.~\eqref{eq:M3_Kdf0} of the main text. Note that, in Eq.~\eqref{eq:Swaveproj_D}, we have included an explicit cutoff at $k_{\sf max} = (s+m^2)/(2\sqrt{s})$. This is done without any additional approximation as $H(p,k)$ has vanishing support for $k > k_{\sf max}$.

We comment here that, in general, projecting $\mathcal M_3^{(u,u)}$ to definite partial waves is not equivalent to doing the same for the symmetrized three-hadron scattering amplitude. However, as discussed around Eqs.~(15) and (16) of Ref.~\cite{Jackura:2018xnx} any definite-angular-momentum component of the the fully symmetric $\mathcal M_3$ can be assembled from known combinations of the various angular-momentum components of $\mathcal M_3^{(u,u)}$. Thus the latter form a basis for constructing the physical three-hadron amplitudes.

\bigskip
To solve Eq.~\eqref{eq:Swaveproj_D} numerically we replace the integral $\int_0^{k_{\sf max}} dk'$ with a discrete sum $\sum_{k'} \Delta k$ containing $N$ terms. Then a discretized version of the equation can be written in a matrix form
\begin{equation}
\label{eq:DmatrixEqn}
\boldsymbol D(N, \epsilon) 
= -
\boldsymbol \cM \cdot \boldsymbol G (\epsilon) \cdot \boldsymbol \cM - \boldsymbol \cM 
\cdot 
\boldsymbol G( \epsilon)
\cdot
\boldsymbol P
\cdot \boldsymbol D(N, \epsilon) \,,
\end{equation}
where we have introduced the following $N \times N$ matrix representations
\begin{align}
\boldsymbol G_{pk} (\epsilon)   = G_{\sf s}( p, k,\epsilon) \,, \qquad %
\boldsymbol \cM_{k'k}   = \delta_{k'k} \, \cM_2( E_{2,k}^{\star}) \,, \qquad %
\boldsymbol P_{k'k}   = \delta_{k' k} \, \frac{k^2 \Delta k}{(2 \pi)^2 \omega_{k}} \,,
\end{align}
as well as $ \boldsymbol D(N, \epsilon)$, which becomes our target quantity in the ordered double limit
\begin{equation}
\mathcal D^{(u,u)}_{\sf s}(p,k) = \lim_{\epsilon \to 0}\lim_{N \to \infty}\boldsymbol D_{pk}(N, \epsilon) \,.
\end{equation}
Here, in a slight abuse of notation, the indices $p\,k$ represent the choices that are closest to physical momenta $p$ and $k$ for a given $N$ value. Eq.~\eqref{eq:DmatrixEqn} can then be solved through a matrix inverse to yield
\begin{equation}
\boldsymbol D(N, \epsilon)
= -
\big [ \mathbb I +
\boldsymbol \cM 
\cdot 
\boldsymbol G( \epsilon)
\cdot
\boldsymbol P
\big]^{-1} \cdot
\boldsymbol \cM \cdot \boldsymbol G (\epsilon) \cdot \boldsymbol \cM \,.
\label{eq:FinalIE}
\end{equation}

Finally we return to the endcap factors appearing on either side of $\Kiso$ in Eq.~\eqref{eq:M3uu}. These are defined in Eqs.~(105) and (106) of Ref.~\cite{Hansen:2014eka} and in the overall $S$-wave approximation they take the form
\begin{equation}
\label{eq:EEdef}
\mathcal E^{(u)}_{\sf s}( p) =\frac{1}{3} - \mathcal M_2(E_{2,p}^\star) \rho(E_{2,p}^\star) - \int \! \frac{dp' p'^2}{(2 \pi)^2 \omega_{p'}} \mathcal D_{\sf s}^{(u,u)}( p, p') \rho( E_{2,p'}^\star) \,, %
\end{equation}
where
\begin{align}
 \rho(E_{2,p}^\star) &\equiv - i \frac{J\big ( E_{2,k}^{\star 2} / [2 m]^2 \big )}{16 \pi  E_{2,k}^{\star}} 
\sqrt{E_{2,p}^{\star2}/4-m^2} \,, 
\label{eq:rhodef}
\end{align}
and it is understood that the the $i \sqrt{-x}$ branch is taken for $x<0$.

As $\Kiso$ is consistent with zero in all fits we have considered, the main purpose in keeping track of these quantities is to estimate the propagation of uncertainties into $\mathcal M_3$. In particular we note
\begin{equation}
\label{eq:M3unc}
\Delta \mathcal M_3^2 = \sum_{\eta, \eta'} [ \partial_{\eta} \mathcal M_3] \, C_{\eta \eta'} \, [\partial_{\eta'} \mathcal M_3] \,, 
\end{equation}
where $C_{\eta \eta'}$ represents the fit-parameter covariance matrix  
and the sums run over all inputs to the scattering amplitude, in particular the scattering length $a_0$ as well as $\Kiso$. In the next section we present numerical solutions for $\mathcal M_3$ based on the parameters extracted from the finite-volume energies.

\bigskip

\subsection{Solutions for $\mathcal M_3$}
\label{sec:NumSolveM3}

\begin{figure}[H]
\centering
\includegraphics[width=0.49\textwidth]{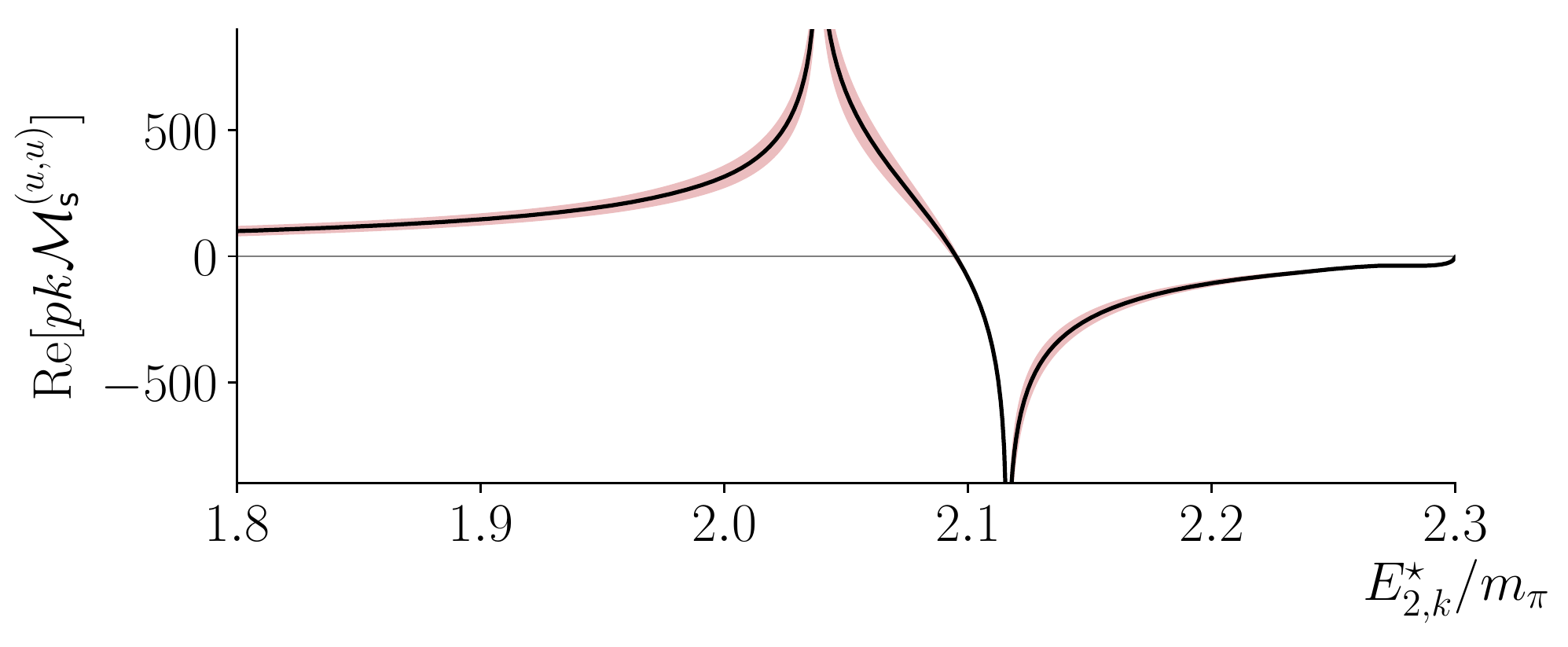}
\includegraphics[width=0.49\textwidth]{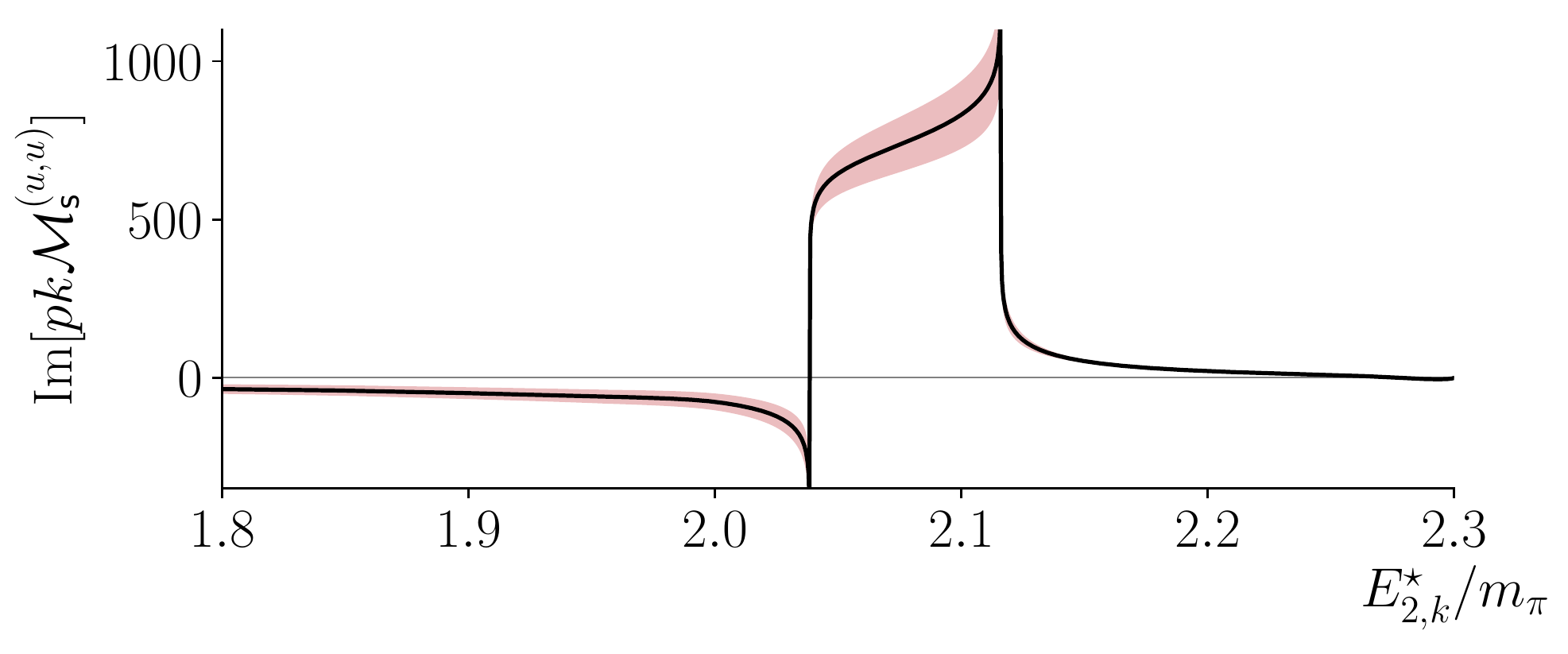}
\caption{Real (left) and imaginary (right) parts of $p \, k \, \mathcal M^{(u,u)}_{\sf s}$, determined by solving Eq.~\eqref{eq:FinalIE} using $m_\pi \A=0.296$ and $\Kiso = 0$ for the central values. As we explain in the text, the uncertainties here follow from propagating the uncertainties on $\Kiso$ and $m_\pi \A$, taken from Fit B${}_{2+3}$ in Table~\ref{tab:simult}.
}
\label{fig:Duu}
\end{figure}

\bigskip

Figure~\ref{fig:Duu} shows the result of solving Eq.~\eqref{eq:FinalIE} and varying $N$ and $\epsilon$ in order to ensure that the ordered double limit has been saturated. Symmetrizing this function over the three incoming and three outgoing momenta yields the full three-hadron scattering amplitude, plotted in Fig.~\ref{fig:3piAmplitudesVSm} of the main text.

To assign an uncertainty estimate to $\mathcal M^{(u,u)}_{\sf s}(p,k)$ we have applied a straightforward adaptation of Eq.~\eqref{eq:M3unc}. Specializing to the case of only $m_\pi \A$ and $\Kiso$ as input parameters, and neglecting the correlations between them, this becomes
\begin{equation}
\label{eq:DeltaMuu}
\big ( \Delta \mathcal M^{(u,u)}_{\sf s}(p,k) \big )^2 = \big ( \Delta [m_\pi \A]  \big )^2 \bigg ( \frac{\partial \mathcal M^{(u,u)}_{\sf s}(p,k) }{\partial (m_\pi \A)} \bigg )^2 +   \big ( \Delta \Kiso/9 \big )^2 + \mathcal O(\Delta \Kiso \rho \mathcal M_2)^2 \,,
\end{equation}
where $\Delta (m_\pi \A)$ and $\Delta \Kiso$ denote the uncertainties on the input parameters and $ \Delta \mathcal M^{(u,u)}_{\sf s}(p,k)$ is the resulting amplitude uncertainty plotted in the figure. In practice one finds that $\Delta \Kiso$ contributes negligibly to the overall uncertainty, simply because the series of $\mathcal M_2$ and $G_{\sf s}$ insertions dominates the value of the amplitude for weakly-interacting systems. For this reason we have taken the leading part of $\mathcal E_{\sf s}^{(u)}$, resulting in the $\Delta \Kiso/9$ term. The additional corrections, indicated by the final term, are negligible for this system. 

The dominant source of uncertainty enters through the scattering length, i.e.~the first term in Eq.~\eqref{eq:DeltaMuu}. To estimate this, we have numerically performed the derivative with respect to $m_\pi \A$. Since $ \mathcal M^{(u,u)}_{\sf s}(p,k)$ is dominated by the contribution proportional to $\mathcal M_2^2$, in this case the overall uncertainty is well approximated by 
\begin{equation}
\frac{ \Delta \mathcal M^{(u,u)}_{\sf s}(p,k) }{\mathcal  M^{(u,u)}_{\sf s}(p,k) } = 2 \frac{ \Delta [m_\pi \A] }{m_\pi \A} \,.
\end{equation}
To produce Fig.~\ref{fig:Duu} we have used $\Delta[m_\pi \A] = 0.016$ and $\Delta \Kiso = 770$. These are taken from B${}_{2+3}$ in Table~\ref{tab:simult} with the uncertainty on $m_\pi \A$ doubled to account for systematic variations between the various fits that have been performed.

\subsection{Operator construction and lists}
\label{app:operators}

Following Ref.~\cite{Dudek:2012gj}, to determine the $\pi\pi$ $I=2$ finite-volume energies we compute correlation functions featuring operators constructed to resemble a $\pi\pi$ structure. Schematically, for an operator in lattice irrep $\Lambda_{12}$ and row $\mu_{12}$ with overall momentum $\vec{k_{12}}$,
\begin{equation}
(\pi\pi)^{[\vec{k_1},\vec{k_2}]\dagger}_{\Lambda_{12} \, \mu_{12}}(\vec{k_{12}}) =
\sum\limits_{\substack{\vec{k_1}, \, \vec{k_2} \\ \vec{k_1}+\vec{k_2} = \vec{k_{12}}}}
\mathcal{C}(\vec{k_{12}}, \Lambda_{12}, \mu_{12}; \vec{k_1}, \Lambda_1; \vec{k_2}, \Lambda_2) \;
\pi^\dagger_{\Lambda_1}(\vec{k_1}) \; \pi^\dagger_{\Lambda_2}(\vec{k_2}) \, ,
\label{equ:2piops}
\end{equation}
where the sum is over all momenta related to $\vec{k_1}$ and $\vec{k_2}$ by allowed lattice rotations, $\mathcal{C}$ is an appropriate generalized Clebsch-Gordan coefficient for $\Lambda_1 \otimes \Lambda_2 \to \Lambda_{12}$, and flavor indices and the projection onto $I=2$ are not written explicitly. Here $\pi_{\Lambda_i}^\dagger(\vec{k_i})$ is the optimal linear combination of operators to interpolate a $\pi$ with momentum $\vec{k_i}$ in irrep%
\footnote{These are all one dimensional and so we omit the irrep row index.}
$\Lambda_i$ using a basis of fermion-bilinear operators featuring various Dirac $\gamma$ matrices and gauge-covariant derivatives -- see Ref.~\cite{Dudek:2012gj} for details. In this work the basis of fermion-bilinear operators used for a $\pi$ operator has up to three derivatives for $\pi$ at rest and up to one derivative for $\pi$ at non-zero momentum, except we use up to two derivatives for $1 \leq |\vec{k_i}|^2 \leq 4$ on the $24^3$ volume. The operators used to compute the $\pi\pi$ spectra shown in Fig.~\ref{fig:FVspec} of the main text are listed in Table~\ref{table:ops:2pi}.

{
\renewcommand{\arraystretch}{1.5}
\begin{table}[tb]
\begin{tabular}{c|l|l}
%\hline
$\vec{P}$ $\Lambda$ & $L/a_s=20$ & $L/a_s=24$ \\
\hline \hline
[000] $A_1^+$
& $\pi_{[000]} \pi_{[000]}$, $\pi_{[100]} \pi_{[100]}$, $\pi_{[110]} \pi_{[110]}$, $\pi_{[111]} \pi_{[111]}$,
& $\pi_{[000]} \pi_{[000]}$, $\pi_{[100]} \pi_{[100]}$, $\pi_{[110]} \pi_{[110]}$, $\pi_{[111]} \pi_{[111]}$, \\
& $\pi_{[200]} \pi_{[200]}$
& $\pi_{[200]} \pi_{[200]}$ \\
\hline
[100] $A_1$
& $\pi_{[000]} \pi_{[100]}$, $\pi_{[100]} \pi_{[110]}$, $\pi_{[110]} \pi_{[111]}$, $\pi_{[100]} \pi_{[200]}$
& $\pi_{[000]} \pi_{[100]}$, $\pi_{[100]} \pi_{[110]}$, $\pi_{[110]} \pi_{[111]}$, $\pi_{[100]} \pi_{[200]}$, \\
&
& $\pi_{[110]} \pi_{[210]}$, $\pi_{[200]} \pi_{[210]}$, $\pi_{[111]} \pi_{[211]}$ \\
\hline
[110] $A_1$
& $\pi_{[000]} \pi_{[110]}$, $\pi_{[100]} \pi_{[100]}$, $\pi_{[100]} \pi_{[111]}$, $\pi_{[110]} \pi_{[110]}$
& $\pi_{[000]} \pi_{[110]}$, $\pi_{[100]} \pi_{[100]}$, $\pi_{[100]} \pi_{[111]}$, $\pi_{[110]} \pi_{[110]}$, \\
& 
& $\pi_{[110]} \pi_{[200]}$, $\pi_{[100]} \pi_{[210]}$, $\pi_{[111]} \pi_{[210]}$, $\pi_{[110]} \pi_{[211]}$ \\
\hline
[111] $A_1$
& $\pi_{[000]} \pi_{[111]}$, $\pi_{[100]} \pi_{[110]}$
& $\pi_{[000]} \pi_{[111]}$, $\pi_{[100]} \pi_{[110]}$, $\pi_{[111]} \pi_{[200]}$, $\pi_{[110]} \pi_{[210]}$, \\
&
& $\pi_{[100]} \pi_{[211]}$ \\
\hline
[200] $A_1$
& $\pi_{[100]} \pi_{[100]}$, $\pi_{[000]} \pi_{[200]}$, $\pi_{[110]} \pi_{[110]}$, $\pi_{[111]} \pi_{[111]}$
& $\pi_{[100]} \pi_{[100]}$, $\pi_{[000]} \pi_{[200]}$, $\pi_{[110]} \pi_{[110]}$, $\pi_{[100]} \pi_{[210]}$, \\
&
& $\pi_{[111]} \pi_{[111]}$, $\pi_{[110]} \pi_{[211]}$, $\pi_{[210]} \pi_{[210]}$ \\
%\hline
\end{tabular}
\caption{The $\pi\pi$ $I=2$ operators, $\pi_{\vec{k_1}} \pi_{\vec{k_2}}$, used to compute the finite-volume energy levels shown in Fig.~\ref{fig:FVspec} of the main text (upper plots) in irrep $\Lambda$ with overall momentum $\vec{P}$. These are constructed from optimized $\pi$ operators with momentum types $\vec{k_1}$ and $\vec{k_2}$; different momentum directions are summed over as in Eq.~(\ref{equ:2piops}). Momenta are displayed using the shorthand notation $[ijk]=\frac{2\pi}{L}(i,j,k)$.}
\label{table:ops:2pi}
\vspace{10pt}
\end{table}
}

In a similar way, operators used to compute $\pi\pi\pi$ $I=3$ energies resemble a $\pi\pi\pi$ structure and are formed by combining a $\pi\pi$ $I=2$ operator with a $\pi$ operator, as detailed in Ref.~\cite{Woss:2019hse}. Schematically, for an operator in lattice irrep $\Lambda$ and row $\mu$ with overall momentum $\vec{P}$,
\begin{equation}
(\pi\pi\pi)^{[\vec{k_{12}}[\vec{k_1},\vec{k_2}],\vec{k_3}]\dagger}_{\Lambda \, \mu}(\vec{P}) =
\sum\limits_{\substack{\vec{k_{12}}, \, \vec{k_3} \\ \vec{k_{12}}+\vec{k_3} = \vec{P}}}
\mathcal{C}(\vec{P}, \Lambda, \mu; \vec{k_{12}}, \Lambda_{12}, \mu_{12}; \vec{k_3}, \Lambda_3) \;
(\pi\pi)^{[\vec{k_1},\vec{k_2}]\dagger}_{\Lambda_{12} \, \mu_{12}}(\vec{k_{12}}) \; \pi^\dagger_{\Lambda_3}(\vec{k_3}) \, ,
\label{equ:3piops}
\end{equation}
where the sum is over all momenta related to $\vec{k_{12}}$ and $\vec{k_3}$ by allowed lattice rotations and, again, flavor indices and the projection onto $I=3$ are not written explicitly.

From Bose symmetry, a $\pi\pi\pi$ system must be symmetric under the interchange of any pair of pions. The pions have no intrinsic spin and we are considering $I=3$ which means that the flavor structure is symmetric under interchange of any pair; therefore, the spatial structure must also be symmetric under the interchange of any pair of pions. The operator construction in Eq.~(\ref{equ:3piops}) gives operators with the correct symmetry properties because the three pions are identical, but it does not make this symmetry manifest and two different sets of ($|\vec{k_1}|$, $|\vec{k_2}|$, $|\vec{k_3}|$, $|\vec{k_{12}}|$, $\Lambda_{12}$) may lead to equivalent operators, or a number of different sets may give linearly-dependent operators.%
\footnote{Strictly we mean the types of momenta (i.e.~the equivalence class of momenta related by rotations in the octahedral group or little group) rather than the magnitudes, but there is no distinction for the momenta we are considering here.} 
To ensure we have an appropriate set of independent operators, for each $\vec{P}$ and $\Lambda$, we construct an operator for every possible set \linebreak ($|\vec{k_1}|$, $|\vec{k_2}|$, $|\vec{k_3}|$, $|\vec{k_{12}}|$, $\Lambda_{12}$) with $|\vec{k_1}|^2 + |\vec{k_2}|^2 + |\vec{k_3}|^2$ less than some cutoff. 
We write Eq.~\eqref{equ:3piops} schematically as,
\begin{equation}
(\pi\pi\pi)_{\Lambda  \mu}(\vec{P})  = \sum \widetilde{\mathcal{C}}(\vec{k_1},\vec{k_2},\vec{k_3}) \, \pi^\dagger(\vec{k_1}) \pi^\dagger(\vec{k_2}) \pi^\dagger(\vec{k_3})  = \widetilde { \mathcal C} \cdot \mathcal V_{[\pi \pi \pi]}\, ,
\end{equation}
where $\widetilde {\mathcal C}$ with no labels represents a row vector of coefficients and $\mathcal V_{[\pi \pi \pi]}$ a column of operators, such that the dot-product reproduces the sum. We then introduce the matrix $R(ijk)$ which acts on $\mathcal V_{[\pi \pi \pi]}$ by mapping a given entry $\pi^\dagger(\vec p_1) \pi^\dagger(\vec p_2) \pi^\dagger(\vec p_3)$ into $\pi^\dagger(\vec p_i) \pi^\dagger(\vec p_j) \pi^\dagger(\vec p_k)$.
This allows us to define the symmetrized vector of Clebsch-Gordan coefficients
\begin{equation}
\widetilde {\mathcal C} \cdot R(123)  + \widetilde {\mathcal C} \cdot R(231)  + \widetilde {\mathcal C} \cdot R(312)  + \widetilde {\mathcal C} \cdot R(132)  + \widetilde {\mathcal C} \cdot R(213)  + \widetilde {\mathcal C} \cdot R(321)   \,,
\end{equation}
where $R(123)$ is just the identity matrix.
The final step is to check that the resulting vector is non-zero and linearly independent from the analogous expressions for the already considered operators.
The resulting sets of independent operators used in this work are listed in Tables~\ref{table:ops:3pi:0}, \ref{table:ops:3pi:1}, \ref{table:ops:3pi:2a}, \ref{table:ops:3pi:2b} and \ref{table:ops:3pi:3}.\footnote{This is not always a unique choice and any independent set of operators could be used to achieve the same results.} In extracting the energies, subsets of the full operator set were also considered, in order to investigate sensitivity to the detailed choice of operator basis.

Finally, in order to give further intuition into the operators used, we also include a diagrammatic representation of the individual pion momentum assignments in the tables. The diagrams portray the integer vectors $\vec d_1$, $\vec d_2$, $\vec d_3$, each given by $\vec d_i = L \vec k_i/(2 \pi)$. The vectors are assigned a color (orange and green for the first two pions, and blue for the third) and the absence of any given color corresponds to a vector of magnitude zero.  
As summarized by Eq.~\eqref{equ:3piops}, our operator construction is based on combining two-pion operators in a definite irrep with the third pion.
An alternative basis is given by summing a given momentum assignment, represented by a given set $\vec d_1 \vec d_2 \vec d_3$, over all rotations in the octahedral group (in the case of $\vec P=[000]$) or else a little group thereof (for non-zero total momentum) weighted by the appropriate Clebsch-Gordan coefficients. The operators reached via this alternative construction are equal to a linear combination of those given by Eq.~\eqref{equ:3piops}. 

For example, on the third line of Table~\ref{table:ops:3pi:1}, two distinct momentum assignments arise from combining the $\pi \pi_{[000]A_1^+}$ with the third $\pi_{[100]}$ operator. In this case the $\pi \pi_{[000]A_1^+}$ is built from individual pions with a unit of back-to-back momentum. When one sums over the coefficients projecting onto $A_1^+$, contributions arise with the back-to-back axis both aligned and perpendicular to the total momentum direction, $\vec P=[001]$. Thus, momentum assignments corresponding to both diagrams shown in line 3 of Table~\ref{table:ops:3pi:1} contribute to the operator on that line. By contrast, on line 4 of Table~\ref{table:ops:3pi:1} only a single momentum configuration contributes, as indicated. This implies that the two operators are independent, since they are built from independent linear combinations of the two momentum configurations. 

Operators 8 through 11 of Table~\ref{table:ops:3pi:1} give a more complicated example. The first two (8 and 9) correspond to two linear combinations of two configuration types, labeled with subscripts 1 and 3, and the next two (10 and 11) are equal to linear combinations of the operators labeled 1, 2, and 4. 

The diagrams in Tables~\ref{table:ops:3pi:0}, \ref{table:ops:3pi:1}, \ref{table:ops:3pi:2a}, \ref{table:ops:3pi:2b} and \ref{table:ops:3pi:3} provide a cross check on the linear-independence of the operators.  Each row corresponds to a linear combination of the displayed momentum configurations and the number of linearly independent operators is equal to the number of distinct diagrams.

\begin{table}[h]
\begin{center}
\setlength\extrarowheight{5pt}
\begin{tabular}{  c | c || c | c | c || c || c  }
%\hline
& \ $\boldsymbol{d_1^2 \, d_2^2 \, d_3^2}$ \ & $\pi_{\vec{k_1}}$ & \ $\pi_{\vec{k_2}}$ & \ $\pi\pi_{\vec{k_{12}} \Lambda_{12}}$ ($I=2$) \ & $\pi_{\vec{k_3}}$ & \ \ momentum configurations\ \  \\[5pt]
\hline \hline
\ 1 \ &\ \ $\textbf{000}$ \ \ & $ \ \pi_{[000]}\ $ & $\ \pi_{[000]}\ $ & $ \ \pi\pi_{[000] A_1^+} \ $ & $\ \pi_{[000]}\ $ &  \begin{tabular}{c} \includegraphics[scale=0.15]{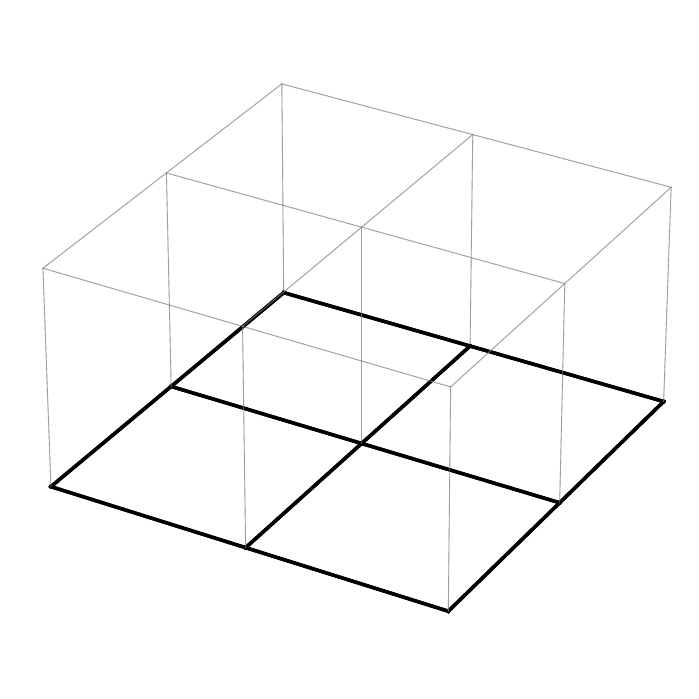}  \vspace{-4pt}  \end{tabular}  \\[7pt] \hline \hline
\ 2 \ &\ \ $\textbf{011}$ \ \ & $ \ \pi_{[100]}\ $ & $\ \pi_{[000]}\ $ & $ \ \pi\pi_{[100] A_1} \ $ & $\ \pi_{[100]}\ $ &    \begin{tabular}{c} \includegraphics[scale=0.15]{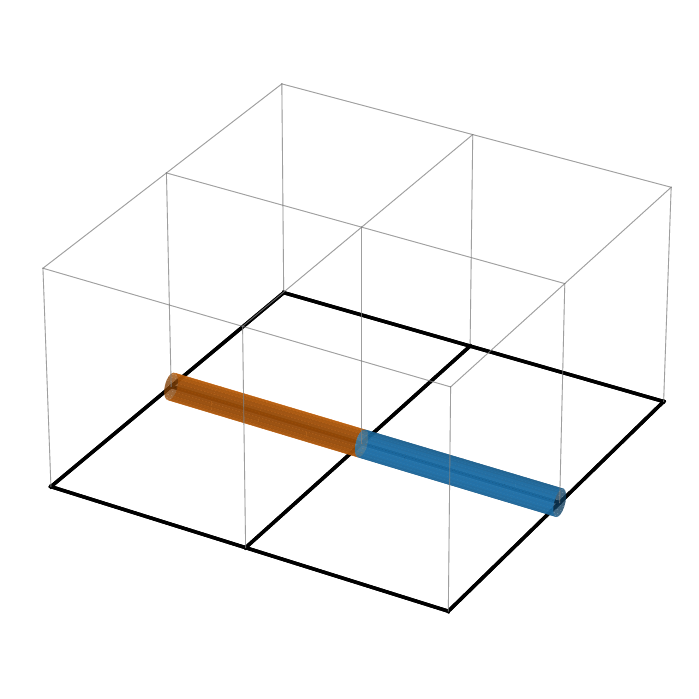} \vspace{-4pt}  \end{tabular}  \\[7pt] \hline \hline
\ 3 \ &\ \ $\textbf{022}$ \ \ & $ \ \pi_{[110]}\ $ & $\ \pi_{[000]}\ $ & $ \ \pi\pi_{[110] A_1} \ $ & $\ \pi_{[110]}\ $  &    \begin{tabular}{c} \includegraphics[scale=0.15]{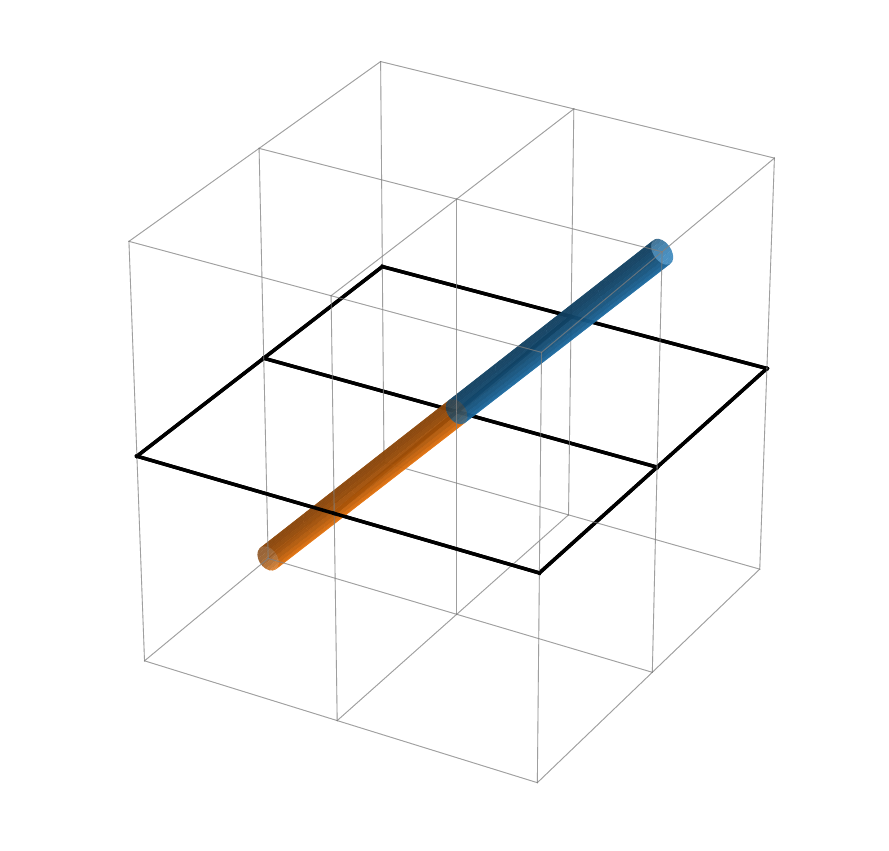} \vspace{-2pt} \end{tabular}  \\[7pt]  \hline
\ 4 \ &\ \ $\textbf{112}$ \ \ & $ \ \pi_{[100]}\ $ & $\ \pi_{[100]}\ $ & $ \ \pi\pi_{[110] A_1} \ $ & $\ \pi_{[110]}\ $ & \begin{tabular}{c} \includegraphics[scale=0.15]{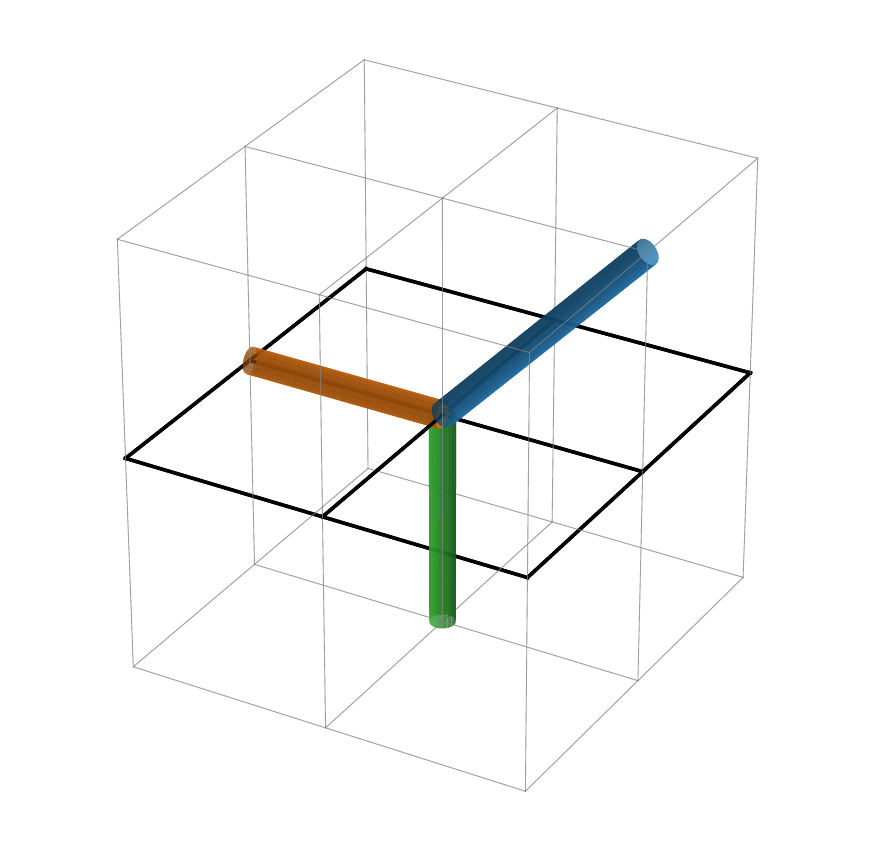} \vspace{-2pt} \end{tabular}  \\[7pt]   \hline \hline
\ {5} \ &\ \ $\textbf{033}$ \ \ & $ \ \pi_{[111]}\ $ & $\ \pi_{[111]}\ $ & $ \ \pi\pi_{[000] A_1^+} \ $ & $\ \pi_{[000]}\ $ &  \begin{tabular}{c} \includegraphics[scale=0.15]{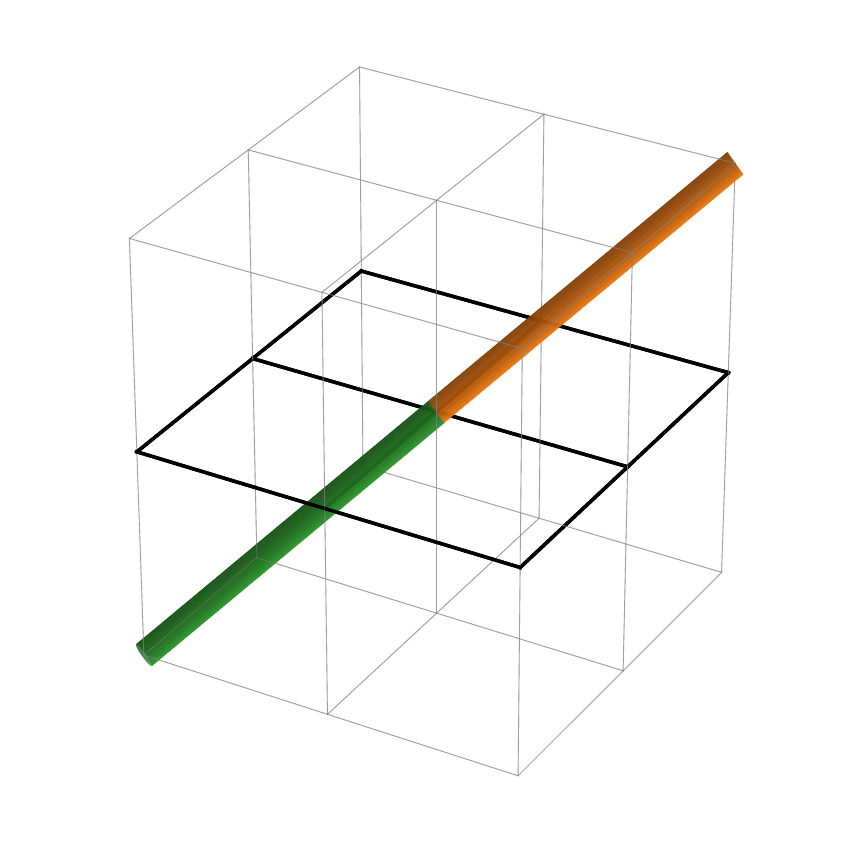} \vspace{-2pt} \end{tabular}  \\[7pt]   \hline 
\ {6} \ &\ \ $\textbf{114}$ \ \ & $ \ \pi_{[200]}\ $ & $\ \pi_{[100]}\ $ & $ \ \pi\pi_{[100] A_1} \ $ & $\ \pi_{[100]}\ $  &  \begin{tabular}{c}  \vspace{-17pt} \\\includegraphics[scale=0.21]{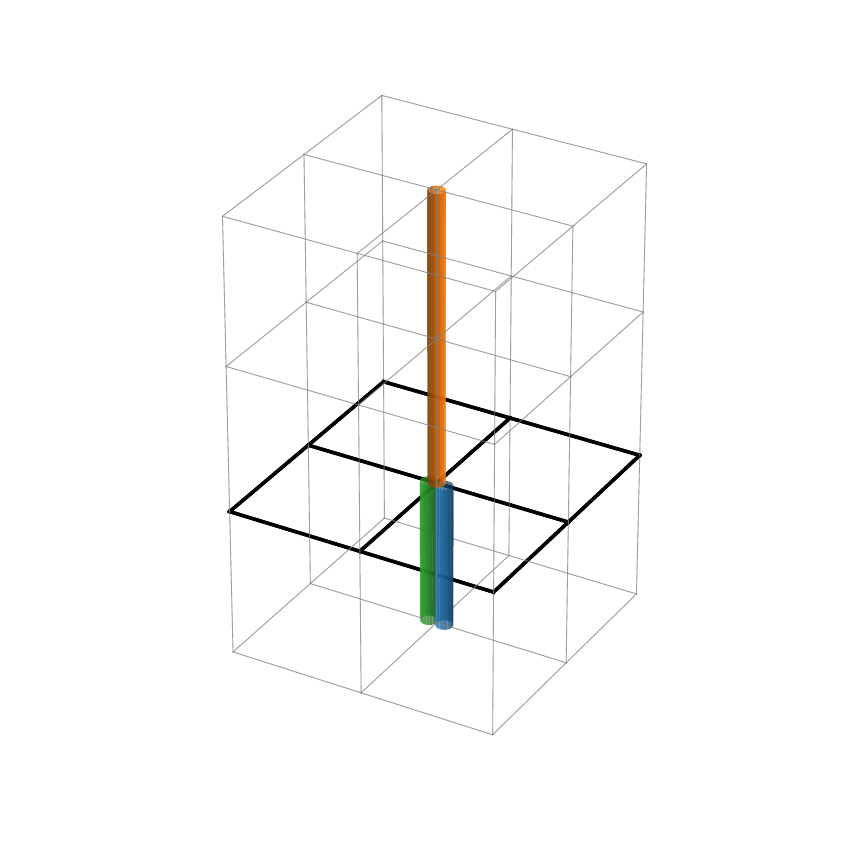}  \end{tabular} \vspace{-5pt}  \\[7pt]  \hline
\ {7} \ &\ \ $\textbf{123}$ \ \ & $ \ \pi_{[111]}\ $ & $\ \pi_{[100]}\ $ & $ \ \pi\pi_{[110] A_1} \ $ & $\ \pi_{[110]}\ $  &   \begin{tabular}{c} \includegraphics[scale=0.15]{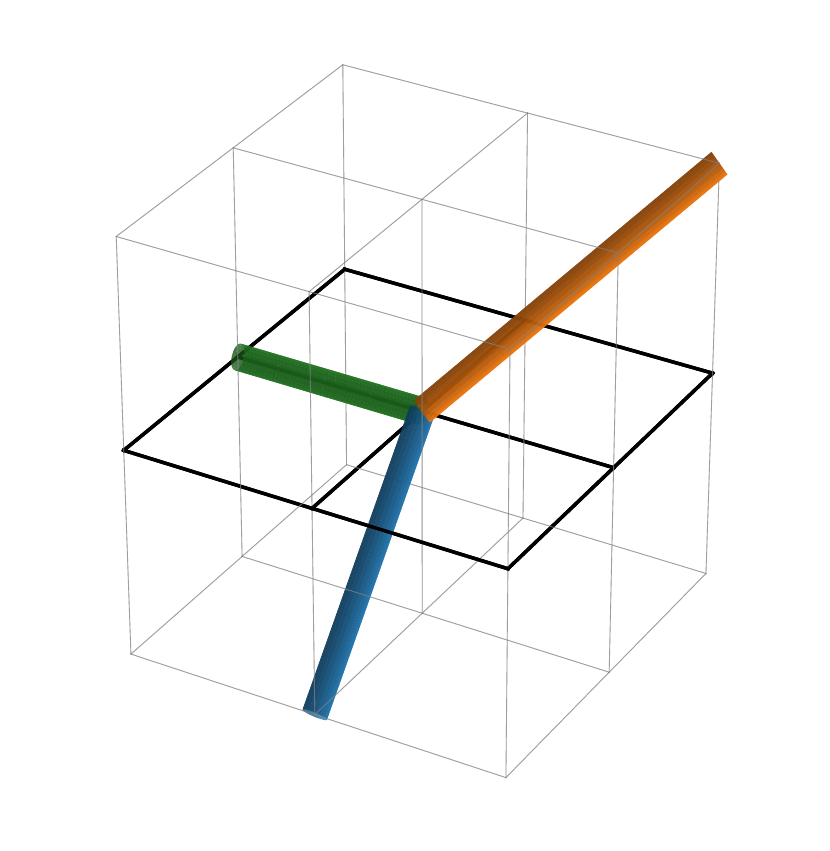} \vspace{-2pt}  \end{tabular}  \\[7pt]   \hline
\ 8 \ &\ \ $\textbf{222}$ \ \ & $ \ \pi_{[110]}\ $ & $\ \pi_{[110]}\ $ & $ \ \pi\pi_{[110] A_1} \ $ & $\ \pi_{[110]}\ $  &   \begin{tabular}{c} \includegraphics[scale=0.15]{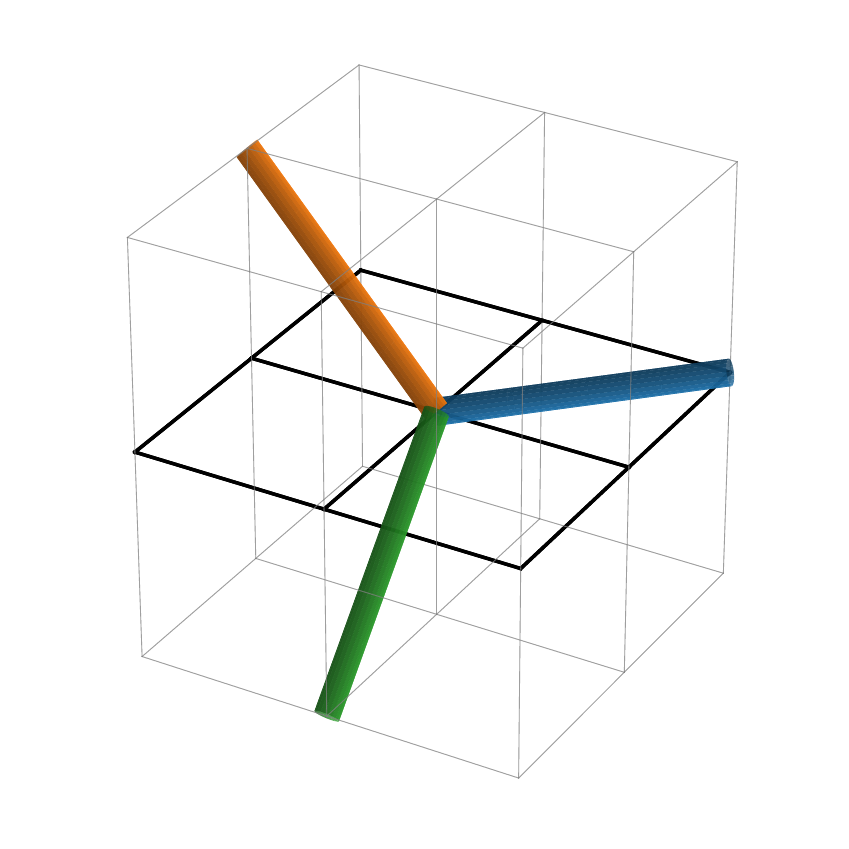} \vspace{-2pt} \end{tabular}  \\[7pt]  \hline \hline
\ 9 \ &\ \ $\textbf{044}$ \ \ & $ \ \pi_{[200]}\ $ & $\ \pi_{[000]}\ $ & $ \ \pi\pi_{[200] A_1} \ $ & $\ \pi_{[200]}\ $  &  \begin{tabular}{c} \vspace{-22pt} \\ \includegraphics[scale=0.29]{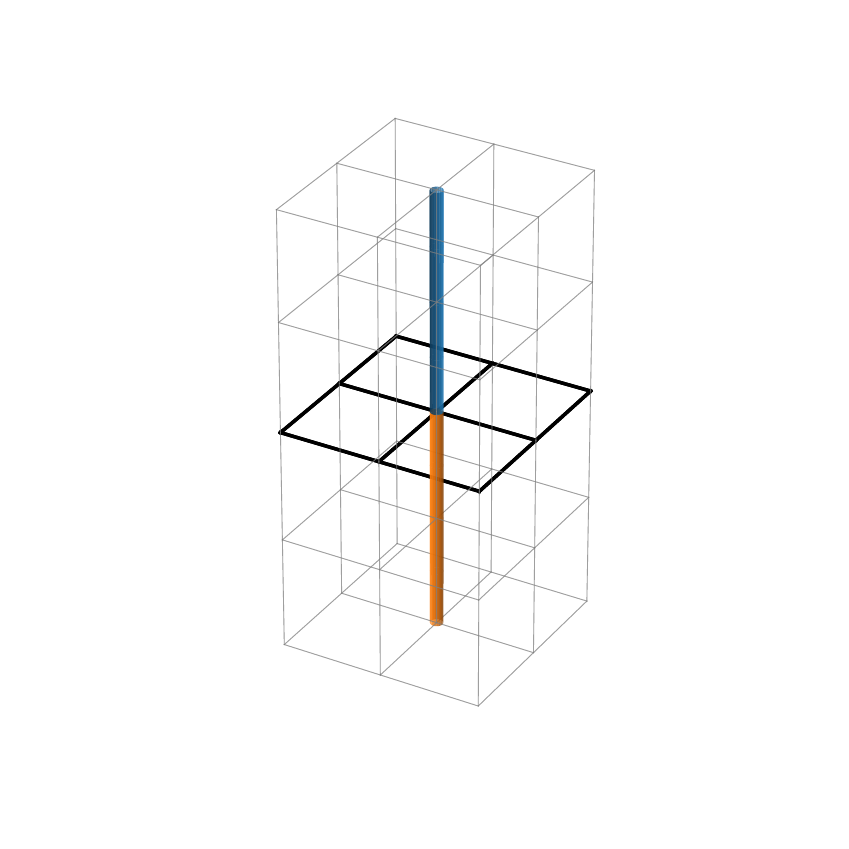} \vspace{-10pt} \end{tabular}  \\[7pt]  \hline
\ {10} \ &\ \ $\textbf{125}$ \ \ & $ \ \pi_{[210]}\ $ & $\ \pi_{[100]}\ $ & $ \ \pi\pi_{[110] A_1} \ $ & $\ \pi_{[110]}\ $   & \begin{tabular}{c}  \vspace{-17pt} \\ \includegraphics[scale=0.23]{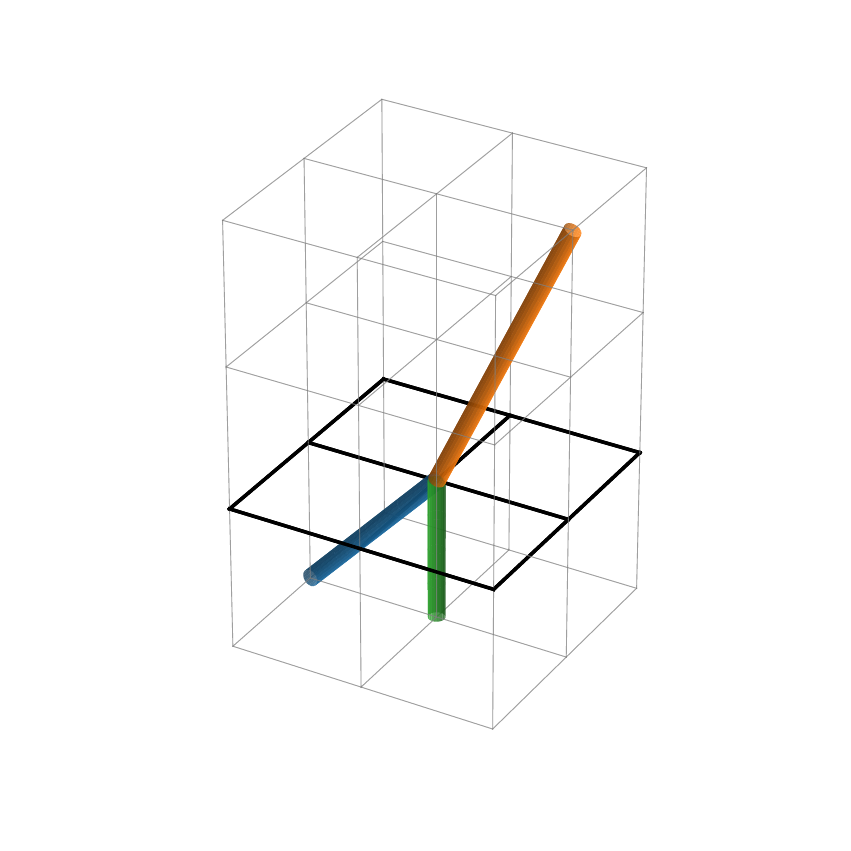} \vspace{-5pt} \end{tabular}  %\\[5pt]  %\hline
\end{tabular}
\end{center}
\caption{The $\pi\pi\pi$ $I=3$ operators used to compute the finite-volume energy levels shown in Fig.~\ref{fig:FVspec} of the main text (lower left plot) in irrep $A_1^-$ with overall momentum $\vec{P}=[000]$, labeled by $\vec d_1^2 \, \vec d_2^2 \, \vec d_3^2$ where $\vec{d_i} = \vec{k_i}(L/2\pi)$. Different momentum directions for the momentum types $\vec{k_1}$, $\vec{k_2}$, $\vec{k_3}$ and $\vec{k_{12}}$ are summed over as in Eq.~(\ref{equ:3piops}). Operators 1 to 8 are used on the $20^3$ volume and operators 1 to 10 are used on the $24^3$ volume. The momentum configuration diagrams in the rightmost column are explained in the text. Operators separated by a single horizontal line correspond to states that are degenerate in the non-relativistic, non-interacting theory.}
\label{table:ops:3pi:0}
\end{table}

\begin{table}
{
\centering
\begin{center}
\setlength\extrarowheight{5pt}
\begin{tabular}{  c | c || c | c | c || c || c  }
%\hline
& \ $\boldsymbol{d_1^2 \, d_2^2 \, d_3^2}$ \ & $\pi_{\vec{k_1}}$ & \ $\pi_{\vec{k_2}}$ & \ $\pi\pi_{\vec{k_{12}} \Lambda_{12}}$ ($I=2$) \ & $\pi_{\vec{k_3}}$ & \ \ momentum configurations\ \  \\[5pt]
\hline \hline
\   1 \ &\ \ $\textbf{001}$ \ \ & $ \ \pi_{[000]}\ $ & $\ \pi_{[000]}\ $ & $ \ \pi\pi_{[000] A_1^+} \ $ & $\ \pi_{[100]} \ $ & \begin{tabular}{c} \includegraphics[scale=0.15]{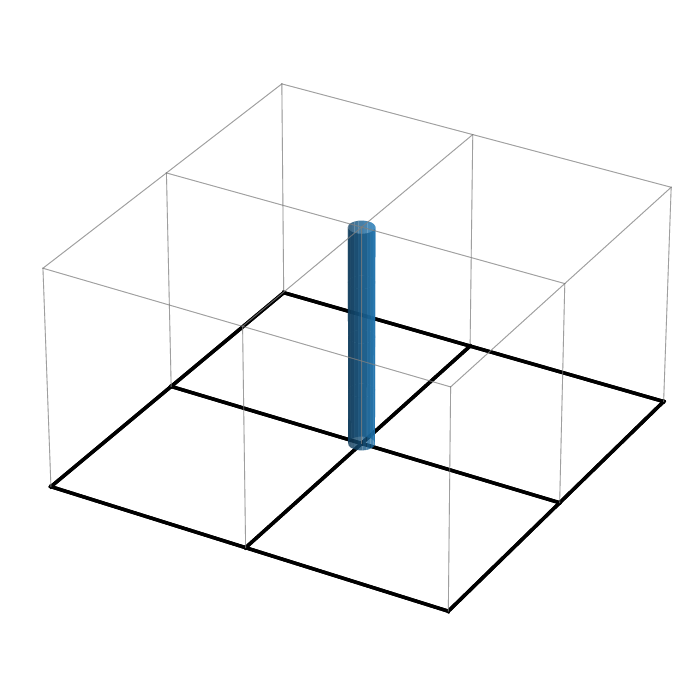} \vspace{-5pt} \end{tabular} \\[7pt]\hline \hline
\   2 \ &\ \ $\textbf{012}$ \ \ & $ \ \pi_{[100]}\ $ & $\ \pi_{[000]}\ $ & $ \ \pi\pi_{[100] A_1} \ $ & $\ \pi_{[110]} \ $ & \begin{tabular}{c} \includegraphics[scale=0.15]{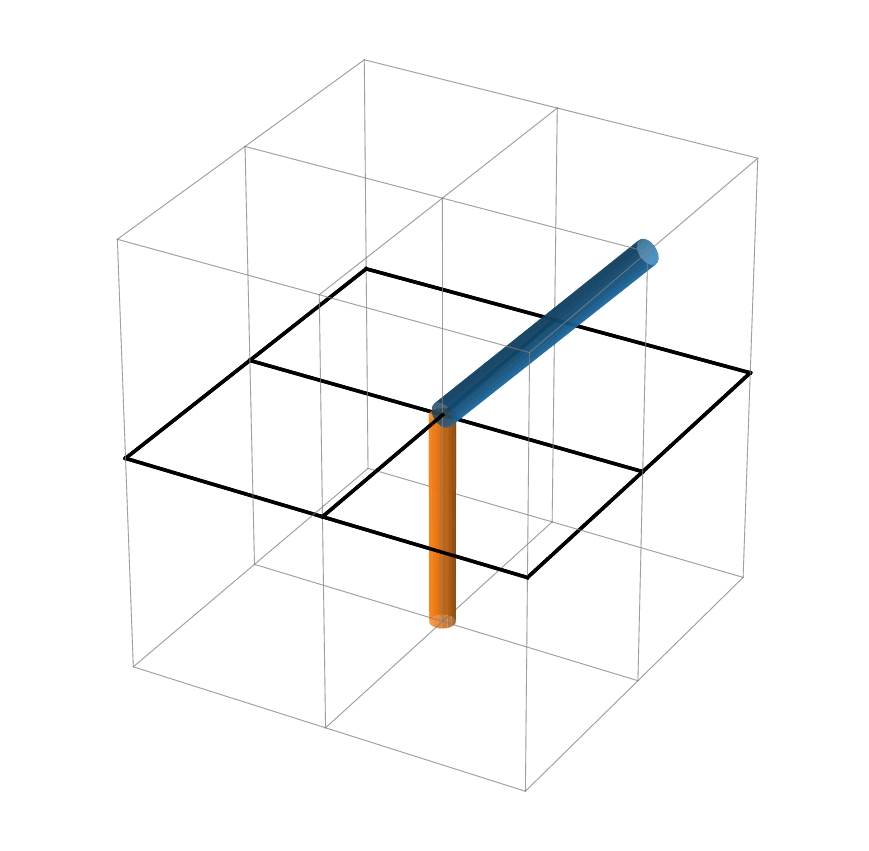} \vspace{-2pt} \end{tabular} \\[7pt]
\hline
\   3 \ &\ \ $\textbf{111}$ \ \ & $ \ \pi_{[100]}\ $ & $\ \pi_{[100]}\ $ & $ \ \pi\pi_{[000] A_1^+} \ $ & $\ \pi_{[100]} \ $ & \begin{tabular}{c} \includegraphics[scale=0.15]{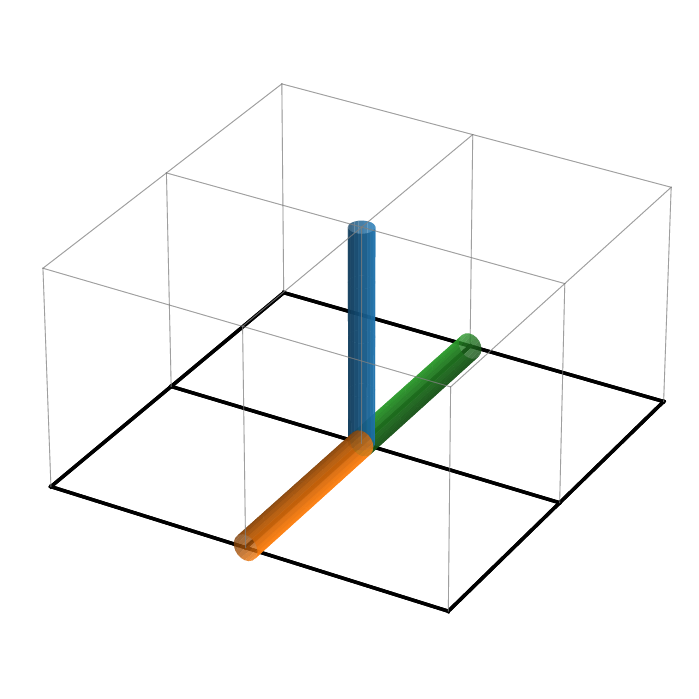} \vspace{-5pt} ${}^{\!\!\!\!\!\!1}$ \end{tabular} \ \begin{tabular}{c} \includegraphics[scale=0.15]{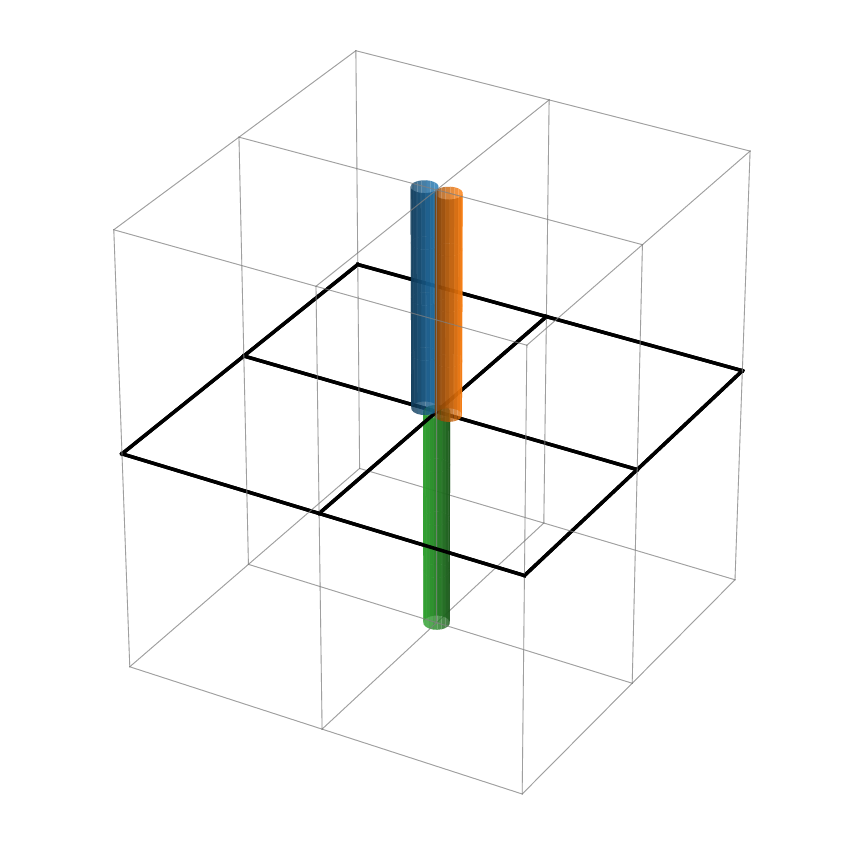}${}^{\!\!\!\!\!2}$ \end{tabular} \\[7pt]
\   4 \ &\ \ $\textbf{111}$ \ \ & $ \ \pi_{[100]}\ $ & $\ \pi_{[100]}\ $ & $ \ \pi\pi_{[110] A_1} \ $ & $\ \pi_{[100]} \ $ & \begin{tabular}{c} \includegraphics[scale=0.15]{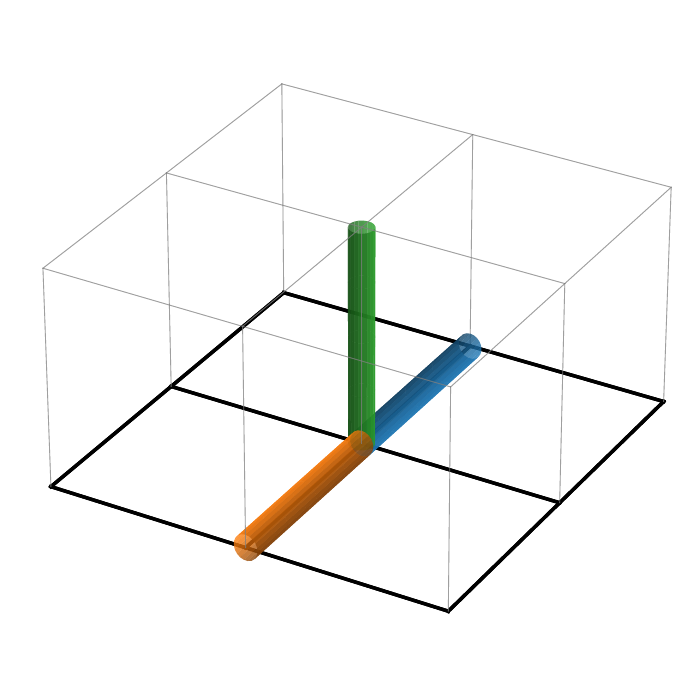} \vspace{-2pt} ${}^{\!\!\!\!\!\!1}$ \end{tabular} \\[7pt]\hline \hline
\   5 \ &\ \ $\textbf{014}$ \ \ & $ \ \pi_{[100]}\ $ & $\ \pi_{[000]}\ $ & $ \ \pi\pi_{[100] A_1} \ $ & $\ \pi_{[200]} \ $ & \begin{tabular}{c} \vspace{-17pt}   \\ \ \includegraphics[scale=0.23]{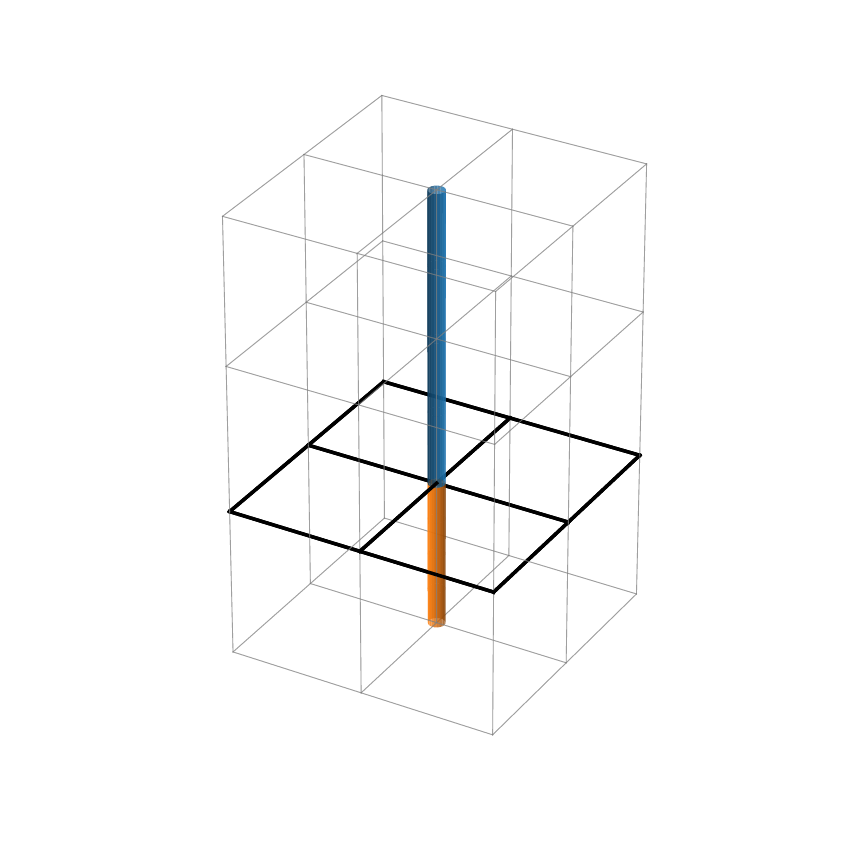} \end{tabular} \vspace{-5pt}  \\[7pt] \hline
\   6 \ &\ \ $\textbf{023}$ \ \ & $ \ \pi_{[110]}\ $ & $\ \pi_{[000]}\ $ & $ \ \pi\pi_{[110] A_1} \ $ & $\ \pi_{[111]} \ $ & \begin{tabular}{c} \includegraphics[scale=0.15]{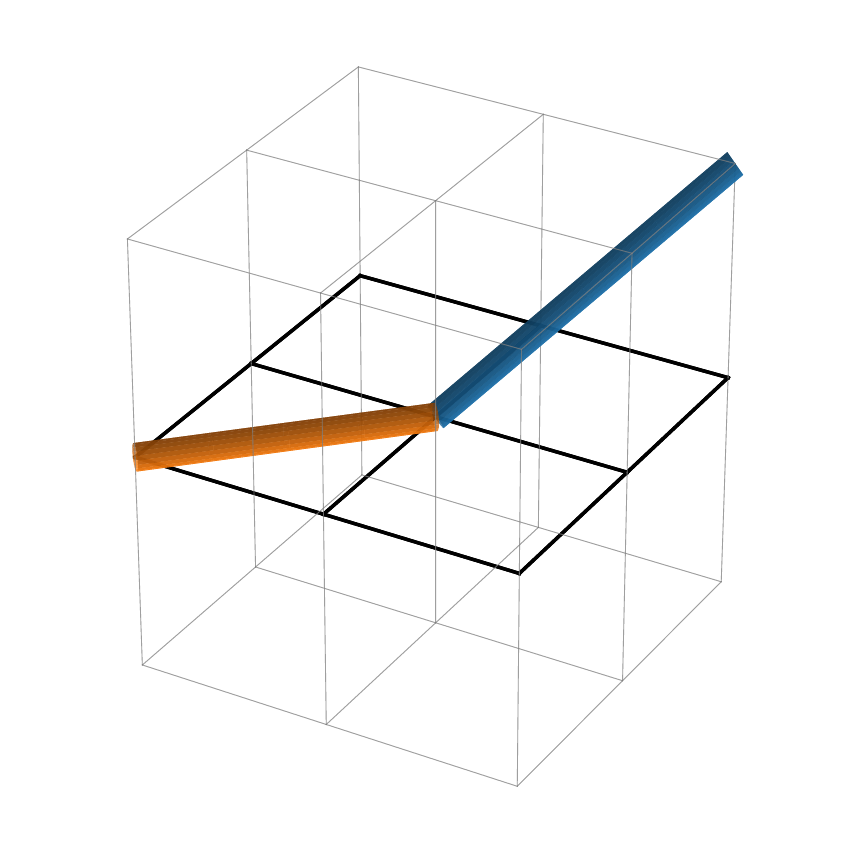} \vspace{-2pt} \end{tabular} \\[7pt]
\hline
\   7 \ &\ \ $\textbf{113}$ \ \ & $ \ \pi_{[100]}\ $ & $\ \pi_{[100]}\ $ & $ \ \pi\pi_{[110] A_1} \ $ & $\ \pi_{[111]} \ $ & \begin{tabular}{c} \includegraphics[scale=0.15]{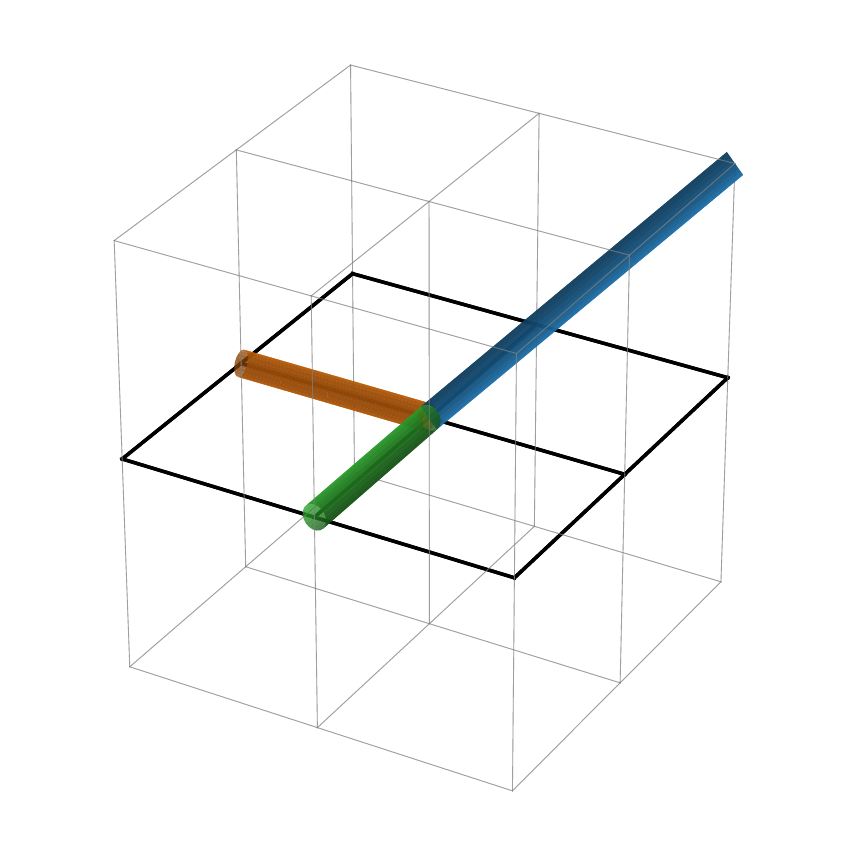} \vspace{-2pt} \end{tabular} \\[7pt] \hline  
\   8 \ &\ \ $\textbf{122}$ \ \ & $ \ \pi_{[110]}\ $ & $\ \pi_{[100]}\ $ & $ \ \pi\pi_{[111] A_1} \ $ & $\ \pi_{[110]} \ $ & \multirow{2}{*}{\begin{tabular}{c} 
\includegraphics[scale=0.15]{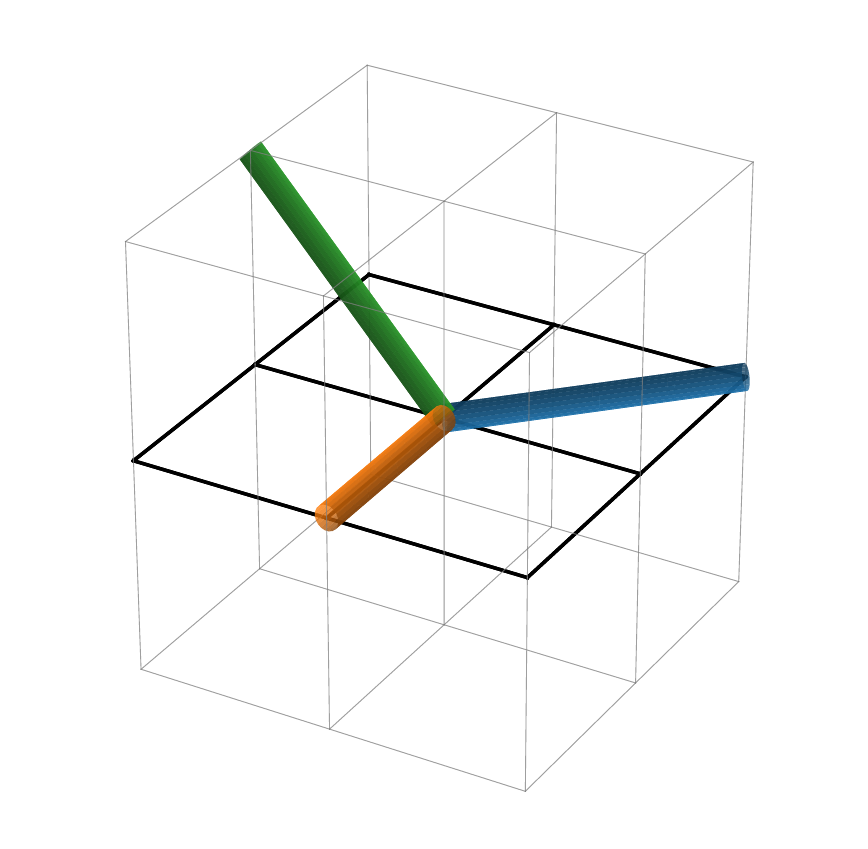}${}^{\!\!\!\!\!\!1}$ \ \includegraphics[scale=0.15]{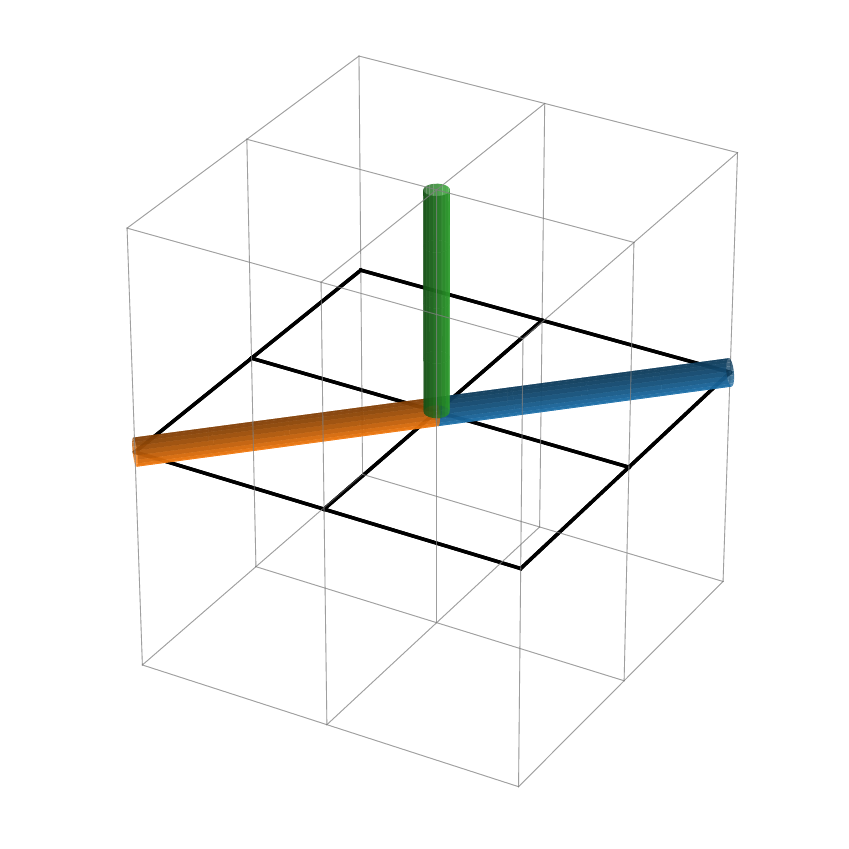}${}^{\!\!\!\!\!\!3}$ \end{tabular}} \vspace{-5pt}\\[0pt] 
\   9 \ &\ \ $\textbf{122}$ \ \ & $ \ \pi_{[110]}\ $ & $\ \pi_{[100]}\ $ & $ \ \pi\pi_{[111] E_2} \ $ & $\ \pi_{[110]} \ $ & \\[9pt] 
\   {10} \ &\ \ $\textbf{122}$ \ \ & $ \ \pi_{[110]}\ $ & $\ \pi_{[100]}\ $ & $ \ \pi\pi_{[100] A_1} \ $ & $\ \pi_{[110]} \ $ & \multirow{3}{*}{\begin{tabular}{c} \\[-10pt] \includegraphics[scale=0.15]{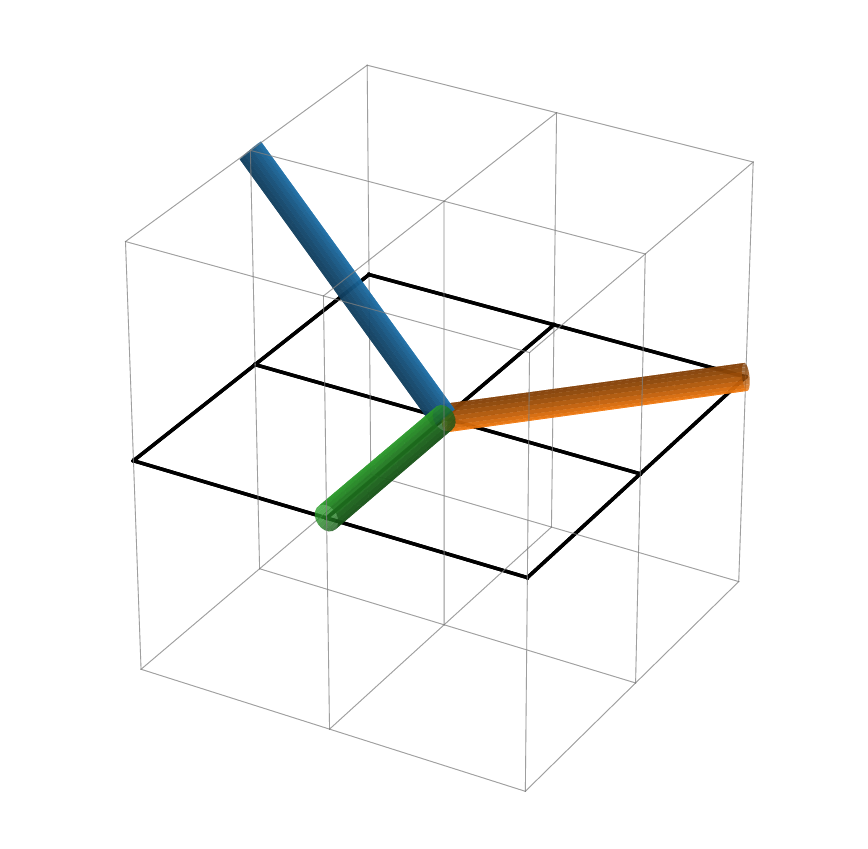}${}^{\!\!\!\!\!\!1}$ \ \includegraphics[scale=0.15]{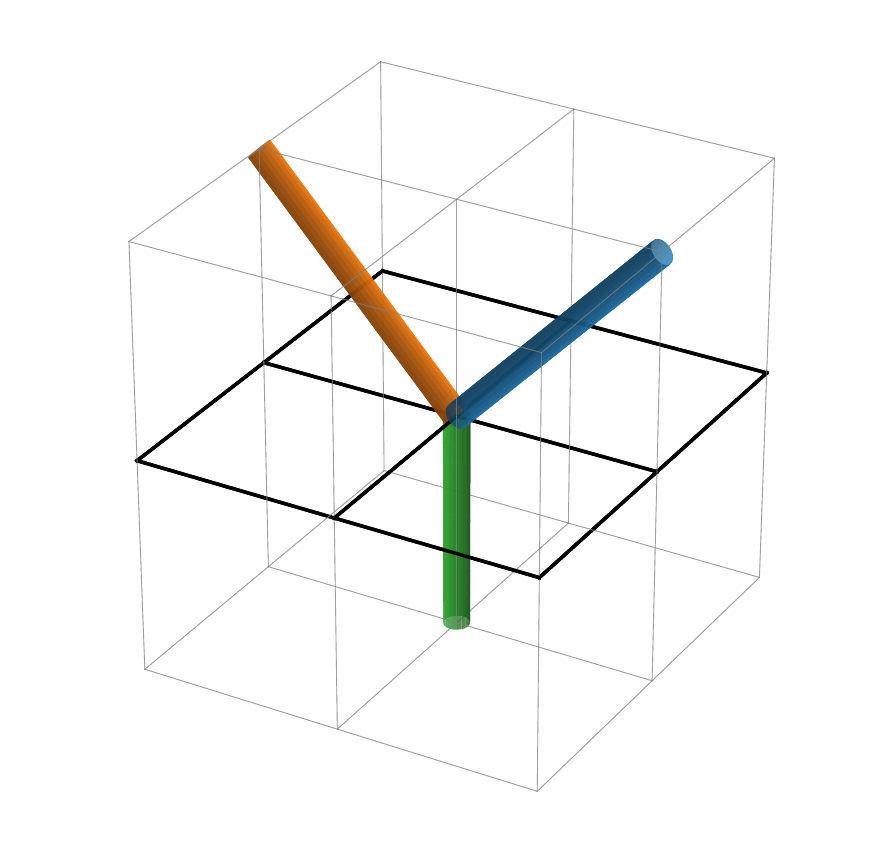}${}^{\!\!\!\!\!\!2}$ \ \includegraphics[scale=0.15]{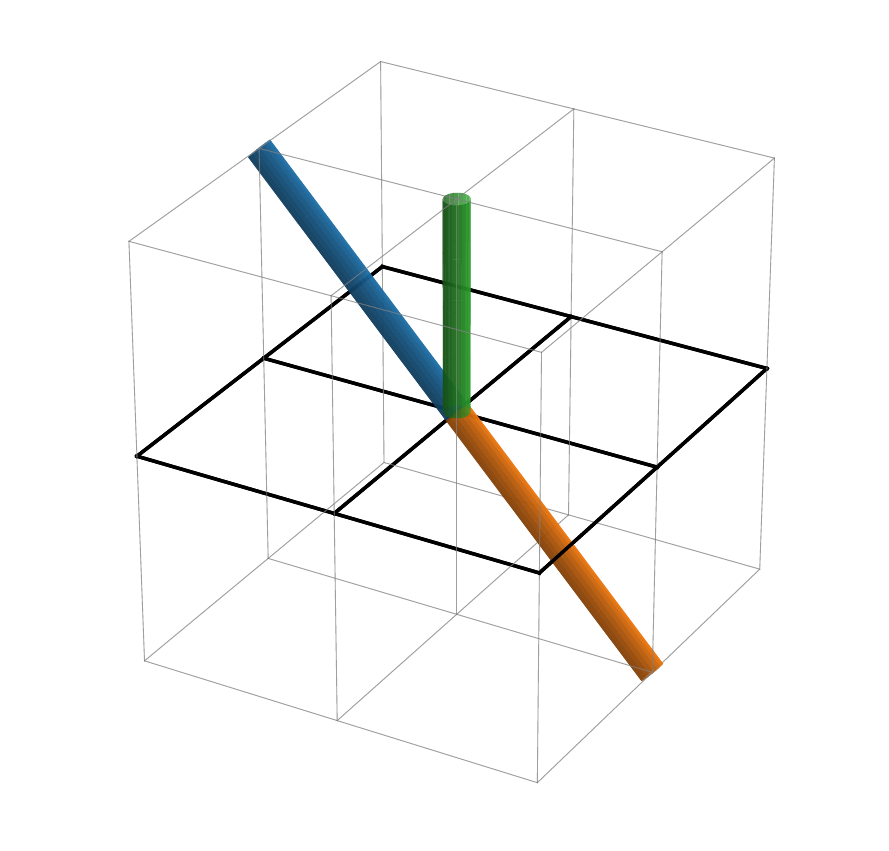}${}^{\!\!\!\!\!\!4}$ \vspace{-5pt} \end{tabular}}\\[15pt]
\   {11} \ &\ \ $\textbf{122}$ \ \ & $ \ \pi_{[110]}\ $ & $\ \pi_{[100]}\ $ & $ \ \pi\pi_{[100] B_1} \ $ & $\ \pi_{[110]} \ $ & \\[7pt]\hline \hline
\  {12} \ &\ \ $\textbf{025}$ \ \ & $ \ \pi_{[110]}\ $ & $\ \pi_{[000]}\ $ & $ \ \pi\pi_{[110] A_1} \ $ & $\ \pi_{[210]} \ $ & \begin{tabular}{c} \includegraphics[scale=0.23]{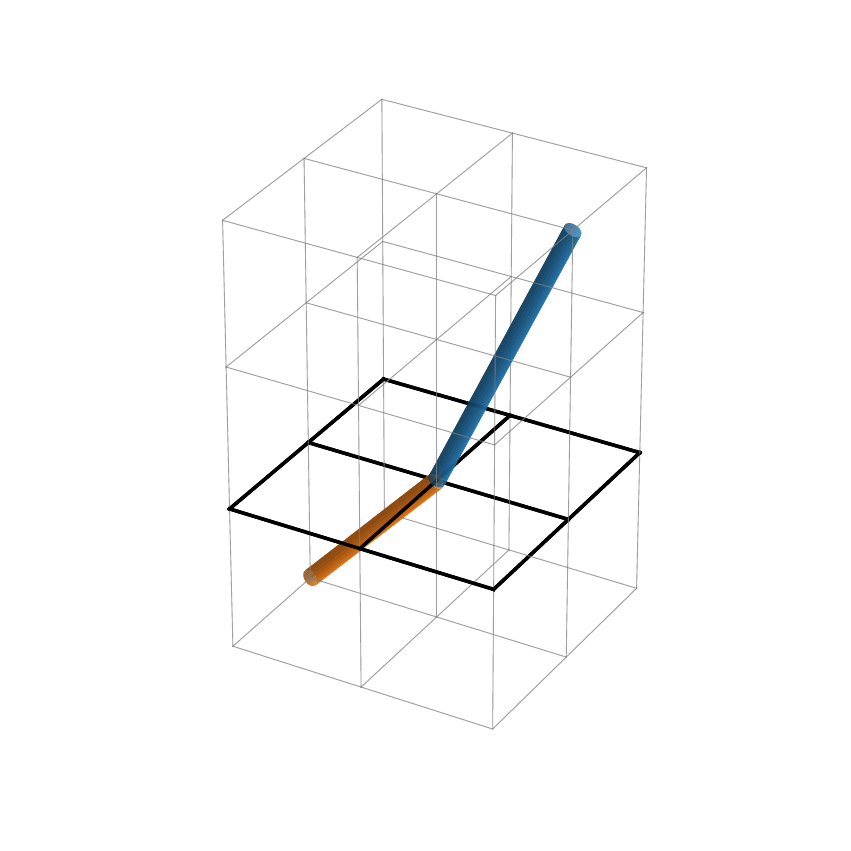} \vspace{-5pt} \end{tabular} \\[5pt] \hline
\  {13} \ &\ \ $\textbf{115}$ \ \ & $ \ \pi_{[210]}\ $ & $\ \pi_{[100]}\ $ & $ \ \pi\pi_{[110] A_1} \ $ & $\ \pi_{[100]} \ $ & \multirow{2}{*}{\begin{tabular}{c} \includegraphics[scale=0.23]{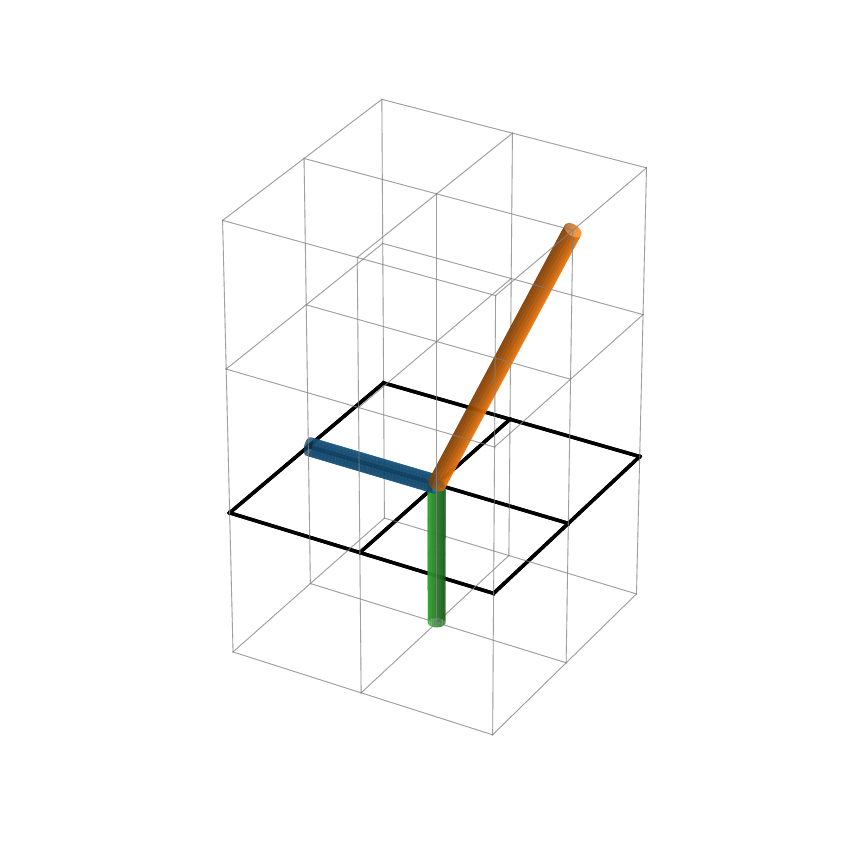} \!\!\!\!\! \includegraphics[scale=0.23]{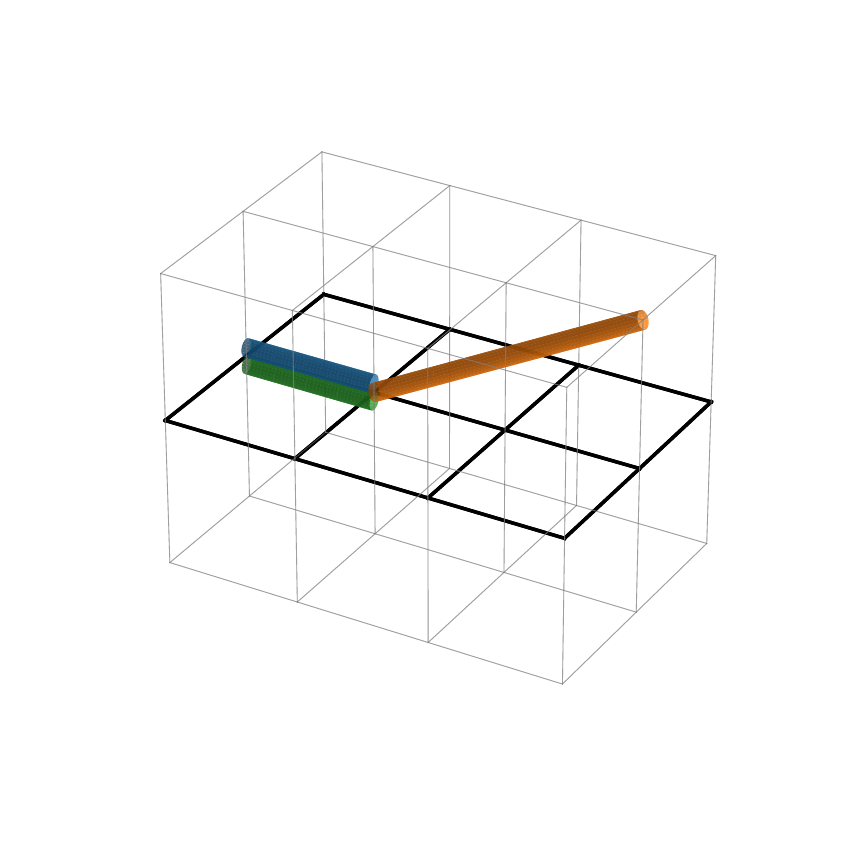}   \end{tabular}}\\[15pt]
\  {14} \ &\ \ $\textbf{115}$ \ \ & $ \ \pi_{[210]}\ $ & $\ \pi_{[100]}\ $ & $ \ \pi\pi_{[110] B_1} \ $ & $\ \pi_{[100]} \ $ & \\[9pt] \hline
\  {15} \ &\ \ $\textbf{124}$ \ \ & $ \ \pi_{[200]}\ $ & $\ \pi_{[100]}\ $ & $ \ \pi\pi_{[210] A_1} \ $ & $\ \pi_{[110]} \ $ & \begin{tabular}{c} \includegraphics[scale=0.23]{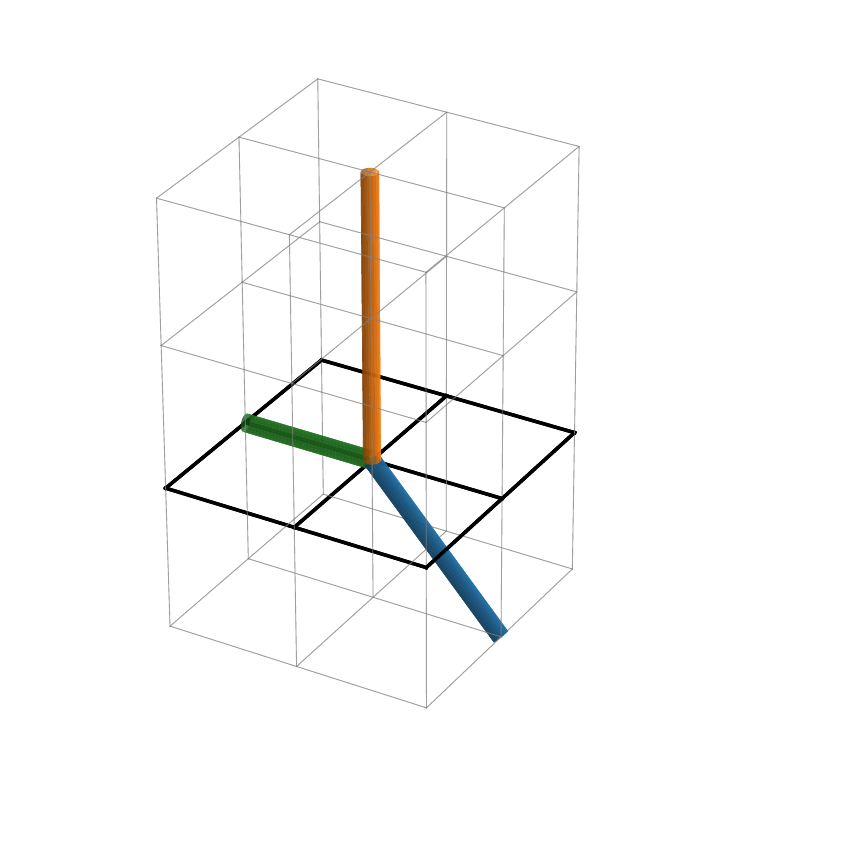} \end{tabular} \vspace{-15pt} \\[0pt] 
\  {16} \ &\ \ $\textbf{124}$ \ \ & $ \ \pi_{[200]}\ $ & $\ \pi_{[100]}\ $ & $ \ \pi\pi_{[100] A_1} \ $ & $\ \pi_{[110]} \ $ & \begin{tabular}{c} \includegraphics[scale=0.23]{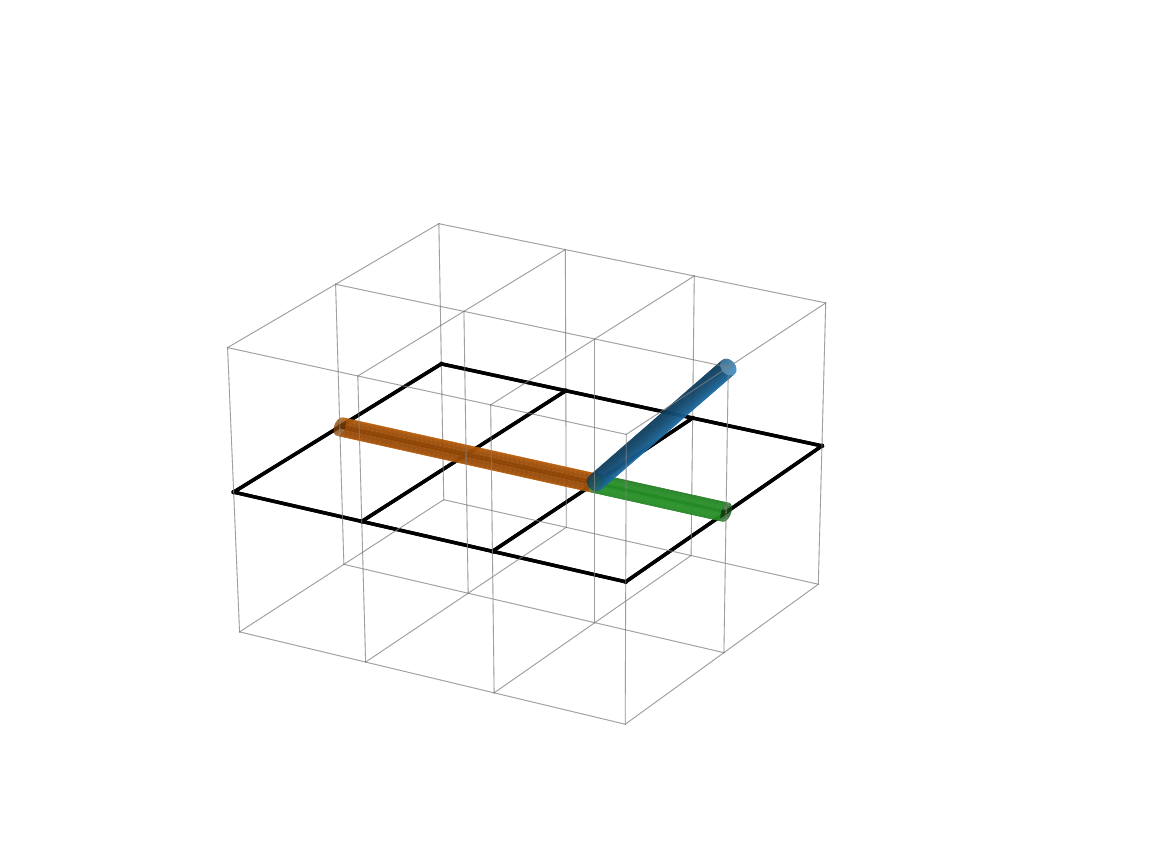} \vspace{-5pt} \end{tabular} \\[0pt]
%\hline
\end{tabular}
\end{center}
}
\caption{As Table~\ref{table:ops:3pi:0} but for the $A_2$ irrep with overall momentum $\vec{P}=[001]$. Operators 1 to 8 are used on the $20^3$ volume and operators 1 to 16 are used on the $24^3$ volume. Operators separated by a single horizontal line correspond to states that are degenerate in the non-relativistic, non-interacting theory. Operators with no horizontal line correspond to states that are degenerate in the relativistic, non-interacting theory but are split by the interactions.}
\label{table:ops:3pi:1}
\end{table}

\begin{table}
\begin{center}
\setlength\extrarowheight{5pt}
\begin{tabular}{   c | c || c | c | c || c || c   }
%\hline
& \ $\boldsymbol{d_1^2 \, d_2^2 \, d_3^2}$ \ & $\pi_{\vec{k_1}}$ & \ $\pi_{\vec{k_2}}$ & \ $\pi\pi_{\vec{k_{12}} \Lambda_{12}}$ ($I=2$) \ & $\pi_{\vec{k_3}}$ & \ \ momentum configurations\ \  \\[5pt]
\hline \hline
\ 1 \ &\ \ $\textbf{002}$ \ \ & $ \ \pi_{[000]}\ $ & $\ \pi_{[000]}\ $ & $ \ \pi\pi_{[000] A_1^+} \ $ & $\ \pi_{[110]} \ $ & \begin{tabular}{c} \includegraphics[scale=0.15]{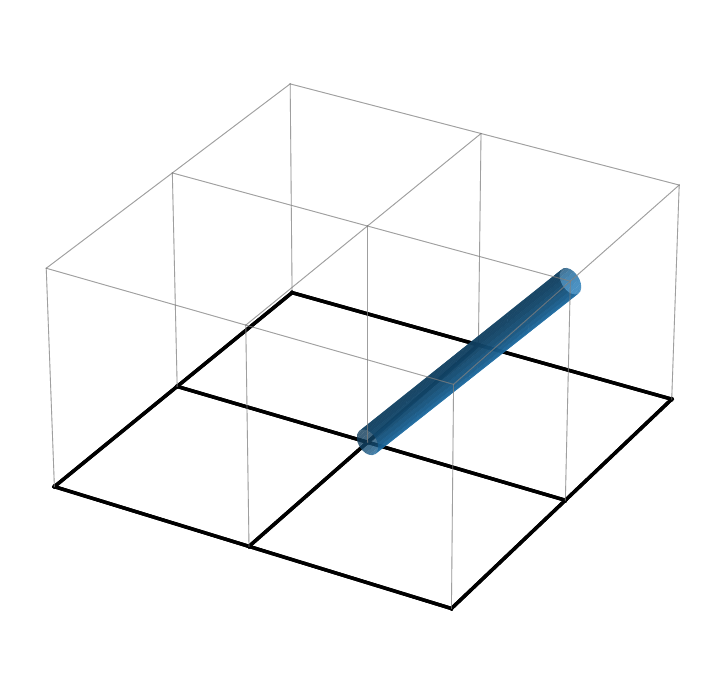} \vspace{-5pt} \end{tabular} \\[5pt] \hline
\  2 \ &\ \ $\textbf{011}$ \ \ & $ \ \pi_{[100]}\ $ & $\ \pi_{[000]}\ $ & $ \ \pi\pi_{[100] A_1} \ $ & $\ \pi_{[100]} \ $ & \begin{tabular}{c} \includegraphics[scale=0.15]{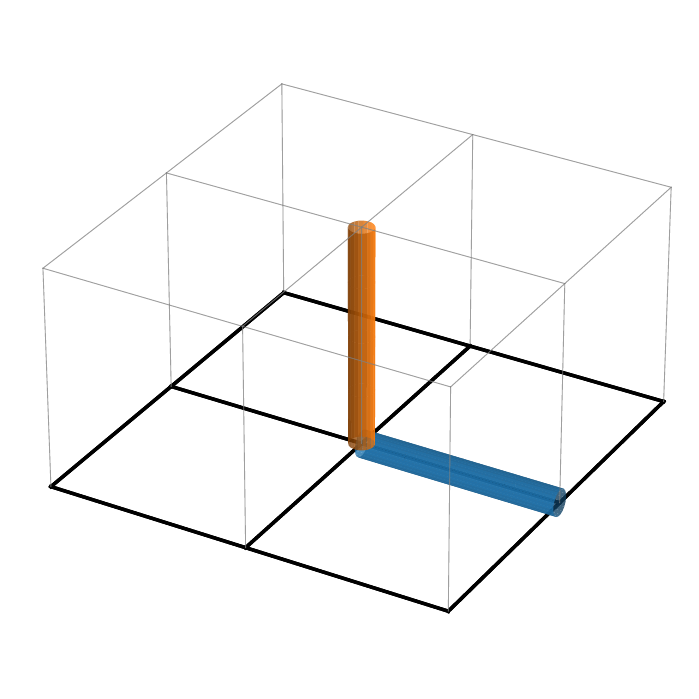} \vspace{-5pt}  \end{tabular} \\[5pt]\hline \hline
\  3 \ &\ \ $\textbf{013}$ \ \ & $ \ \pi_{[100]}\ $ & $\ \pi_{[000]}\ $ & $ \ \pi\pi_{[100] A_1} \ $ & $\ \pi_{[111]} \ $ & \begin{tabular}{c} \includegraphics[scale=0.15]{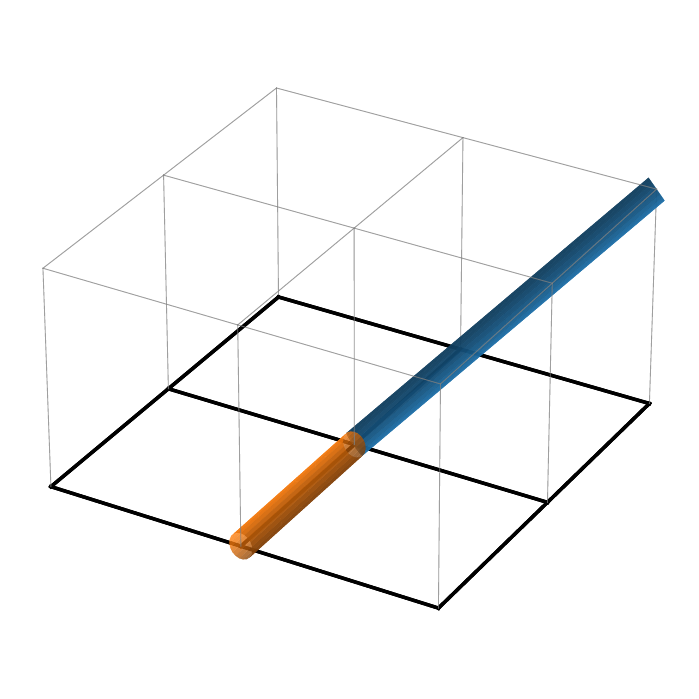} \vspace{-5pt}  \end{tabular} \\[5pt] \hline
\ \ 4 \ &\ \ $\textbf{022}$ \ \ & $ \ \pi_{[110]}\ $ & $\ \pi_{[000]}\ $ & $ \ \pi\pi_{[110] A_1} \ $ & $\ \pi_{[110]} \ $ & \begin{tabular}{c} \includegraphics[scale=0.15]{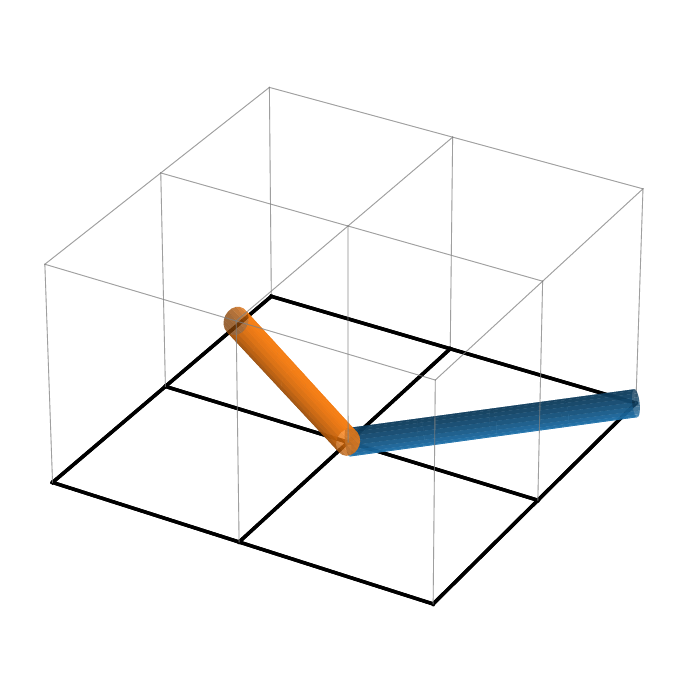} \vspace{-5pt}  \end{tabular} \\[5pt] \hline \hline
\  5 \ &\ \ $\textbf{112}$ \ \ & $ \ \pi_{[100]}\ $ & $\ \pi_{[100]}\ $ & $ \ \pi\pi_{[000] A_1^+} \ $ & $\ \pi_{[110]} \ $ & \begin{tabular}{c} \includegraphics[scale=0.15]{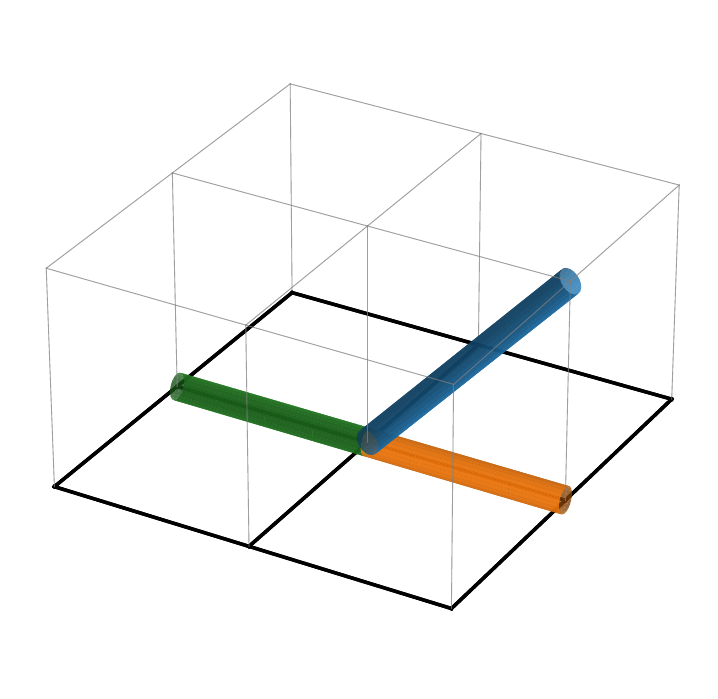}${}^{\!\!\!\!1}$ \end{tabular} \ \begin{tabular}{c} \includegraphics[scale=0.15]{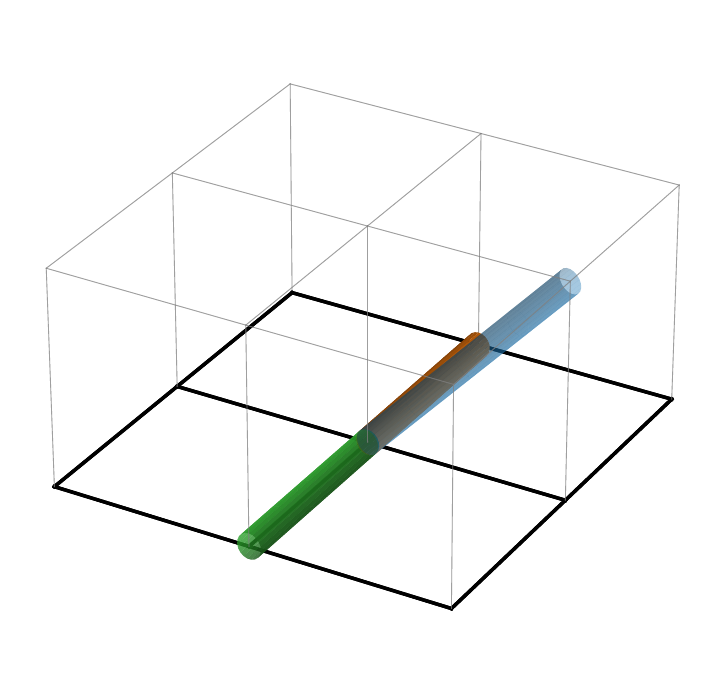}${}^{\!\!\!\!2}$ \end{tabular} \\[5pt]
\  6 \ &\ \ $\textbf{112}$ \ \ & $ \ \pi_{[100]}\ $ & $\ \pi_{[100]}\ $ & $ \ \pi\pi_{[110] A_1} \ $ & $\ \pi_{[110]} \ $ & \begin{tabular}{c} \includegraphics[scale=0.15]{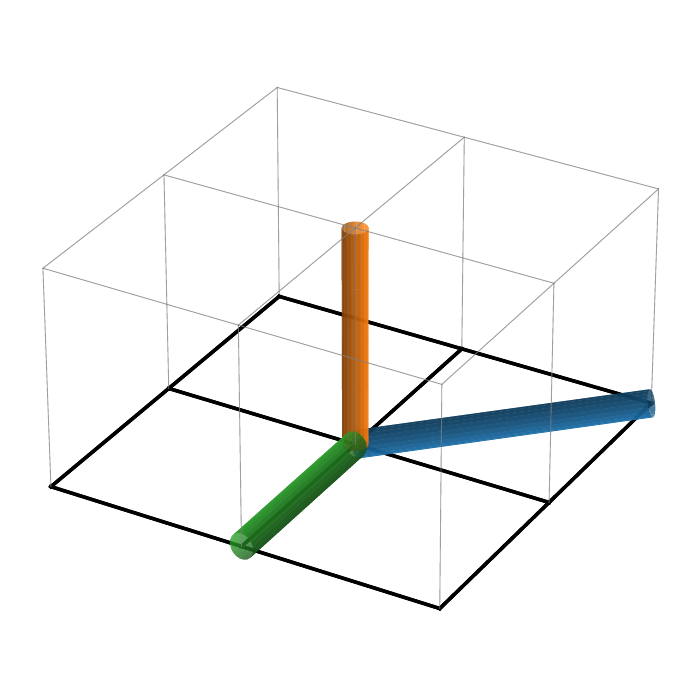} \end{tabular} \\[7pt]
\ 7 \ &\ \ $\textbf{112}$ \ \ & $ \ \pi_{[100]}\ $ & $\ \pi_{[100]}\ $ & $ \ \pi\pi_{[200] A_1} \ $ & $\ \pi_{[110]} \ $ & \begin{tabular}{c} \includegraphics[scale=0.15]{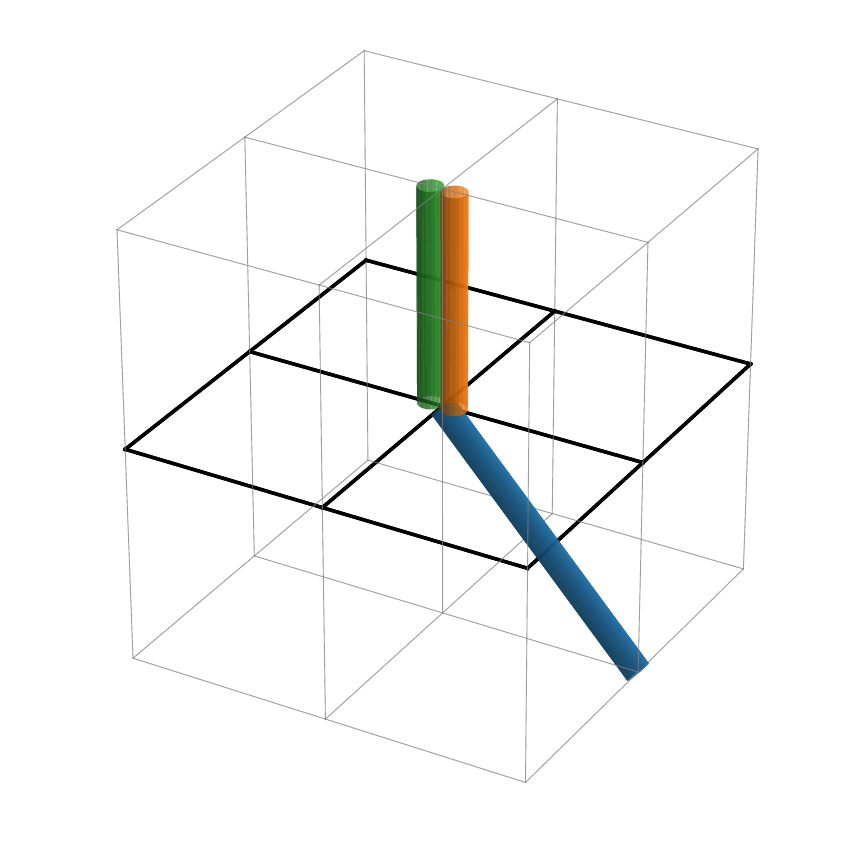} \end{tabular} \\[5pt]
\ 8 \ &\ \ $\textbf{112}$ \ \ & $ \ \pi_{[100]}\ $ & $\ \pi_{[100]}\ $ & $ \ \pi\pi_{[000] E^+} \ $ & $\ \pi_{[110]} \ $ & \begin{tabular}{c} \includegraphics[scale=0.15]{figs/opfigs/OpFigP011ddd112A.pdf}${}^{\!\!\!\!1}$ \end{tabular} \ \begin{tabular}{c} \includegraphics[scale=0.15]{figs/opfigs/OpFigP011ddd112B.pdf}${}^{\!\!\!\!2}$ \end{tabular} \\[5pt]
%\hline
\end{tabular}
\end{center}
\caption{As Table~\ref{table:ops:3pi:0} but for the $A_2$ irrep with overall momentum $\vec{P}=[011]$ (continued in Table~\ref{table:ops:3pi:2b}). Operators 1 to 11 are used on the $20^3$ volume and operators 1 to 20 are used on the $24^3$ volume.}
\label{table:ops:3pi:2a}
\end{table}

\begin{table}
\begin{center}
\setlength\extrarowheight{5pt}
\begin{tabular}{    c | c || c | c | c || c || c  }
%\hline
& \ $\boldsymbol{d_1^2 \, d_2^2 \, d_3^2}$ \ & $\pi_{\vec{k_1}}$ & \ $\pi_{\vec{k_2}}$ & \ $\pi\pi_{\vec{k_{12}} \Lambda_{12}}$ ($I=2$) \ & $\pi_{\vec{k_3}}$ & \ \ momentum configurations\ \  \\[5pt]
\hline \hline
\ {9} \ &\ \ $\textbf{015}$ \ \ & $ \ \pi_{[100]}\ $ & $\ \pi_{[000]}\ $ & $ \ \pi\pi_{[100] A_1} \ $ & $\ \pi_{[210]} \ $ & \begin{tabular}{c} \includegraphics[scale=0.23]{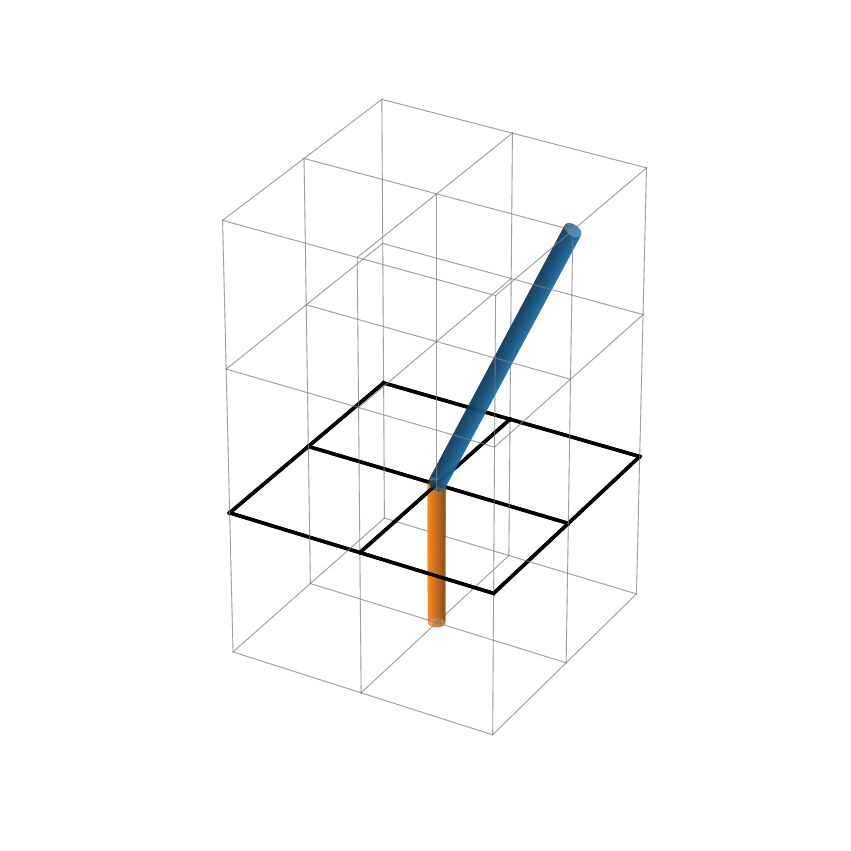} \end{tabular} \vspace{-5pt} \\[5pt] \hline
\ {10} \ &\ \ $\textbf{024}$ \ \ & $ \ \pi_{[110]}\ $ & $\ \pi_{[000]}\ $ & $ \ \pi\pi_{[110] A_1} \ $ & $\ \pi_{[200]} \ $ & \begin{tabular}{c} \includegraphics[scale=0.23]{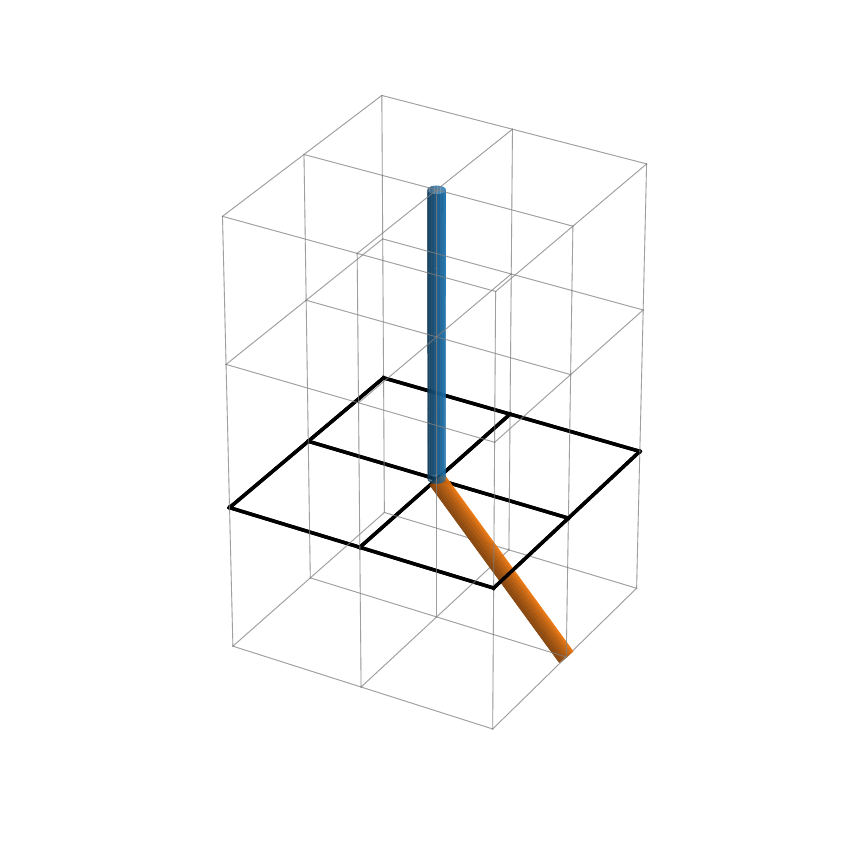} \end{tabular} \vspace{-5pt} \\[5pt] \hline
\ {11} \ &\ \ $\textbf{114}$ \ \ & $ \ \pi_{[200]}\ $ & $\ \pi_{[100]}\ $ & $ \ \pi\pi_{[210] A_1} \ $ & $\ \pi_{[100]} \ $ & \begin{tabular}{c} \includegraphics[scale=0.23]{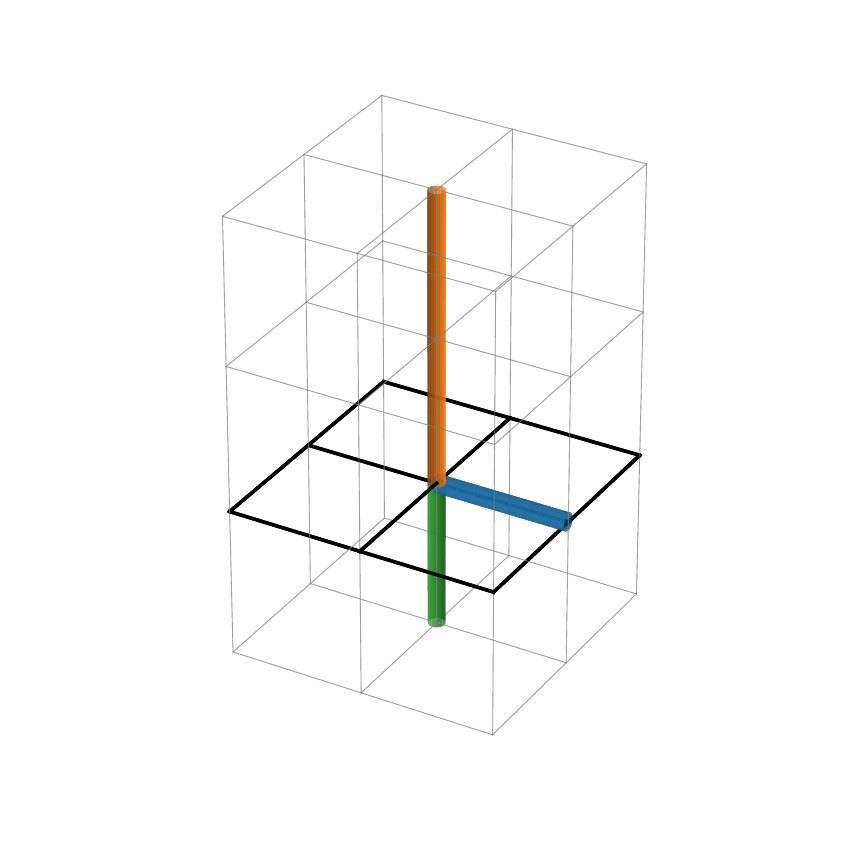} \end{tabular} \vspace{-5pt} \\[5pt] \hline
\ {12} \ &\ \ $\textbf{123}$ \ \ & $ \ \pi_{[111]}\ $ & $\ \pi_{[100]}\ $ & $ \ \pi\pi_{[211] A_1} \ $ & $\ \pi_{[110]} \ $ & \begin{tabular}{c} \includegraphics[scale=0.15]{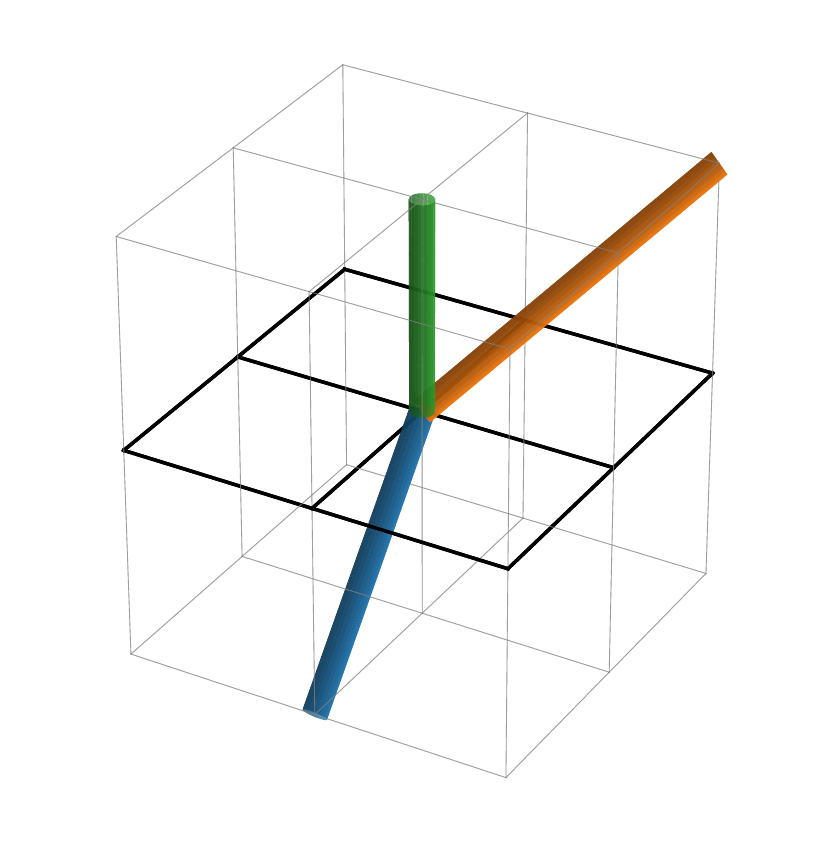} \end{tabular} \\[5pt]
\ {13} \ &\ \ $\textbf{123}$ \ \ & $ \ \pi_{[111]}\ $ & $\ \pi_{[100]}\ $ & $ \ \pi\pi_{[110] A_1} \ $ & $\ \pi_{[110]} \ $ & \begin{tabular}{c} \includegraphics[scale=0.15]{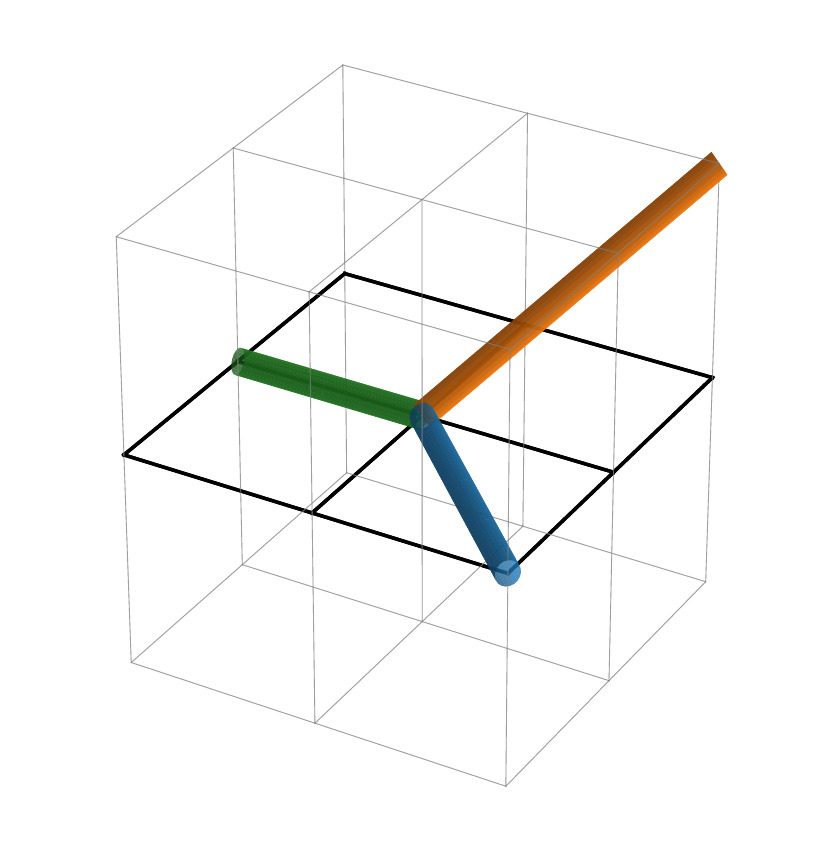}$^{\!\!\!\!\!\!1}$ \end{tabular} \ \begin{tabular}{c} \includegraphics[scale=0.15]{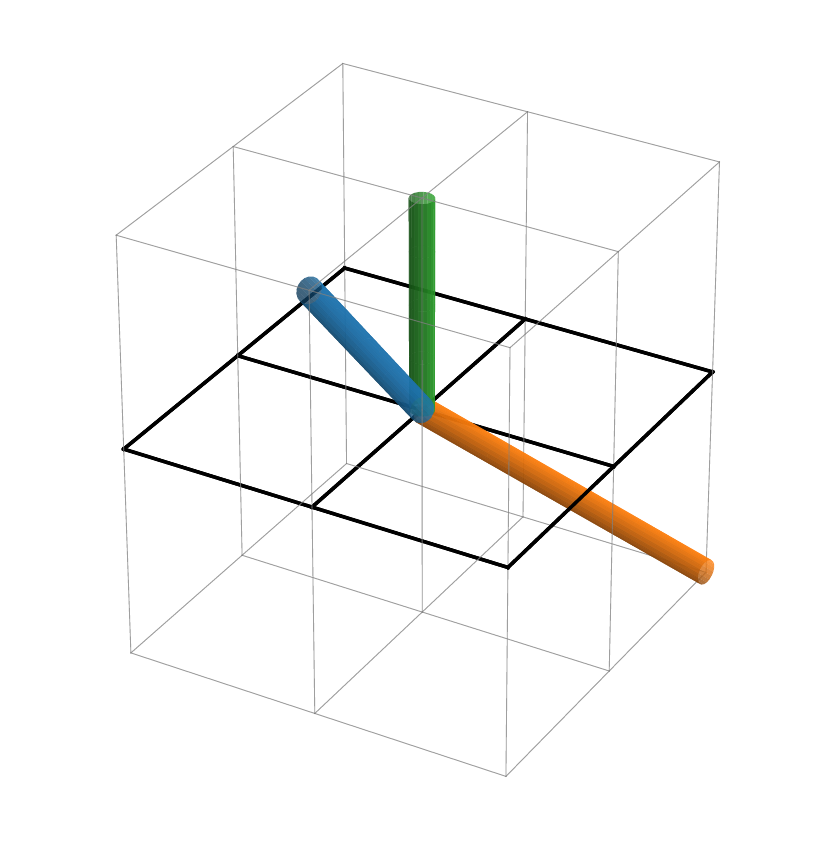}$^{\!\!\!\!\!\!2}$ \end{tabular} \\[20pt]
\ {14} \ &\ \ $\textbf{123}$ \ \ & $ \ \pi_{[111]}\ $ & $\ \pi_{[100]}\ $ & $ \ \pi\pi_{[110] B_2} \ $ & $\ \pi_{[110]} \ $ & \begin{tabular}{c} \includegraphics[scale=0.15]{figs/opfigs/OpFigP011ddd123B.pdf}$^{\!\!\!\!\!\!1}$ \end{tabular} \ \begin{tabular}{c} \includegraphics[scale=0.15]{figs/opfigs/OpFigP011ddd123C.pdf}$^{\!\!\!\!\!\!2}$ \end{tabular} \\[20pt]
\hline
\ {15} \ &\ \ $\textbf{222}$ \ \ & $ \ \pi_{[110]}\ $ & $\ \pi_{[110]}\ $ & $ \ \pi\pi_{[000] A_1^+} \ $ & $\ \pi_{[110]} \ $ & \begin{tabular}{c} \includegraphics[scale=0.15]{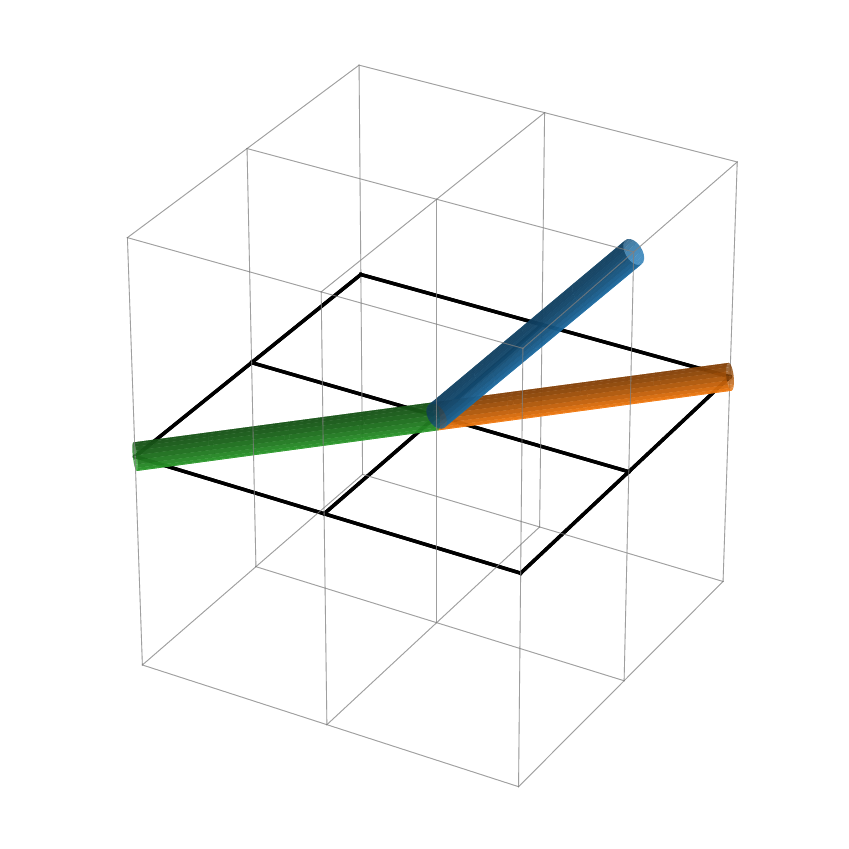}$^{\!\!\!\!\!\!1}$ \end{tabular} \!\!\! \begin{tabular}{c} \includegraphics[scale=0.15]{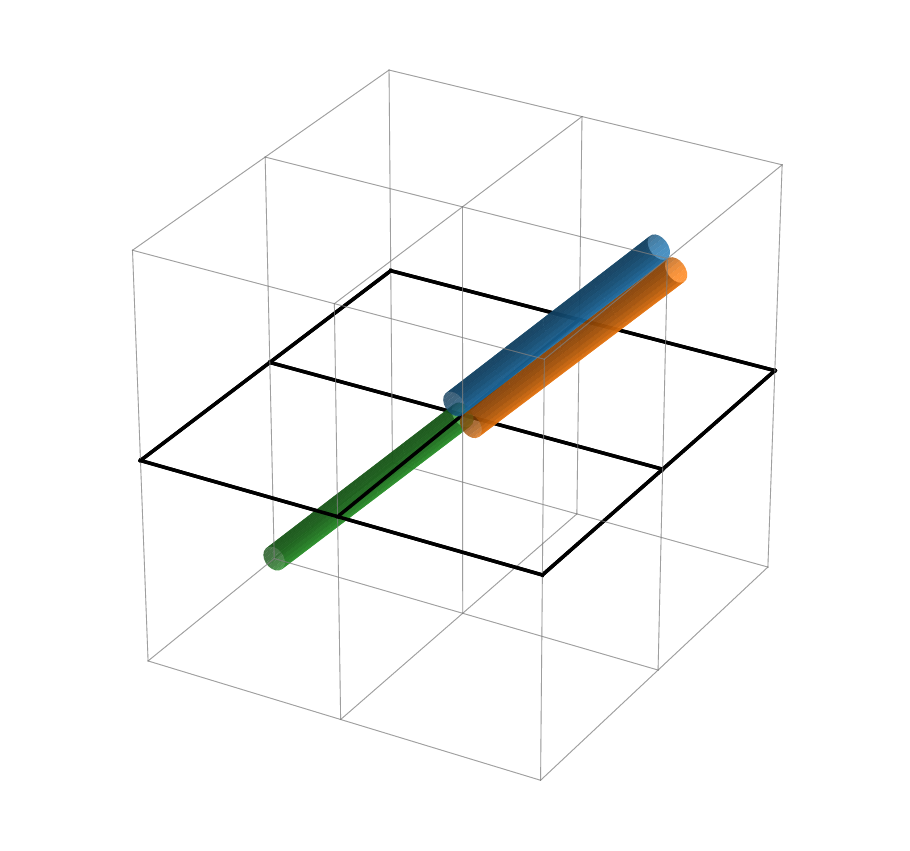}$^{\!\!\!\!\!\!2}$ \end{tabular} \!\!\! \begin{tabular}{c} \includegraphics[scale=0.15]{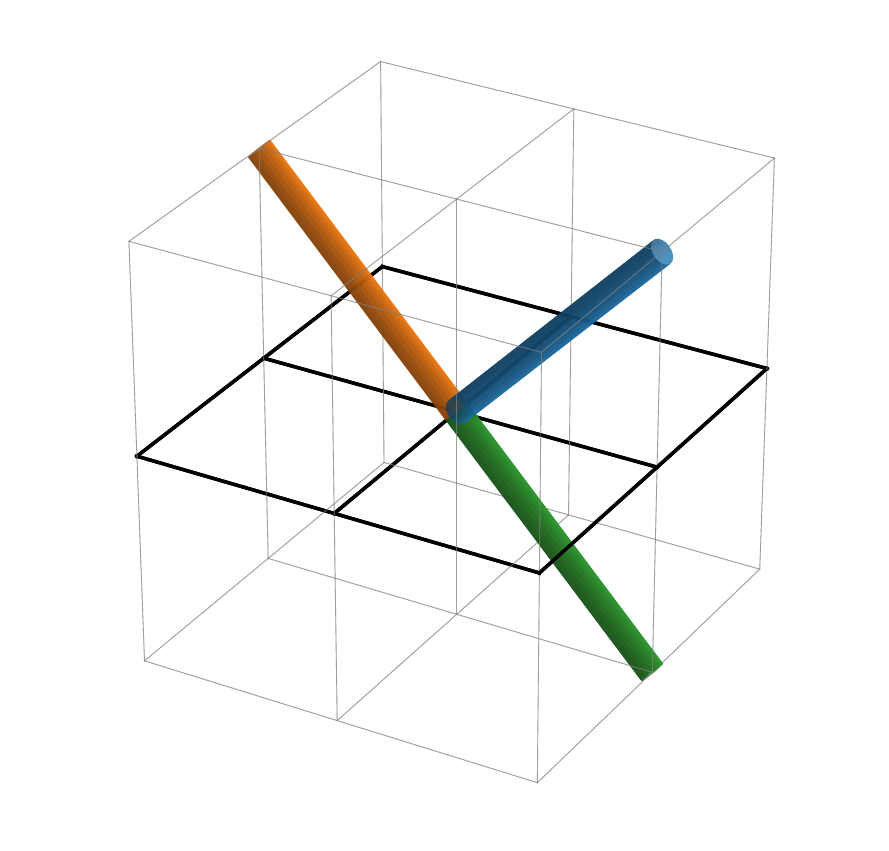}$^{\!\!\!\!\!\!3}$ \end{tabular} \\[5pt]
\ {16} \ &\ \ $\textbf{222}$ \ \ & $ \ \pi_{[110]}\ $ & $\ \pi_{[110]}\ $ & $ \ \pi\pi_{[211] A_1} \ $ & $\ \pi_{[110]} \ $ & \begin{tabular}{c} \includegraphics[scale=0.15]{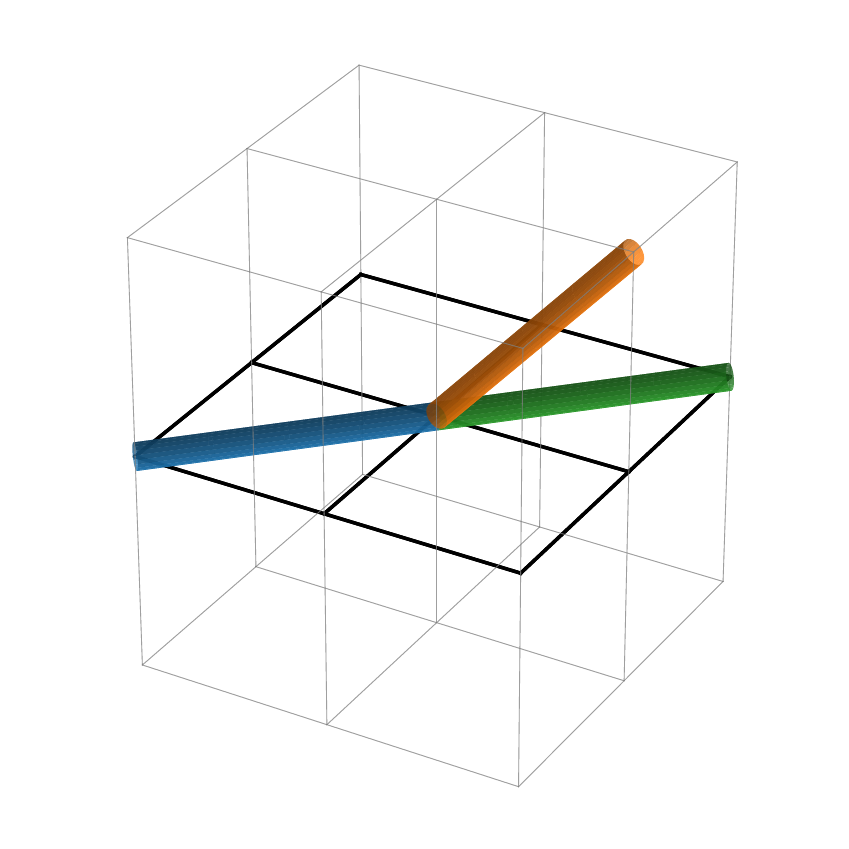}$^{\!\!\!\!\!\!1}$ \end{tabular} \\[5pt]
\ {17} \ &\ \ $\textbf{222}$ \ \ & $ \ \pi_{[110]}\ $ & $\ \pi_{[110]}\ $ & $ \ \pi\pi_{[110] A_1} \ $ & $\ \pi_{[110]} \ $ & \begin{tabular}{c} \includegraphics[scale=0.15]{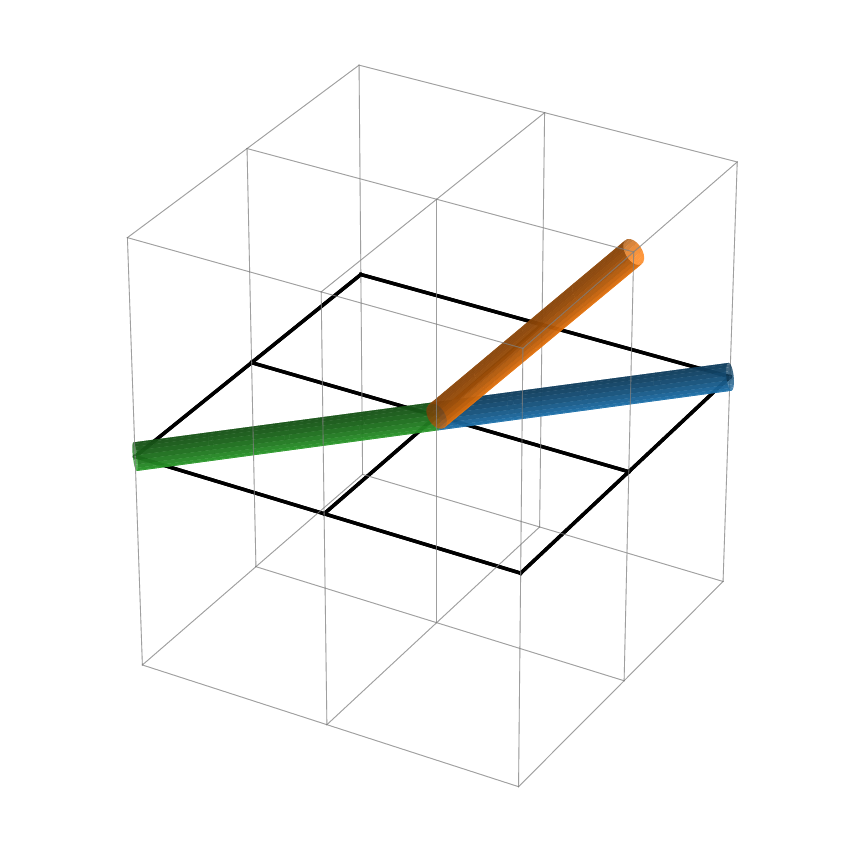}$^{\!\!\!\!\!\!1}$ \end{tabular} \ \begin{tabular}{c} \includegraphics[scale=0.15]{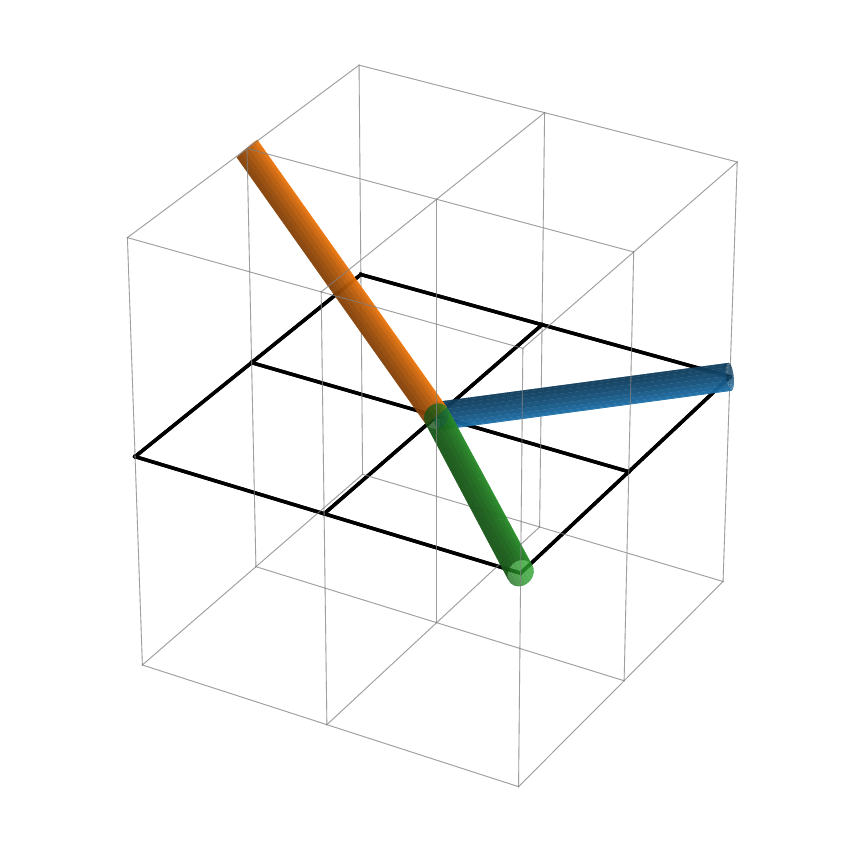}$^{\!\!\!\!\!\!4}$ \end{tabular} \\[5pt]
\ {18} \ &\ \ $\textbf{222}$ \ \ & $ \ \pi_{[110]}\ $ & $\ \pi_{[110]}\ $ & $ \ \pi\pi_{[200] A_1} \ $ & $\ \pi_{[110]} \ $ & \begin{tabular}{c} \includegraphics[scale=0.15]{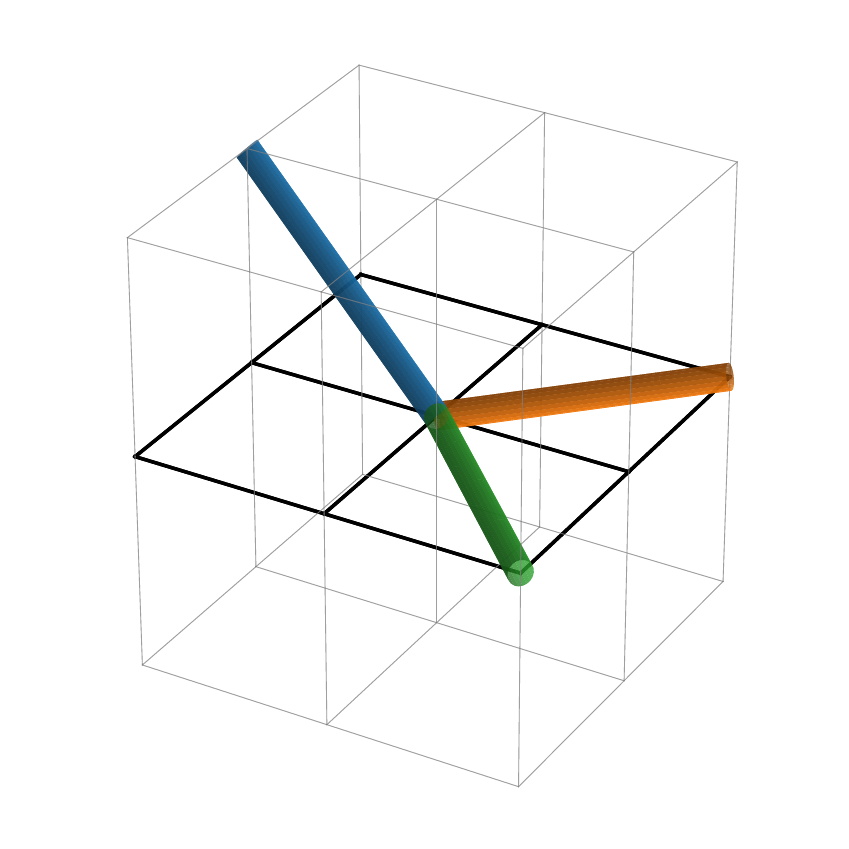}$^{\!\!\!\!\!\!4}$ \end{tabular} \ \begin{tabular}{c} \includegraphics[scale=0.15]{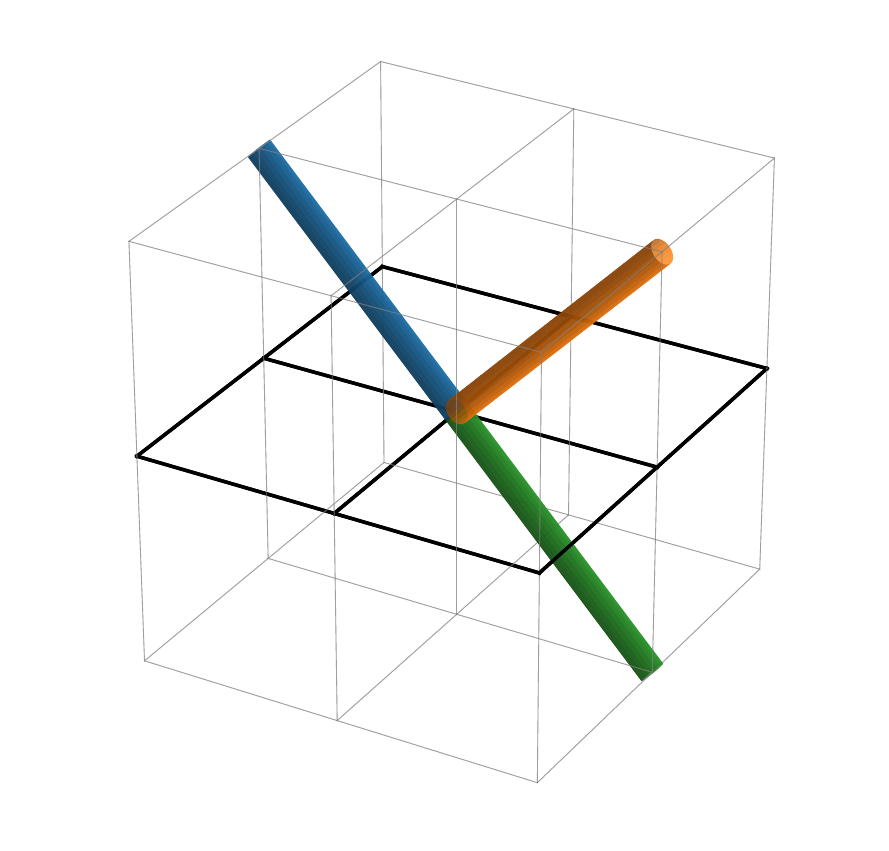}$^{\!\!\!\!\!\!3}$ \end{tabular} \\[20pt] \hline \hline
\ {19} \ &\ \ $\textbf{026}$ \ \ & $ \ \pi_{[110]}\ $ & $\ \pi_{[000]}\ $ & $ \ \pi\pi_{[110] A_1} \ $ & $\ \pi_{[211]} \ $ &  \ \begin{tabular}{c} \includegraphics[scale=0.23]{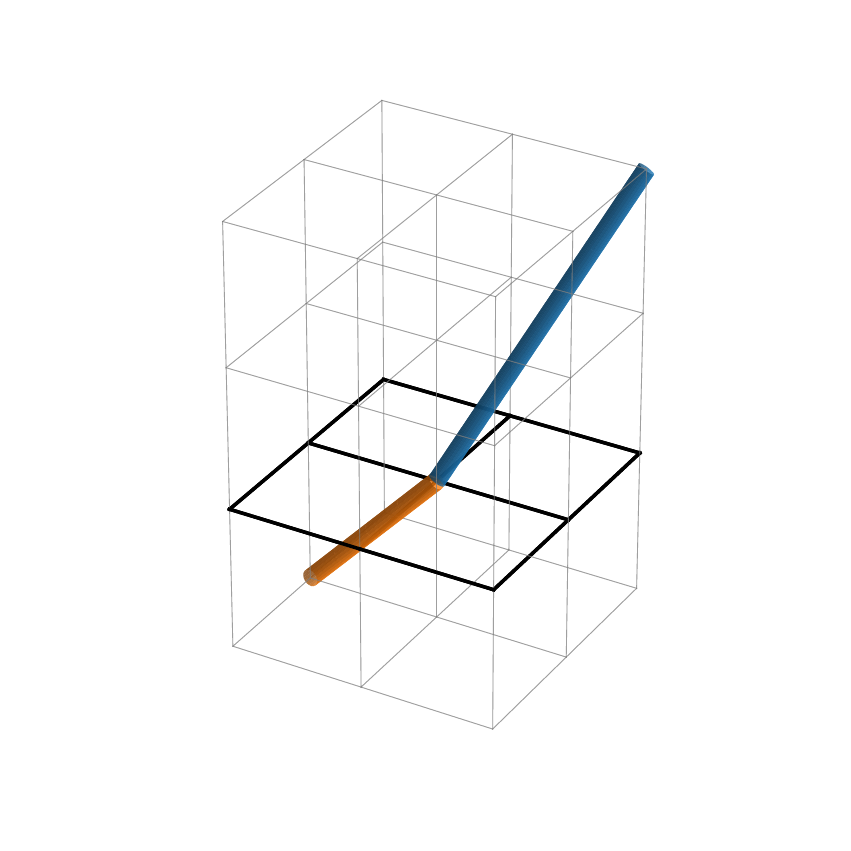} \end{tabular}  \vspace{-5pt} \\[5pt] \hline
\ {20} \ &\ \ $\textbf{035}$ \ \ & $ \ \pi_{[111]}\ $ & $\ \pi_{[000]}\ $ & $ \ \pi\pi_{[111] A_1} \ $ & $\ \pi_{[210]} \ $ &  \begin{tabular}{c} \vspace{-20pt} \\ \includegraphics[scale=0.23]{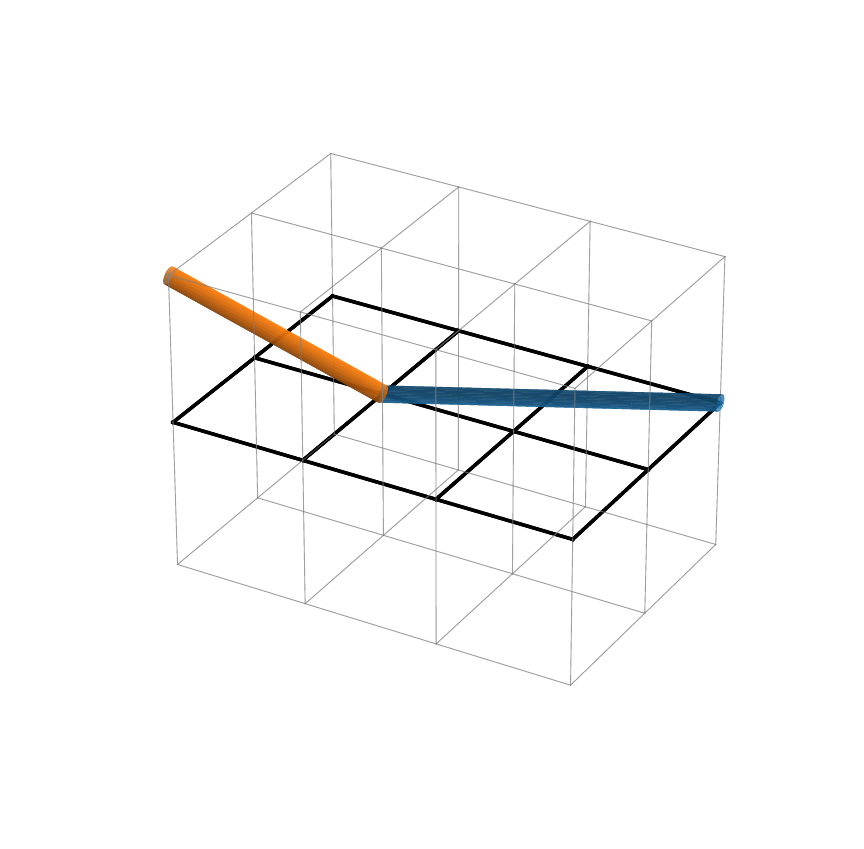}  \vspace{-10pt} \end{tabular}  \\[5pt]
%\hline
\end{tabular}
\end{center}
\caption{Continuation of Table~\ref{table:ops:3pi:2a}.}
\label{table:ops:3pi:2b}
\end{table}

\begin{table}
\begin{center}
\setlength\extrarowheight{5pt}
\begin{tabular}{    c | c || c | c | c || c || c  }
%\hline
& \ $\boldsymbol{d_1^2 \, d_2^2 \, d_3^2}$ \ & $\pi_{\vec{k_1}}$ & \ $\pi_{\vec{k_2}}$ & \ $\pi\pi_{\vec{k_{12}} \Lambda_{12}}$ ($I=2$) \ & $\pi_{\vec{k_3}}$ & \ \ momentum configurations\ \  \\[5pt]
\hline \hline
\ 1 \ &\ \ $\textbf{003}$ \ \ & $ \ \pi_{[000]}\ $ & $\ \pi_{[000]}\ $ & $ \ \pi\pi_{[000] A_1^+} \ $ & $\ \pi_{[111]} \ $ &  \begin{tabular}{c} \includegraphics[scale=0.15]{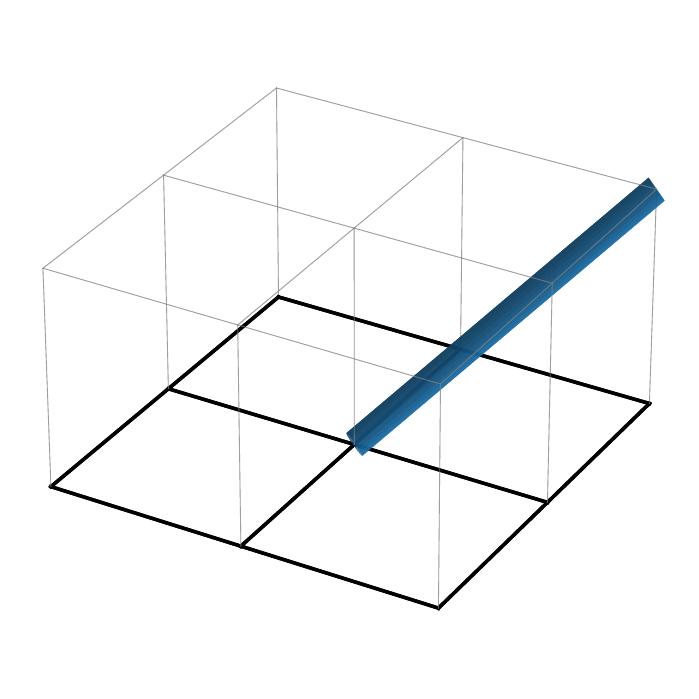} \vspace{-5pt} \end{tabular}   \\[5pt] \hline
\ 2 \ &\ \ $\textbf{012}$ \ \ & $ \ \pi_{[110]}\ $ & $\ \pi_{[000]}\ $ & $ \ \pi\pi_{[110] A_1} \ $ & $\ \pi_{[100]} \ $ & \begin{tabular}{c} \includegraphics[scale=0.15]{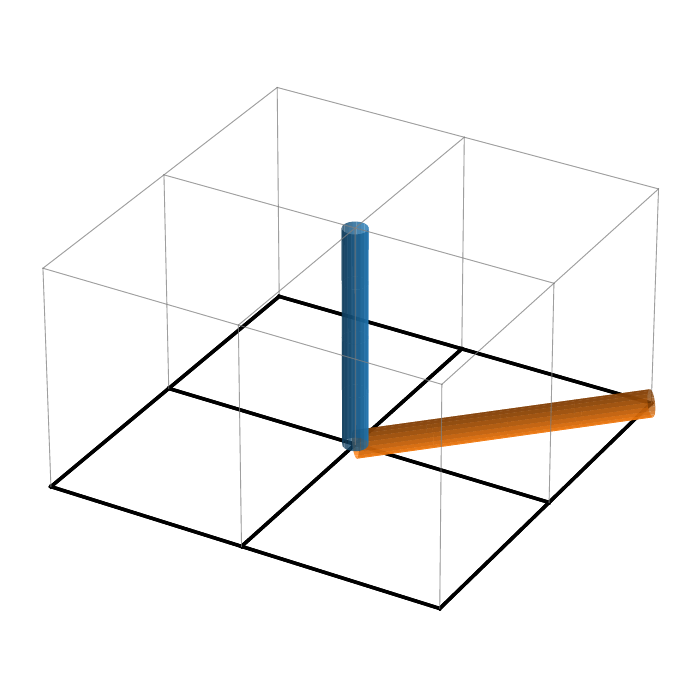} \vspace{-5pt} \end{tabular}  \\[5pt] \hline
\ 3 \ &\ \ $\textbf{111}$ \ \ & $ \ \pi_{[100]}\ $ & $\ \pi_{[100]}\ $ & $ \ \pi\pi_{[110] A_1} \ $ & $\ \pi_{[100]} \ $ &  \begin{tabular}{c} \includegraphics[scale=0.15]{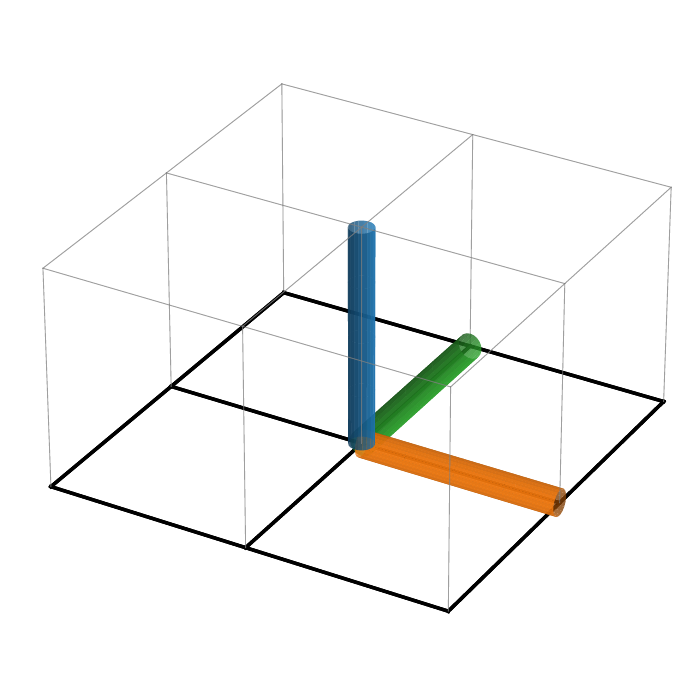} \vspace{-5pt} \end{tabular}  \\[5pt]\hline \hline
\ 4 \ &\ \ $\textbf{113}$ \ \ & $ \ \pi_{[100]}\ $ & $\ \pi_{[100]}\ $ & $ \ \pi\pi_{[000] A_1^+} \ $ & $\ \pi_{[111]} \ $ &  \begin{tabular}{c} \includegraphics[scale=0.15]{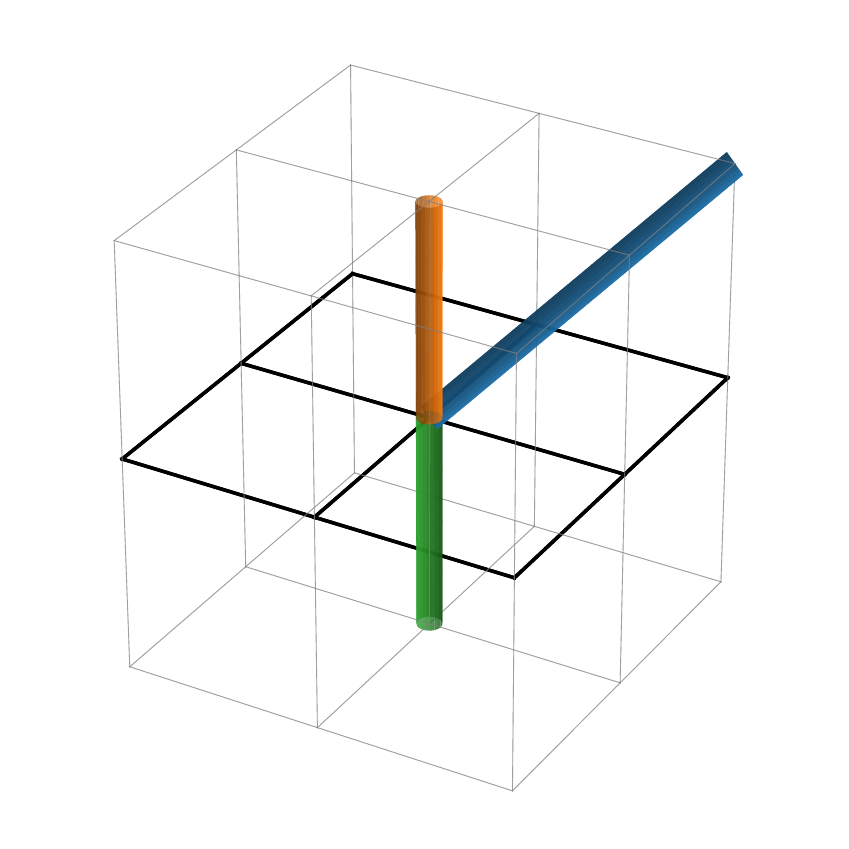} \vspace{-2pt} \end{tabular}  \\[5pt]
\ 5 \ &\ \ $\textbf{113}$ \ \ & $ \ \pi_{[100]}\ $ & $\ \pi_{[100]}\ $ & $ \ \pi\pi_{[200] A_1} \ $ & $\ \pi_{[111]} \ $ &  \begin{tabular}{c} \includegraphics[scale=0.15]{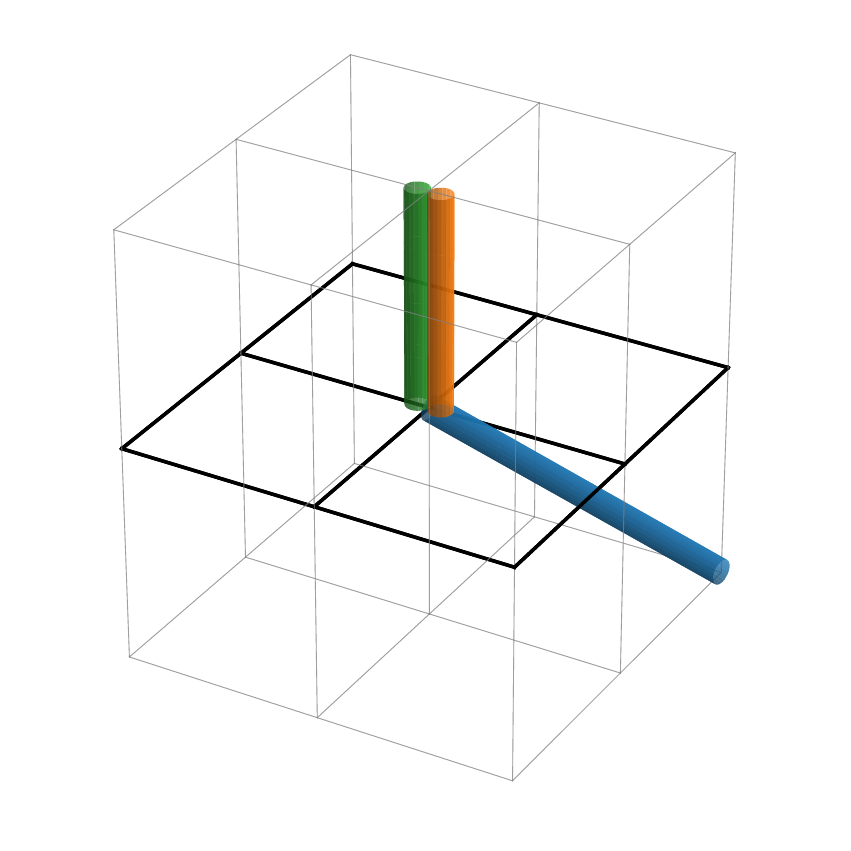} \vspace{-2pt} \end{tabular}  \\[5pt] \hline
\ 6 \ &\ \ $\textbf{122}$ \ \ & $ \ \pi_{[110]}\ $ & $\ \pi_{[100]}\ $ & $ \ \pi\pi_{[100] A_1} \ $ & $\ \pi_{[110]} \ $ &  \begin{tabular}{c} \includegraphics[scale=0.15]{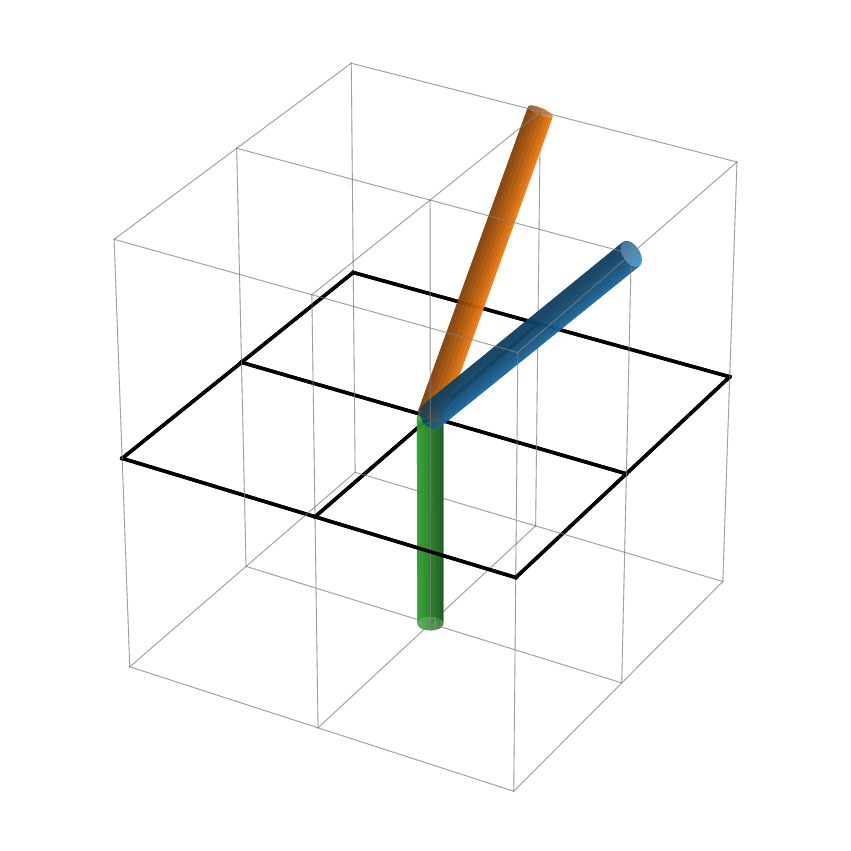}$^{\!\!\!\!\!\!1}$\vspace{-2pt} \end{tabular} \begin{tabular}{c} \includegraphics[scale=0.15]{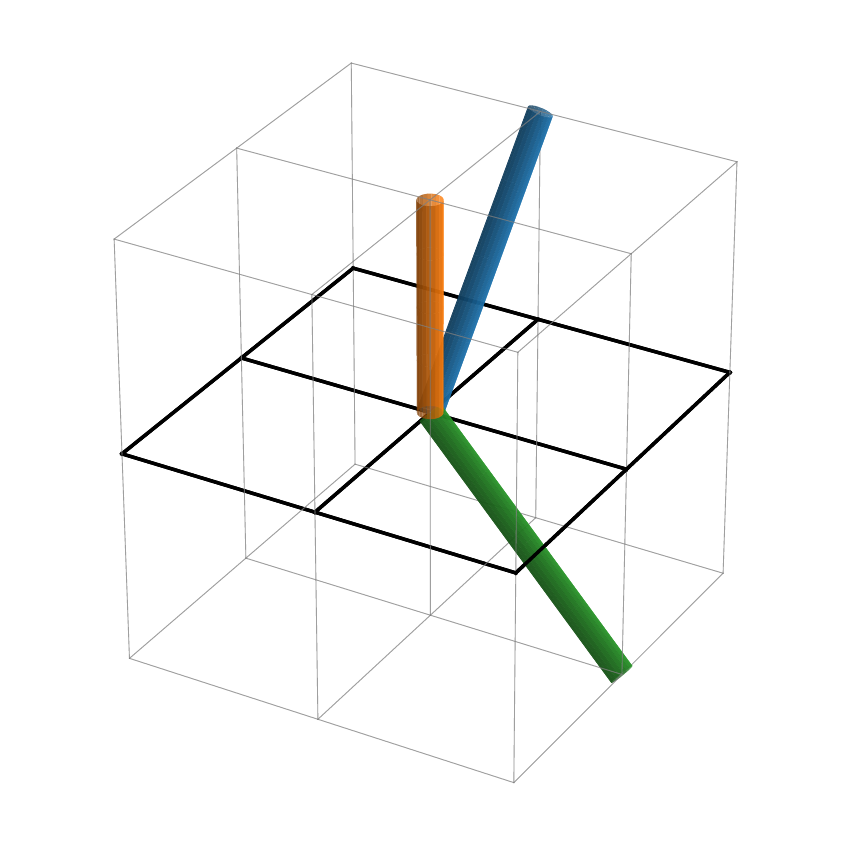}$^{\!\!\!\!\!\!2}$\vspace{-2pt} \end{tabular}  \\[5pt]
\ {7} \ &\ \ $\textbf{122}$ \ \ & $ \ \pi_{[110]}\ $ & $\ \pi_{[110]}\ $ & $ \ \pi\pi_{[110] A_1} \ $ & $\ \pi_{[100]} \ $ &  \begin{tabular}{c} \includegraphics[scale=0.15]{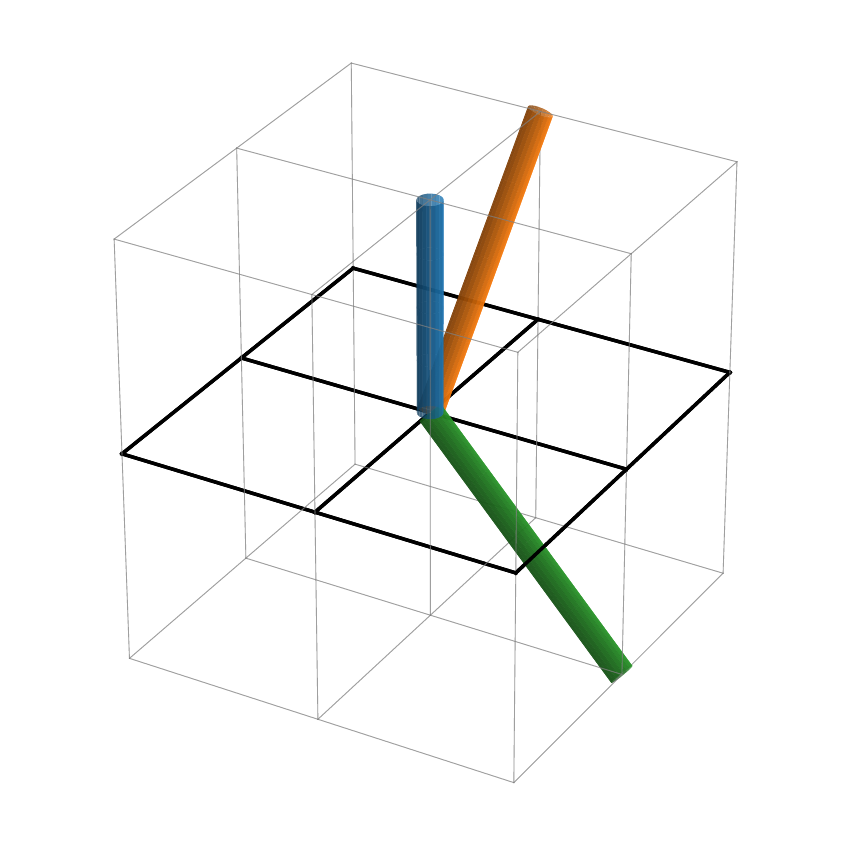}$^{\!\!\!\!\!\!2}$\vspace{-2pt} \end{tabular}  \\[5pt] %\hline 
\end{tabular}
\end{center}
\caption{As Table~\ref{table:ops:3pi:0} but for the $A_2$ irrep with overall momentum $\vec{P}=[111]$. Operators 1 to 7 are used on both the $20^3$ and $24^3$ volumes.}
\label{table:ops:3pi:3}
\end{table}

\end{document}